\def\(#1){(\call{#1})}
\def\call#1{{#1}}


\font\twelverm=cmr10 scaled 1200    \font\twelvei=cmmi10 scaled 1200
\font\twelvesy=cmsy10 scaled 1200   \font\twelveex=cmex10 scaled 1200
\font\twelvebf=cmbx10 scaled 1200   \font\twelvesl=cmsl10 scaled 1200
\font\twelvett=cmtt10 scaled 1200   \font\twelveit=cmti10 scaled 1200

\skewchar\twelvei='177   \skewchar\twelvesy='60


\def\twelvepoint{\normalbaselineskip=12.4pt
  \abovedisplayskip 12.4pt plus 3pt minus 9pt
  \belowdisplayskip 12.4pt plus 3pt minus 9pt
  \abovedisplayshortskip 0pt plus 3pt
  \belowdisplayshortskip 7.2pt plus 3pt minus 4pt
  \smallskipamount=3.6pt plus1.2pt minus1.2pt
  \medskipamount=7.2pt plus2.4pt minus2.4pt
  \bigskipamount=14.4pt plus4.8pt minus4.8pt
  \def\rm{\fam0\twelverm}          \def\it{\fam\itfam\twelveit}%
  \def\sl{\fam\slfam\twelvesl}     \def\bf{\fam\bffam\twelvebf}%
  \def\mit{\fam 1}                 \def\cal{\fam 2}%
  \def\tt{\twelvett}
  \def\nullspace{\nulldelimiterspace=0pt \mathsurround=0pt }
  \def\big##1{{\hbox{$\left##1\vbox to 10.2pt{}\right.\nullspace$}}}
  \def\Big##1{{\hbox{$\left##1\vbox to 13.8pt{}\right.\nullspace$}}}
  \def\bigg##1{{\hbox{$\left##1\vbox to 17.4pt{}\right.\nullspace$}}}
  \def\Bigg##1{{\hbox{$\left##1\vbox to 21.0pt{}\right.\nullspace$}}}
  \textfont0=\twelverm   \scriptfont0=\tenrm   \scriptscriptfont0=\sevenrm
  \textfont1=\twelvei    \scriptfont1=\teni    \scriptscriptfont1=\seveni
  \textfont2=\twelvesy   \scriptfont2=\tensy   \scriptscriptfont2=\sevensy
  \textfont3=\twelveex   \scriptfont3=\twelveex  \scriptscriptfont3=\twelveex
  \textfont\itfam=\twelveit
  \textfont\slfam=\twelvesl
  \textfont\bffam=\twelvebf \scriptfont\bffam=\tenbf
  \scriptscriptfont\bffam=\sevenbf
  \normalbaselines\rm}



\def\beginlinemode{\endmode
  \begingroup\parskip=0pt \obeylines\def\\{\par}\def\endmode{\par\endgroup}}
\def\beginparmode{\endmode
  \begingroup \def\endmode{\par\endgroup}}
\let\endmode=\par
{\obeylines\gdef\
{}}
\def\singlespace{\baselineskip=\normalbaselineskip}

\def\oneandahalfspace{\baselineskip=\normalbaselineskip
  \multiply\baselineskip by 3 \divide\baselineskip by 2}
\def\doublespace{\baselineskip=\normalbaselineskip \multiply\baselineskip by 2}

\newcount\firstpageno
\firstpageno=2
\footline={\ifnum\pageno<\firstpageno{\hfil}\else{\hfil\twelverm\folio\hfil}\fi}
\let\rawfootnote=\footnote		
\def\footnote#1#2{{\rm\singlespace\parindent=0pt\rawfootnote{#1}{#2}}}
\def\raggedcenter{\leftskip=4em plus 12em \rightskip=\leftskip
  \parindent=0pt \parfillskip=0pt \spaceskip=.3333em \xspaceskip=.5em
  \pretolerance=9999 \tolerance=9999
  \hyphenpenalty=9999 \exhyphenpenalty=9999 }
\def\dateline{\rightline{\ifcase\month\or
  January\or February\or March\or April\or May\or June\or
  July\or August\or September\or October\or November\or December\fi
  \space\number\year}}
\def\received{\vskip 3pt plus 0.2fill
 \centerline{\sl (Received\space\ifcase\month\or
  January\or February\or March\or April\or May\or June\or
  July\or August\or September\or October\or November\or December\fi
  \qquad, \number\year)}}


\hsize=6.5truein
\vsize=8.9truein
\parskip=\medskipamount
\twelvepoint		
\doublespace		
\overfullrule=0pt	



\def\title			
  {\null\vskip 3pt plus 0.2fill
   \beginlinemode \doublespace \raggedcenter \bf}

\def\author			
  {\vskip 3pt plus 0.2fill \beginlinemode
   \singlespace \raggedcenter}

\def\affil			
  {\vskip 3pt plus 0.1fill \beginlinemode
   \oneandahalfspace \raggedcenter \sl}

\def\abstract			
  {\vskip 3pt plus 0.3fill \beginparmode
   \doublespace \narrower ABSTRACT: }

\def\endtitlepage		
  {\endpage			
   \body}

\def\body			
  {\beginparmode}		

\def\head#1{			
  \filbreak\vskip 0.5truein	
  {\immediate\write16{#1}
   \raggedcenter \uppercase{#1}\par}
   \nobreak\vskip 0.25truein\nobreak}

\def\refto#1{$^{#1}$}		

\def\references			
  {\head{References}		
   \beginparmode
   \frenchspacing \parindent=0pt \leftskip=1truecm
   \parskip=8pt plus 3pt \everypar{\hangindent=\parindent}}

\gdef\refis#1{\indent\hbox to 0pt{\hss#1.~}}	

\gdef\journal#1, #2, #3, 1#4#5#6{		
    {\sl #1~}{\bf #2}, #3, (1#4#5#6)}		

\gdef\journ2 #1, #2, #3, 1#4#5#6{		
    {\sl #1~}{\bf #2}: #3, (1#4#5#6)}		

\def\refstylenp{		
  \gdef\refto##1{ [##1]}				
  \gdef\refis##1{\indent\hbox to 0pt{\hss##1)~}}	
  \gdef\journal##1, ##2, ##3, ##4 {			
     {\sl ##1~}{\bf ##2~}(##3) ##4 }}

\def\refstyleprnp{		
  \gdef\refto##1{ [##1]}				
  \gdef\refis##1{\indent\hbox to 0pt{\hss##1)~}}	
  \gdef\journal##1, ##2, ##3, 1##4##5##6{		
    {\sl ##1~}{\bf ##2~}(1##4##5##6) ##3}}

\def\endreferences{\body}

\def\figurecaptions		
  {\endpage
   \beginparmode
   \head{Figure Captions}
}

\def\endpage			
  {\vfill\eject}

\def\endpaper			
  {\endmode\vfill\supereject}

\def\endit
  {\endpaper\end}


\def\ref#1{Ref. #1}			
\def\Ref#1{Ref. #1}			

\def\frac#1#2{{\textstyle #1 \over \textstyle #2}}

\def\half{{\textstyle {1 \over 2}}}
\def\eg{{\it e.g.,\ }}

\def\sla{\raise.15ex\hbox{$/$}\kern-.57em}
\def\leaderfill{\leaders\hbox to 1em{\hss.\hss}\hfill}
\def\twiddle{\lower.9ex\rlap{$\kern-.1em\scriptstyle\sim$}}
\def\bigtwiddle{\lower1.ex\rlap{$\sim$}}
\def\gtwid{\mathrel{\raise.3ex\hbox{$>$\kern-.75em\lower1ex\hbox{$\sim$}}}}
\def\ltwid{\mathrel{\raise.3ex\hbox{$<$\kern-.75em\lower1ex\hbox{$\sim$}}}}
\def\square{\kern1pt\vbox{\hrule height 1.2pt\hbox{\vrule width 1.2pt\hskip 3pt
   \vbox{\vskip 6pt}\hskip 3pt\vrule width 0.6pt}\hrule height 0.6pt}\kern1pt}

\catcode`@=11
\newcount\r@fcount \r@fcount=0
\newcount\r@fcurr
\immediate\newwrite\reffile
\newif\ifr@ffile\r@ffilefalse
\def\w@rnwrite#1{\ifr@ffile\immediate\write\reffile{#1}\fi\message{#1}}

\def\writer@f#1>>{}
\def\referencefile{
  \r@ffiletrue\immediate\openout\reffile=\jobname.ref%
  \def\writer@f##1>>{\ifr@ffile\immediate\write\reffile%
    {\noexpand\refis{##1} = \csname r@fnum##1\endcsname = %
     \expandafter\expandafter\expandafter\strip@t\expandafter%
     \meaning\csname r@ftext\csname r@fnum##1\endcsname\endcsname}\fi}%
  \def\strip@t##1>>{}}

\def\citeall#1{\xdef#1##1{#1{\noexpand\cite{##1}}}}
\def\cite#1{\each@rg\citer@nge{#1}}	

\def\each@rg#1#2{{\let\thecsname=#1\expandafter\first@rg#2,\end,}}
\def\first@rg#1,{\thecsname{#1}\apply@rg}	
\def\apply@rg#1,{\ifx\end#1\let\next=\relax
\else,\thecsname{#1}\let\next=\apply@rg\fi\next}

\def\citer@nge#1{\citedor@nge#1-\end-}	
\def\citer@ngeat#1\end-{#1}
\def\citedor@nge#1-#2-{\ifx\end#2\r@featspace#1 
  \else\citel@@p{#1}{#2}\citer@ngeat\fi}	
\def\citel@@p#1#2{\ifnum#1>#2{\errmessage{Reference range #1-#2\space is bad.}%
    \errhelp{If you cite a series of references by the notation M-N, then M and
    N must be integers, and N must be greater than or equal to M.}}\else%
 {\count0=#1\count1=#2\advance\count1 by1\relax\expandafter\r@fcite\the\count0,%
  \loop\advance\count0 by1\relax
    \ifnum\count0<\count1,\expandafter\r@fcite\the\count0,%
  \repeat}\fi}

\def\r@featspace#1#2 {\r@fcite#1#2,}	
\def\r@fcite#1,{\ifuncit@d{#1}
    \newr@f{#1}%
    \expandafter\gdef\csname r@ftext\number\r@fcount\endcsname%
                     {\message{Reference #1 to be supplied.}%
                      \writer@f#1>>#1 to be supplied.\par}%
 \fi%
 \csname r@fnum#1\endcsname}
\def\ifuncit@d#1{\expandafter\ifx\csname r@fnum#1\endcsname\relax}%
\def\newr@f#1{\global\advance\r@fcount by1%
    \expandafter\xdef\csname r@fnum#1\endcsname{\number\r@fcount}}

\let\r@fis=\refis			
\def\refis#1#2#3\par{\ifuncit@d{#1}
   \newr@f{#1}%
   \w@rnwrite{Reference #1=\number\r@fcount\space is not cited up to now.}\fi%
  \expandafter\gdef\csname r@ftext\csname r@fnum#1\endcsname\endcsname%
  {\writer@f#1>>#2#3\par}}

\def\ignoreuncited{
   \def\refis##1##2##3\par{\ifuncit@d{##1}%
     \else\expandafter\gdef\csname r@ftext\csname r@fnum##1\endcsname\endcsname%
     {\writer@f##1>>##2##3\par}\fi}}

\def\r@ferr{\endreferences\errmessage{I was expecting to see
\noexpand\endreferences before now;  I have inserted it here.}}
\let\r@ferences=\references
\def\references{\r@ferences\def\endmode{\r@ferr\par\endgroup}}

\let\endr@ferences=\endreferences
\def\endreferences{\r@fcurr=0
  {\loop\ifnum\r@fcurr<\r@fcount
    \advance\r@fcurr by 1\relax\expandafter\r@fis\expandafter{\number\r@fcurr}%
    \csname r@ftext\number\r@fcurr\endcsname%
  \repeat}\gdef\r@ferr{}\endr@ferences}


\let\r@fend=\endpaper\gdef\endpaper{\ifr@ffile
\immediate\write16{Cross References written on []\jobname.REF.}\fi\r@fend}

\catcode`@=12

\citeall\refto		
\citeall\ref		%
\citeall\Ref		%
\catcode`@=11
\newcount\tagnumber\tagnumber=0

\immediate\newwrite\eqnfile
\newif\if@qnfile\@qnfilefalse
\def\write@qn#1{}
\def\writenew@qn#1{}
\def\w@rnwrite#1{\write@qn{#1}\message{#1}}
\def\@rrwrite#1{\write@qn{#1}\errmessage{#1}}

\def\taghead#1{\gdef\t@ghead{#1}\global\tagnumber=0}
\def\t@ghead{}

\expandafter\def\csname @qnnum-3\endcsname
  {{\t@ghead\advance\tagnumber by -3\relax\number\tagnumber}}
\expandafter\def\csname @qnnum-2\endcsname
  {{\t@ghead\advance\tagnumber by -2\relax\number\tagnumber}}
\expandafter\def\csname @qnnum-1\endcsname
  {{\t@ghead\advance\tagnumber by -1\relax\number\tagnumber}}
\expandafter\def\csname @qnnum0\endcsname
  {\t@ghead\number\tagnumber}
\expandafter\def\csname @qnnum+1\endcsname
  {{\t@ghead\advance\tagnumber by 1\relax\number\tagnumber}}
\expandafter\def\csname @qnnum+2\endcsname
  {{\t@ghead\advance\tagnumber by 2\relax\number\tagnumber}}
\expandafter\def\csname @qnnum+3\endcsname
  {{\t@ghead\advance\tagnumber by 3\relax\number\tagnumber}}

\def\equationfile{%
  \@qnfiletrue\immediate\openout\eqnfile=\jobname.eqn%
  \def\write@qn##1{\if@qnfile\immediate\write\eqnfile{##1}\fi}
  \def\writenew@qn##1{\if@qnfile\immediate\write\eqnfile
    {\noexpand\tag{##1} = (\t@ghead\number\tagnumber)}\fi}
}

\def\callall#1{\xdef#1##1{#1{\noexpand\call{##1}}}}
\def\call#1{\each@rg\callr@nge{#1}}

\def\each@rg#1#2{{\let\thecsname=#1\expandafter\first@rg#2,\end,}}
\def\first@rg#1,{\thecsname{#1}\apply@rg}
\def\apply@rg#1,{\ifx\end#1\let\next=\relax%
\else,\thecsname{#1}\let\next=\apply@rg\fi\next}

\def\callr@nge#1{\calldor@nge#1-\end-}
\def\callr@ngeat#1\end-{#1}
\def\calldor@nge#1-#2-{\ifx\end#2\@qneatspace#1 %
  \else\calll@@p{#1}{#2}\callr@ngeat\fi}
\def\calll@@p#1#2{\ifnum#1>#2{\@rrwrite{Equation range #1-#2\space is bad.}
\errhelp{If you call a series of equations by the notation M-N, then M and
N must be integers, and N must be greater than or equal to M.}}\else%
 {\count0=#1\count1=#2\advance\count1 by1\relax\expandafter\@qncall\the\count0,%
  \loop\advance\count0 by1\relax%
    \ifnum\count0<\count1,\expandafter\@qncall\the\count0,%
  \repeat}\fi}

\def\@qneatspace#1#2 {\@qncall#1#2,}
\def\@qncall#1,{\ifunc@lled{#1}{\def\next{#1}\ifx\next\empty\else
  \w@rnwrite{Equation number \noexpand\(>>#1<<) has not been defined yet.}
  >>#1<<\fi}\else\csname @qnnum#1\endcsname\fi}

\let\eqnono=\eqno
\def\eqno(#1){\tag#1}
\def\tag#1$${\eqnono(\displayt@g#1 )$$}

\def\aligntag#1$${\gdef\tag##1\\{&(\displayt@g##1 )\cr}\eqalignno{#1\\}$$
  \gdef\tag##1$${\eqnono(\displayt@g##1 )$$}}

\def\eqalignno#1{\displ@y \tabskip\centering
  \halign to\displaywidth{\hfil$\displaystyle{##}$\tabskip\z@skip
    &$\displaystyle{{}##}$\hfil\tabskip\centering
    &\llap{$\displayt@gpar##$}\tabskip\z@skip\crcr
    #1\crcr}}

\def\displayt@gpar(#1){(\displayt@g#1 )}

\def\displayt@g#1 {\rm\ifunc@lled{#1}\global\advance\tagnumber by1
        {\def\next{#1}\ifx\next\empty\else\expandafter
        \xdef\csname @qnnum#1\endcsname{\t@ghead\number\tagnumber}\fi}%
  \writenew@qn{#1}\t@ghead\number\tagnumber\else
        {\edef\next{\t@ghead\number\tagnumber}%
        \expandafter\ifx\csname @qnnum#1\endcsname\next\else
        \w@rnwrite{Equation \noexpand\tag{#1} is a duplicate number.}\fi}%
  \csname @qnnum#1\endcsname\fi}

\def\ifunc@lled#1{\expandafter\ifx\csname @qnnum#1\endcsname\relax}

\let\@qnend=\end\gdef\end{\if@qnfile
\immediate\write16{Equation numbers written on []\jobname.EQN.}\fi\@qnend}

\catcode`@=12

\input miniltx
\input graphicx.sty


\def\cf{{\it cf.~}}
\def\references{\head{\noindent\bf References}\beginparmode\frenchspacing
               \leftskip=1truecm\parskip=8pt plus 3pt 
		\everypar{\hangindent=3em}}

\line{\hfill UCSBTH-91-15}
\singlespace
\vskip .13 in
\centerline{\bf CLASSICAL EQUATIONS FOR QUANTUM SYSTEMS} 
\vskip .26 in
\centerline{\bf Murray Gell-Mann}
\centerline{\sl Lauritsen Laboratory,
  California Institute of Technology}
\centerline{\sl  Pasadena, CA 91125}
\vskip .08 in
\centerline{and}
\vskip .08 in
\centerline{\bf James B. Hartle}
\centerline{\sl Department of Physics,
 University of California}
\centerline{\sl  Santa Barbara, CA 93106}
\vskip .26 in

{
\midinsert
\centerline{\bf ABSTRACT}

The origin of the phenomenological deterministic laws that 
approximately govern the quasiclassical domain of familiar experience is
considered in the context of the quantum mechanics of closed systems
such as the universe as a whole.  A formulation of quantum mechanics is
used that predicts probabilities for the individual members of a set of
alternative coarse-grained histories that {\it decohere},
which means that
there is negligible quantum interference between the individual
histories in the set.  We investigate the requirements for coarse
grainings to yield decoherent sets of histories that are
quasiclassical, i.e. such that the individual histories
obey, with high probability, effective classical equations of motion
interrupted continually by small fluctuations and occasionally by large
ones.  We discuss these requirements generally but study them
specifically for coarse grainings of the type that follows a
distinguished subset of a complete set of variables while ignoring the
rest.  More coarse graining is needed to achieve decoherence than would
be suggested by naive arguments based on the uncertainty principle. 
Even coarser graining is required in  the
distinguished variables for them to have the necessary inertia to approach
classical predictability in the presence of the noise consisting of the
fluctuations that typical mechanisms of decoherence produce. We describe
the derivation of phenomenological equations of motion explicitly for a
particular class of models.  Those models assume 
configuration space and a fundamental Lagrangian that is the difference
between a kinetic energy quadratic in the velocities and a
potential energy. The distinguished variables are
taken to be
a fixed subset
of co\"ordinates of configuration space. 
  The initial density matrix of the closed system is assumed
to factor into a product of a density matrix in the distinguished subset
and another in the rest of the coordinates.  With
these restrictions, we improve
the derivation from quantum mechanics of the phenomenological equations
of motion governing a quasiclassical domain in the following respects:
Probabilities of the correlations in time that define equations of
motion are explicitly considered.  Fully non-linear cases are studied.
Methods are exhibited for finding the form of the phenomenological
equations of motion even when these are only distantly related to those
of the fundamental action.  The demonstration of the connection between
quantum-mechanical causality and causality in classical phenomenological
equations of motion is generalized. The connections among decoherence,
noise, dissipation, and the amount of coarse graining necessary to
achieve classical predictability are investigated quantitatively. Routes
to removing the restrictions on the models in order
 to deal with more realistic
coarse grainings are
described.
 
\endinsert
}
\vfill\eject
\doublespace

\centerline{\bf I.  Introduction}
\vskip .13 in
In a universe governed at a fundamental level by quantum-mechanical
laws, characterized by indeterminacy and distributed probabilities, what
is the origin of the phenomenological, deterministic laws that
approximately  govern
the quasiclassical domain of everyday experience?  What features and
limitations of these classical laws can be traced to their underlying
quantum-mechanical origin? This paper addresses such questions in the
context of the quantum mechanics of closed systems --- most
realistically and generally the universe as a whole.

It is a familiar notion in the quantum mechanics of simple ``measured''
systems 
 that some coarseness in their description is needed if
they are to approximate classical behavior.  The Heisenberg uncertainty
principle, for example, limits the accuracy with which position and
momentum can be specified simultaneously.  Successions of such suitably
crude measurements of position and
momentum can be correlated by the classical equations of motion
following from the fundamental action of the system, provided there is a
suitable initial state --- typically a narrow wave packet.

In discussing the quasiclassical domain of familiar experience, however,
we are dealing with a much more general situation than is envisioned by
elementary analyses of the above type.  We are concerned, first of all,
with the classical features that may be exhibited by the behavior of the
closed system irrespective of whether those features are receiving attention
from ``observers''. 
We should be able to deal, for example, with the
classical behavior of the moon whether or not any ``observer'' is  
looking at it.
Second, we are necessarily concerned, not just with classical behavior
exhibited by correlations among events at a few times of our choosing, but
also with
the classical behavior of whole orbits in as refined a description of
the system  as is
possible.  Third, we are concerned with phenomenological equations of
motion, the form of which may be only very indirectly related to that of the
fundamental action.  The fundamental action, after all, may be that of 
heterotic superstring  
theory,  
 while the equations of motion (such as the Navier-Stokes equation) governing
the familiar quasiclassical domain involve such quantities as the
averages of the densities of field energy and momenta over volumes very
much larger than the Planck scale!  Further, the applicability
of effective classical
equations of motion may be branch-dependent, that is, contingent on
events that have happened.  The classical equations governing the
motion of the moon, for example, are contingent on its actually having
formed in the early history of the solar system.  In such general
situations, simply identifying the form of the phenomenological
 classical equations of
motion becomes an important problem.

It is a characteristic feature of the general situations described above
that a much coarser 
graining is needed for
quasiclassical behavior would be naively suggested by arguments
based on the uncertainty principle.  As we shall argue below, a large
amount of coarse graining is needed to accomplish decoherence, which
is an important ingredient of quasiclassical behavior as well as a
sufficient (stronger than necessary) condition
for assigning probabilities to the coarse-grained histories of the closed
system. Further coarse graining is then necessary to achieve
the ``inertia'' required for 
approximate predictability in the presence of the noise from the
fluctuations that typical 
mechanisms of
decoherence involve. All this coarse graining has
important consequences for the form of the effective classical
equations of motion.  Their form may be as much influenced by the character
of the coarse graining and the mechanisms of decoherence as by the
fundamental equations of motion.  The effective classical equations of
motion necessarily include  phenomena like dissipation
arising from the mechanisms that produce decoherence.  This
paper is concerned with general methods of deriving the form of the
phenomenological classical equations of motion 
and with the description of the noise that causes
deviations from those equations  
and from classical predictability.
As a result of the coarse
graining, the noise includes the effects of classical (typically
statistical-mechanical) fluctuations as well as quantum fluctuations,
and these effects are mixed.  The resulting indeterminacy, as indicated
above, goes far
beyond the elementary indeterminacy of the Heisenberg uncertainty
principle.  An accurate framework for prediction may be
achieved by
incorporating a generalized
Langevin force that represents this noise into the
classical equation of motion.  Our paper is therefore concerned 
with the derivation of the general form and distribution of such
forces as well as with the equation itself.

It is known, of course, that even in the classical deterministic limit
one can encounter in non-linear systems the phenomenon of chaos, in
which the sensitivity of the
outcome to the initial conditions is exponential in time.  In
the presence of chaos, even small fluctuations (including quantum
fluctuations) can be amplified to
produce large uncertainties in later behavior.  A treatment of this
combined effect of classical chaos and of indeterminacy arising from
quantum mechanics, including the associated coarse graining, has often
proved elusive in discussions of quantum chaos, but is amenable to
analysis by the methods we shall describe [\cite{1}].
\vskip .26 in
\centerline{\bf II. Decoherence, Inertia, and Equations of Motion}
\vskip .13 in
In this section we give a qualitative discussion of the role of
decoherence in  the derivation of phenomenological classical equations of
motion.  This will serve to review some aspects of the quantum mechanics of
closed systems, motivate the subsequent mathematical derivations, 
and
make connections with earlier work known to us.  For the quantum
mechanics of closed systems, we follow our
discussion in [\cite {2,33,3}], where references to the earlier literature
may be found.

Most generally, quantum mechanics predicts the probabilities of the
individual members of a set of alternative, coarse-grained, time
histories of a closed system.  By a coarse-grained history we mean, for
example, one for which not every variable is specified and 
those that are specified are not fixed at every time or with
arbitrary precision. It is
evident why coarse-grained histories are of interest to us as observers
of the universe.  Our observations fix only a tiny fraction of the
variables describing the universe and fix those only very imprecisely.  As
observers we therefore necessarily deal with a very coarse-grained
description of the closed system in which we live.   However, from the
theoretical point of view, it is not necessary that the description be
so very coarse-grained or that the coarse graining be so dependent on
us.  There is a more
fundamental reason for interest in sets of coarse-grained alternative
histories:  In the quantum mechanics of closed systems, probabilities may
be assigned only to those sets of histories for which there is
negligible quantum mechanical interference between the individual
histories in the set (given the system's Hamiltonian and
initial quantum state) [\cite {5, 6, 2}]. We shall define various kinds of
decoherence\footnote{$^1$}{The term ``decoherence'' is used in
several different ways in
the literature.  We have used the term to refer to a
property of a set of alternative histories of a closed system.
Specifically a set of coarse-grained histories decoheres when there is
negligible interference between the individual histories in the set as
measured by one of the several conditions discussed in this Section.  In
the literature the term ``decoherence'' has also been used to refer to
the decay in time of the off-diagonal elements of a reduced density
matrix defined by a coarse-graining of variables at a single moment of
time, for example, the density matrix defined by (4.21).  These two
notions are not exactly the same.  A
reduced density matrix can be defined for those special coarse grainings
that distinguish a fixed set of co\"ordinates.  However, the vanishing
of the off-diagonal
elements of this reduced density matrix at a succession of times is not
identical with the decoherence of the corresponding histories, as will
be discussed in Section IV.  Yet the two notions
are not unconnected.  In the accessible, although unrealistic, model coarse
grainings of the kind studied in this paper, where both notions are
defined, typical {\it mechanisms} of decoherence ensure the validity of
both.  (See, e.g. [\cite {3}], Section II.6.4)~~ A general notion of
mechanism of decoherence can be defined [\cite {19}] that generalizes
the reduced density matrix definition of decoherence in the context of
the
decoherence of histories and characterizes more precisely how they are
connected.\hfill\break
\indent It would be clearer to use the terms ``decoherence of
histories'' when referring to one notion and ``decoherence of density
matrices'' when referring to the other.  In this paper by
``decoherence'' we always mean the decoherence of sets of histories as
defined precisely in this section.} [\cite{33}],
 all of which imply the vanishing of this
interference and for all of which coarse graining is necessary.
  An arbitrarily fine-grained description of the universe
would reveal the phase correlations between histories, while  in a
coarse-grained description they may be absent.  The probabilities of
decoherent sets of coarse-grained alternative histories constitute the
useful predictions in
quantum mechanics.

Among the coarse-grained decohering sets of alternative histories of
this universe must be the sets that describe the quasiclassical domain
of familiar experience.  These consist of histories that, for the most
part, are defined by ranges of values of ``quasiclassical operators''
correlated in time with high probability by classical phenomenological
laws.  We have discussed elsewhere the problem  of distinguishing such
quasiclassical domains from all other decohering sets of histories and
in particular the problem of deriving the form of the ``quasiclassical
operators'' that characterize them [\cite {2,33}].  Here,  we want to focus 
on a more
specific and less general question.  That is the question of the
derivation and form of the classical phenomenological equations of
motion {\it assuming} coarse graining is given that leads to
decoherence.

A simple form of coarse graining consists of averaging over some
variables and following the remaining ones.  Let us call the two classes of
variables ``ignored'' and ``distinguished'' respectively.  Widely occurring
mechanisms of decoherence involve the rapid dispersal of 
quantum-mechanical phase information among the ignored variables as they
interact with the distinguished ones.  Thus, for example, sets of histories 
that
distinguish the center-of-mass positions of bodies as light as a grain of
dust can be very efficiently ``decohered'' by the collisions of the bodies
with the omnipresent $3^\circ$K cosmic background radiation [\cite {7}].  Such
interactions can be expected to produce
deviations from the predictability that characterizes classical
behavior.  That is, they produce noise. For quasiclassical behavior,
such that the distinguished variables mostly resist the noise and follow
approximately classical predictable paths, a high inertia is required for
the distinguished variables.  In general, therefore, we
expect that a {\it much} coarser graining is necessary to achieve long
stretches of 
predictable behavior than is needed to achieve mere decoherence, and
decoherence, as we remarked, requires much coarser graining than is
needed for mere consistency with the uncertainty principle.  Furthermore,
mechanisms that produce decoherence naturally lead to processes, 
such as dissipation, that
are necessarily 
included in the equations of motion that describe predictable
behavior.  There are thus  connections among decoherence, noise, 
dissipation, and the amount of coarse graining necessary  to achieve
classical predictability.  This paper explores those
connections.

The habitually decohering quasiclassical operators that characterize our
everyday quasiclassical domain include such ``hydrodynamic'' variables
as the averages, over suitable volumes, of densities of energy,
momentum, and other conserved or approximately conserved quantities.
Such coarse grainings are not of the simple type described above in
which the co\"ordinates of a configuration space are separated once and
for all into a set ``distinguished'' by the coarse graining and a set
that is ``ignored''.  First, the coarse grainings corresponding to these
averages are not defined by ranges of {\it co\"ordinates}.  In addition,
realistic coarse grainings are in general branch-dependent, meaning in
this case, that the volumes over which the averages can be usefully taken
are contingent on prior events in specific histories [\cite{2}].
However, since the simple types of coarse graining are more easily
analysed than the realistic ones, we shall begin our discussion with a
class of model problems that are based on distinguished and ignored
co\"ordinates and later return to how to generalize our results to more
realistic situations.

Our central result is a derivation of the classical
equation of motion, including effective forces and noise, for a specific
type of coarse graining of the histories of a class of quantum systems.
Each system is assumed to have a Lagrangian that is the
difference between a kinetic energy quadratic in the velocities and
a potential energy independent of velocities but allowed to be
fully non-linear.
Coarse grainings are considered that distinguish a fixed subset of the
co\"ordinates of the system's configuration space while ignoring the
rest.  The initial density matrix of the closed system is assumed to
factor into a product of a density matrix in the distinguished variables
and another density matrix in the ignored variables.  We show that when
such sets of coarse-grained histories decohere, the quantum-mechanical
probabilities of the individual coarse-grained histories can be
represented as the probabilities of the histories of a
{\it classical} system evolving from distributed initial
conditions under the action of a stochastic force.  The initial
conditions follow a Wigner distribution derivable from the initial
density matrix.  The distribution of the ``total force'' 
 (including the
inertial term ``$-m{\bf a}$'') is a kind of generalization to histories
of the 
Wigner distribution; 
 it is derivable from the decoherence functional that measures
quantum coherence.  Neither of these distributions is in general
positive.  The effective classical equation of motion is the condition
that the expected value of the total force vanish.  When the noise
arising
from the stochastic force is almost negligible, so that the deviations
from the effective classical equation of motion are small, then we
achieve classical predictability.  The noise, whether small or not, can
be treated by incorporating a generalized Langevin force into the
effective equation of motion.  The distribution of the ``total force''
can also be regarded as a distribution of this Langevin force; it is, in
general, non-Gaussian and, as mentioned above, not necessarily positive,
as is to be expected from the quantum-mechanical nature of the problem.

Our characterization of the effective classical equation of motion and
the Langevin noise distribution leads to a method of identifying the
equation of motion and a systematic expansion procedure for calculating
the noise distribution.  These techniques are not restricted to linear
problems or Gaussian noise.
The key to the method is that the decoherence functional, which depends
on a pair of coarse-grained histories, can be expressed as a path
integral over a quantity $\exp (i A)$, where we can expand {\it A} in a
power series in the difference between the distinguished co\"ordinates in
one history and those in the other.  For the familiar mechanisms of
decoherence, to which we alluded above, the higher order terms in the
expansion are expected to be small, and we can start by retaining only
the linear and quadratic terms, which permit us to treat nonlinear
equations of motion with Gaussian noise.  The higher order terms then
give the non-Gaussian corrections to the distribution of the Langevin
force.

As mentioned above, we would like to treat 
still more general and more realistic problems,
in which we escape the limitation to coarse graining that begins with
distinguished and ignored co\"ordinate variables, as well as the
restriction to factored initial density matrices and the prohibition of
velocity-dependent potentials.  We discuss at length some ideas of how
to free ourselves from these limitations.

The details of the models 
are described in Section III.  The various types of
decoherence are discussed in Section IV.  Section V introduces
distribution functionals for the ``total force'' described above and the
representation of quantum-mechanical probabilities in terms of them.  In
Section VI the equation of motion and distribution of noise are
calculated for the well-known case of linear systems.  The explicit
generalization to non-linear cases is given in Section VII. The
comparison of these results with the corresponding classical analyses is
discussed in Section VIII.  Section IX describes routes to more general
coarse grainings and Section X contains some brief conclusions.

There have been, of course, a great many discussions of the derivation
of classical behavior from quantum mechanics, and it is perhaps
appropriate to offer a few comments on the similarities and differences
between the present discussion and that great body of literature.  As
already mentioned, we aim beyond the elementary discussions of the
classical behavior of {\it measured} systems based on Ehrenfest's theorem, the
WKB approximation, or the Wigner distribution.  Such analyses do not usually
treat noise, cover the {\it effective} classical equations of motion
including such phenomena as dissipation, consider coarse grainings
(other than very obvious ones) 
or deal seriously with  
the probabilities of time histories by which roughly predictable
quasiclassical behavior is inevitably defined.  (Analyses based on the
steepest-descent approximation to Feynman's path integral {\it do} consider
histories but do not typically address the other issues.)

Derivations of the equations of hydrodynamics from statistical physics,
as, for example, in [\cite{8,9,10, 41}], necessarily include phenomena like
dissipation.  However, those accounts known to us derive equations of
motion for the {\it expectation values} of hydrodynamic variables.  In
quantum mechanics a system may be said to obey a classical equation of
motion when the probability is high for the correlations {\it in time}
that the equation of motion requires. 
For example, the center of mass of the earth can be said to obey
Newton's law of motion when the probability is
high that successive determinations of the position of the center of
mass of the earth will be correlated according to that law.
  A complete derivation of classical equations of motion for
quantum systems therefore requires the consideration of the
probabilities for time histories, not just the study of the evolution of
expected values.  That is important, because the requirements of
decoherence include the restriction on
which sets of alternative histories may be assigned
probabilities, while there is no such restriction on which expected
values may be studied.  Furthermore, it is through the study of the
probabilities for histories that the probabilities for the inevitable
deviations from classical predictability are most directly 
assessed.  That is, only
through a study of the probabilities of time histories can we accurately
characterize the mixed quantum and classical-statistical noise that
characterizes those deviations.

The derivation of classical-statistical equations incorporating both
classical determinism and stochastic noise has been extensively
discussed for linear systems.  Evolution equations for probability
distributions on phase space were derived (in certain approximations)
 from the evolution equation
for the Wigner distribution by Caldeira and
Leggett [\cite {11}] and more recently by Hu, Paz, and Zhang [\cite{34}]
and by Zurek [\cite{35}].  Langevin equations have also been extensively
discussed for linear systems.
Caldeira and Leggett [\cite {11}], for example, discuss such equations,
using techniques developed by Feynman and Vernon [\cite {12}], and review
earlier efforts.  The extensive investigation of linear systems in the
quantum optics literature is reviewed in [\cite {13}] and  treatments
from the point of view of statistical mechanics can be found in [\cite
{14, 41}].  None of that work, however, explicitly considers the
probabilities of histories, by which classical behavior is necessarily
defined.\footnote{$^2$}{Recently, Dowker and
Halliwell [\cite{40}]
have studied the decoherence of histories in explicit linear
models and also, in effect, derived their classical equations of
motion, although not including a description of dissipation and noise.}
  Neither is there explicit consideration of the decoherence of
histories, which is a prerequisite for the calculation of probabilities
in the quantum mechanics of a closed system.  By using decoherence and
the probabilities of histories we not only recover the standard Langevin
equations for linear systems but also we can generalize these results to
non-linear cases. 
\taghead{3.}
\vskip .26 in
\centerline{\bf III. Model Systems and Model Coarse Graining}
\vskip .13 in
This paper is concerned with classical predictibility in a certain class
of model quantum-mechanical systems described by particular classes of
model coarse grainings.  In this Section we shall specify these models
and coarse grainings and review the theoretical framework in which the
decoherence and the probabilities of histories are defined. We shall
give a simplified version of this theoretical framework and later
discuss how
it is related to more general ideas in Section
IV.  That way the reader only interested in the models and not
in these general connections can
proceed immediately from Section III to Section V.
We shall be brief.  For greater detail the
reader may consult [\cite {2, 33}],  and the references 
to the earlier
literature found therein.

We consider the quantum mechanics of a closed system
in the approximation that gross fluctuations in the
geometry of spacetime may be neglected.  A background spacetime is
thus fixed and, in particular, there is a well defined time, $t$. The
usual apparatus of Hilbert space, states, operators, etc.~then may 
 be applied in the quantum mechanical process of prediction.
The fundamental dynamics of the system is governed by an action $S$ or
its equivalent Hamiltonian $H$.  The initial condition is specified by a
density matrix $\rho$.

The most refined possible description of a closed system makes use of a
set of fine-grained histories.  The most familiar set of fine-grained
histories are the possible paths, $q^\beta(t)$, in a configuration space
of generalized co\"ordinates that completely describe the system.  These
paths are the {\it single-vaued} functions $q^\beta(t)$ on a fixed time
interval, say $[0,T]$. For example, in a system of scalar fields each
$q^\beta(t)$ might be the value of a field at a different spatial point
considered as a function of time.  (Our description is thus not
restricted to non-relativistic physics.)~~Configuration space
fine-grained histories are the starting point for sum-over-histories
formulations of quantum mechanics and for the model coarse grainings we
shall mostly consider in this paper.  More general possibilities are
discussed in the next Section.

A partition of an entire set of fine-grained histories into
exhaustive and exclusive classes defines a set of {\it coarse-grained
histories};  each class is an individual coarse-grained history.
The individual coarse-grained histories in an exhaustive set may be
grouped into new exclusive sets.  That is an operation of
further {\it coarse
graining}, yielding a coarser-grained set of alternative histories.  The
inverse operation of dividing a set of coarse-grained histories into
smaller classes of the fine-grained histories is an operation called
{\it fine graining}.  Sets of coarse-grained histories are partially
ordered with respect to the operations of fine and coarse graining.
A rich variety of coarse grainings is possible.
As we mentioned before, 
the histories of the quasiclassical
domain of everyday experience, for example, are defined by coarse
grainings utilizing 
ranges of values of averages over  suitable spatial 
regions of such ``hydrodynamic''
variables as the densities of energy, momentum, charges, and currents.
Such realistic coarse grainings are in general branch-dependent, that
is, contingent on which of many possible events have happened.  

Because of their branch dependence, and because of their  indirect 
relation to the
fundamental fields, realistic coarse grainings are not as theoretically
tractable as some model coarse grainings that can be studied. In this
paper we shall study a  
familiar and instructive class of model coarse grainings
in which 
 the co\"ordinates,
$q^\beta$, of configuration space are divided into ones $x^a$ that are
distinguished by the coarse graining
 and the remaining ones $Q^A$, which are ignored.  (For
example, in a simplified model of a universe of particles, the $x^a$ might label
 the center-of-mass 
positions and 
orientations of a group
of massive bodies such as the planets  and the $Q^A$ would then be  
all the rest of the co\"ordinates, 
including the internal
co\"ordinates of the bodies' constituents and the co\"ordinates of 
gas molecules,  etc.~that
interact with the planets.)  Coarse-grained histories of this type are labeled
by partitions of the paths $x^a(t)$.  An individual coarse-grained
history consists of a path $x^a(t)$ 
 along with {\it all} possible paths $Q^A(t)$.

Further coarse graining of the classes
of histories labeled by the paths $x^a(t)$ can be defined by sets of
intervals exhausting the whole range of the $x^a$ at a discrete sequence of
times $t_1<t_2<\cdots<t_n$.  These correspond to intervals on the entire
configuration space of $q^\beta$ that are unrestricted in the $Q^A$ but
are 
subject to the stated restrictions on the $x^a$. We denote the exhaustive 
sets of such 
intervals at the successive times by $\{\Delta^1_{\alpha_1}\}$,
$\{\Delta^2_{\alpha_2}\}$, $\cdots$,
$\{\Delta^n_{\alpha_n}\}$.  The index $k$ on
$\{\Delta^k_{\alpha_k}\}$ labels the particular set, $\alpha_k$
labels the particular interval in the set,  and $t_k$ is the time.
An individual
coarse-grained history in such a set consists of the paths,
$q^\beta(t)$, that thread
a {\it particular} sequence of regions, e.g. $\Delta^1_{\alpha_1}$ at $t_1$,
$\Delta^2_{\alpha_2}$ at $t_2$, etc., and the whole set of
coarse-grained histories is exhausted as the different possible ways
which paths may pass through the regions are enumerated.  A particular
coarse-grained history thus corresponds to a sequence
$(\alpha_1,\cdots,\alpha_n)$, which we shall often abbreviate as just
$\alpha$.

The coherence between individual histories in a coarse-grained set is
measured by the decoherence functional.  This is a complex functional
defined on all pairs of coarse-grained histories in an exhaustive set of
such histories. The sets of histories under discussion are coarse
grainings of fine-grained histories that are paths $q^\beta(t)$ on the
time interval $[0,T]$.  The
sum-over-histories formulation of quantum mechanics is, therefore,
convenient for introducing the decoherence functional; more general
formulations are discussed in the next Section.  
For the coarse grainings under discussion, the decoherence functional
is
$$
D(\alpha^\prime, \alpha)=\int_{\alpha^\prime}\delta q^\prime\int_\alpha\delta
q\delta(q^\prime_f-q_f)
\exp\biggl\{i\Bigl(S[q^\prime(\tau)] -
S[q(\tau)]\Bigr)/\hbar\biggr\}\rho(q^\prime_0,q_0)\ .
\tag one
$$
The path integral over $q^\beta(t)$ is over all paths that start at 
$q^\beta_0$ at
$t=0$, pass through the intervals $\Delta^1_{\alpha_1}$,
$\Delta^2_{\alpha_2}, \cdots, \Delta^n_{\alpha_n}$ of the
$x^a$ 
at times $t_1<\cdots<t_n$, 
and wind up at $q^\beta_f$ at time $T$.  The path integral
over $q^{\prime\beta}(t)$ is similarly defined.  The integrals include an 
integral
over the initial $q^\beta_0$ and final $q^\beta_f$.  Here, $\rho(q^\prime_0,
 q_0)$
 is the
initial density matrix in the $q^\beta$ representation and $S[q(\tau)]$ is the
fundamental action.  The measure for the path integrals is the standard
one induced by the Liouville measure in phase space.  It is described
explicitly in Appendix (a).  Equation \(one) has been compressed by
omitting the indices on the $q^\beta(t)$ and by denoting the entire sequence
$(\alpha_1, \cdots, \alpha_n)$ at $(t_1, \cdots, t_n)$ by a single index
$\alpha$.  We shall employ similar conventions in the rest of the paper.

The coarse grainings under consideration distinguish only the $x^a$ in
the division\hfill\break
\noindent $q^\beta=(x^a, Q^A)$.  Following an analysis of Feynman and
Vernon [\cite {12}], the integrals over the $Q^A$ may therefore be 
carried out over
their whole ranges unrestricted by the particular coarse-grained
histories considered.   Suppose that the action may be decomposed as
$$
S[q(\tau)] = S_{\rm free} [x(\tau)] + S_0 [Q(\tau)] + S_{\rm int} [x(\tau),
Q(\tau)]\ .\tag two
$$
(The use of the subscript
``free'' does not mean that there is no potential energy term in $S_{\rm
free}$.  There is, in general.  Rather it means that the action 
of the $x$'s is free
of any interaction with the $Q$'s.)
The integral over the $Q$'s defines $W$ --- a functional 
of the paths
$x^\prime(t)$ and $x(t)$ and a function of their initial endpoints
$x^\prime_0$ and $x_0$ --- as follows:  
$$
\exp \left(iW\bigl[x^\prime(\tau), x(\tau); x^\prime_0, x_0\bigr)/\hbar\right) 
\tilde\rho(x^\prime_0,
x_0) \equiv \int 
\delta Q^\prime 
\int \delta
Q 
\delta (Q^\prime_f-Q_f)
$$
$$\times \exp\biggl\{i\Bigl(S_0[Q^\prime(\tau)] + S_{\rm int} 
[x^\prime(\tau),
Q^\prime(\tau)] - S_0[Q(\tau)]-S_{\rm int}[x(\tau), Q(\tau)]\Bigr)/\hbar\biggr\}
$$
$$
\times\rho(x^\prime_0, Q^\prime_0; x_0, Q_0)\ .\tag three
$$
Here we have introduced the reduced  density 
matrix $\tilde\rho= S p\ \rho$ associated 
with
the coarse graining,
$$
\tilde\rho(x^\prime_0, x_0)
 \equiv \int\ dQ_0\ \rho(x^\prime_0, Q_0; x_0,
Q_0)\ .\tag four
$$

The functional $W$ is only a slight generalization of the Feynman-Vernon
influence phase and we shall continue to call it that.  It 
depends on the endpoints $x^\prime_0$ and $x_0$ {\it
implicitly} through the paths $x^\prime(t)$ and $x(t)$ because the
actions on the right hand side of \(three) 
 are functionals of these paths.  There is
also an {\it explicit} dependence on $x^\prime_0$ and $x_0$ arising from
the dependence of $\rho$ on these variables in \(three) and \(four).  We
use the notation $W[x^\prime(\tau), x(\tau); x^\prime_0, x_0)$ to
indicate this dependence, the square bracket
to indicate dependence on functions and the round bracket to indicate
dependence on variables, and we maintain this notation
for other cases. 
 The quantity $W$ depends on the time
interval $T$ as well, but we have not indicated this explicitly.

The decoherence functional may then be expressed in terms of $W$ and
$\tilde\rho$ as follows:
$$
D(\alpha^\prime, \alpha) = \int_{\alpha^\prime} \delta x^\prime\int_\alpha
\delta x\ \delta (x^\prime_f-x_f)
$$
$$
\times \exp\biggl\{ i \Bigl(S_{\rm free}[x^\prime(\tau)]- S_{\rm
free}[x(\tau)] + W \bigl[x^\prime(\tau), x(\tau); x^\prime_0, x_0\bigr)
\Bigr)/\hbar\biggr\}
\tilde\rho(x^\prime_0, x_0)\ .\tag five
$$
Thus all the contribution from the ignored
variables is summarized by the functional $W[x^\prime(t), x(t);x^\prime_0,
x_0)$. 

Restrictions on the form of the actions $S_{\rm free} [x]$, $S_0 [Q]$,
and $S_{\rm int} [x, Q]$, as well as on the form of the density matrix
$\rho$, will be needed for the explicit derivation of the equation of
motion, as described in the subsequent sections.  We shall, 
for example, generally make the usual assumption
that $S_{\rm free}[x(\tau)]$ has a simple
``kinetic minus potential'' form
$$
S_{\rm free}[x(\tau)] = \int\nolimits^T_0 dt \left[\half\dot
x^\dagger(t) M\dot x(t) - V\bigl(x(t)\bigr)\right]\ ,
\tag six 
$$
where we have used an obvious matrix notation $x^\dagger Mx =
\Sigma_{ab} x^a M_{ab} x^b$.  Similar assumptions will be made
for $S_0[Q(\tau)]$.  We shall assume that
$S_{\rm int}[x(\tau), Q(\tau)]$ is local in time, that is, of the form
$$
S_{\rm int}\bigl[x(\tau), Q(\tau)\bigr] = \int\nolimits^T_0 dt\ L_{\rm
int} \bigl[x(t), Q(t)\bigr]\ ,\tag seven
$$
with $L_{\rm int}$ independent of velocities.  These assumptions
restrict us only to a widely applicable class of models and it is likely
that similar results can be obtained from weaker assumptions.

A more restrictive assumption concerns the form of the initial $\rho$.
We shall make the conventional assumption [\cite{12, 11}]
that $\rho$ factors into a density matrix in $x$ and
another in $Q$
$$
\rho(x^\prime_0, Q^\prime_0 ; x_0 , Q_0)= \bar\rho (x^\prime_0, x_0)
\rho_B (Q^\prime_0, Q_0)\ ,
\tag eight
$$
so that the variables distinguished by the coarse graining are initially
uncorrelated with those it ignores.  Of course, we do not necessarily
expect initial density matrix of the whole universe to factor
as in \(eight), but, for widespread mechanisms of decoherence that operate
essentially locally in space and time when compared with cosmological
scales, \(eight) is an excellent approximation.  For example, scattering by
the cosmic background radiation can efficiently decohere alternative
positions of the center of mass of a massive body coarse grained on
centimeter scales [\cite {7}].  
The co\"ordinates of the body and radiation may be
correlated in the wave function of the universe, but on the local scales
where the mechanism operates they are effectively uncorrelated, as
described by \(eight).

Factorization has a number of helpful consequences for the form of $W$
and $\tilde \rho$ defined in \(three) and \(four). First, $\tilde\rho$ is
given by
$$
\tilde\rho (x^\prime_0, x_0) = \bar\rho(x^\prime_0, x_0)\ .\tag nine
$$
Second, and most importantly, the influence phase $W$ contains no
explicit dependence on $x^\prime_0$ and
$x_0$ and we may write $W[x^\prime(\tau), x(\tau)]$. This will simplify
the form of the equation of motion, which would otherwise contain terms
arising from the explicit dependence on $x^\prime_0$ and $x_0$.

For most of this paper, therefore, we are considering a class of models
defined by coarse grainings that distinguish a fixed subset of the
co\"ordinates of configuration space, by actions that have the simple
forms \(two), \(six), \(seven), and by an initial density matrix that
factors as in \(eight). 
Especially simple examples of such models are the linear oscillator
models 
studied
by Feynman and Vernon [\cite {12}] and by Caldeira and Leggett [\cite {11}].
 In these models a
distinguished oscillator is coupled linearly to a large number of other
oscillators constituting a thermal bath characterized by a temperature
$T_B$.  The density matrix is assumed to factor as in \(eight).  Let $x(t)$
be the co\"ordinate of the distinguished oscillator and $\omega_R$ its
frequency renormalized by its interactions with the others.  Then, in
the case of a continuum of oscillators, cut off at frequency $\Omega$,
and in the Fokker-Planck limit of $kT_B >>\hbar\Omega>>\hbar\omega_R$,
Caldeira and Leggett find for the influence phase
$$
W[x^\prime(\tau),x(\tau)]=-M\gamma \int^T_0\ dt \left[x^\prime\dot x^\prime - 
x\dot x
+ x^\prime\dot x - x \dot x^\prime\right]
+ i \frac{2M\gamma kT_B}{\hbar}\ \int^T_0 
\ dt\left[x^\prime(t) - x(t)\right]^2
\ ,\tag ten
$$
were $\gamma$ is a coupling constant summarizing the interaction of the
distinguished oscillator with the rest.
\vskip .26 in
\taghead{4.}
\centerline{\bf IV. Decoherence}
\vskip .13 in
\centerline{\sl (a) Decoherence in General}

Quantum mechanics predicts the probabilities for the individual members
of a set of alternative coarse-grained histories only when there is
negligible quantum-mechanical interference between the individual
members of the set [\cite{5, 6, 2}]. Only then do the squares
of amplitudes define probabilities that are consistent with the sum
rules of probability theory.  Sets of histories that exhibit negligible
interference are said to {\it decohere weakly} [\cite{33}].

However, in quantum mechanics we are not interested just in sets of
histories that are consistent in the sense that they can be assigned
probabilities satisfying probability sum rules.  We are interested
also
in sets of histories that constitute the quasiclassical domain of 
 everyday experience.  It is this quasiclassical domain that lies at the root
of the interpretation of quantum mechanics. It is through an
understanding of this domain
that quantum mechanics acquires utility for our experience;
``measurement'' situations arise precisely when variables
become highly correlated with the quasiclassical domain.  
Stronger notions of decoherence are therefore
useful to characterize the realistic mechanisms  of decoherence that lead to
a quasiclassical domain [\cite{33}]. In this section we shall review
several notions of decoherence that we have described in
previous work, and we shall
discuss their connections with each other and with some
other notions of decoherence that have been introduced in the
literature.  We shall be brief.  For greater detail the reader may
consult references [\cite{2}] and [\cite{33}].

We begin by recalling how a set of alternative, coarse-grained histories
of a closed system is described in quantum mechanics.  The simplest
kinds of histories are specified by giving independent sets of
alternatives at a sequence of times $t_1 < t_2 < \cdots < t_n$.
In the Heisenberg picture, alternatives at one
moment of time $t_k$ correspond to a set of projection operators
$\{P^k_{\alpha_k}(t_k)\}$. The index $k$ denotes the set of alternatives
at time $t_k$, while the index $\alpha_k$ denotes the particular
alternative within that set.
These projections represent exclusive alternatives, so they
are orthogonal for different alternatives, and they represent
 an exhaustive set, so 
they sum to unity over all alternatives.  For example, for the coarse
graining by ranges of a distinguished set of variables $x^a$ described
in the preceding Section, the projection $P^k_{\alpha_k}(t_k)$ would
just be the projection onto the range $\Delta^k_{\alpha_k}$ at time
$t_k$. An
individual history in a set 
defined by a sequence of such sets of alternatives corresponds to a 
sequence of particular alternatives $\alpha = (\alpha_1, \cdots,
\alpha_n)$. Each  history is represented by the corresponding
 chain of projection operators
$$
C_\alpha = P^n_{\alpha_n} (t_n) \cdots P^1_{\alpha_1} (t_1)\ .
\tag fourone 
$$
A completely fine-grained set of histories would consist of
one-dimensional projections onto complete sets of states at each and
every time.  Sets of histories defined by sets of projections that are not
all one-dimensional or not at every moment of time are said to be 
   coarse-grained.

In the Heisenberg picture, every exhaustive set of orthogonal projection
operators in Hilbert space $\{P^k_{\alpha_k}\}$ represents, at {\it any}
time, {\it some} set of alternatives for the system.  Of course, the
alternatives corresponding to a given set of projection operators will
have different descriptions in terms of fundamental fields when
different times are assigned to them.  Similarly a sequence of sets of
projection operators define different alternative histories when
different times are assigned to the sets of projections.  However, an
assignment of times leads to meaningful alternative histories only
if
the ordering of the times corresponds to ordering of the projections as
in \(fourone).  If this time ordering is not respected, two inconsistent
sets of alternatives could be assigned the same time and the resulting
alternatives would not be meaningful.

While in quantum mechanics we usually consider
 sets of histories consisting of independent sets of alternatives at
sequences of times, a
more realistic description is achieved by generalizing this notion in
two related ways [\cite{2, 3}].  First, if $\alpha = (\alpha_1, \cdots,
\alpha_n)$ is a history and $\beta = (\beta_1, \cdots, \beta_n)$ is a
distinct history then we may consider the coarser-grained alternative
that the system followed either history $\alpha$ {\it or} history
$\beta$.  The alternative $\alpha$ or $\beta$ is represented by the {\it
sum} of the chains for $\alpha$ and $\beta$
$$
C_{\alpha\ {\rm or}\ \beta} = C_\alpha + C_\beta
\tag fourtwo
$$
but is not itself necessarily a chain of projections of the form
\(fourone). More precisely, if $\{\alpha\}$ is a set of alternative
histories for the closed system defined by sets of
alternative projections at a sequence of
times, then coarser-grained sets are defined by partitions of the
$\{\alpha\}$ into exclusive classes $\{\bar\alpha\}$.  The classes 
correspond to
the individual histories in these coarser-grained sets and are
represented by operators $C_{\bar\alpha}$ that are sums of the
$C_\alpha$
$$
C_{\bar\alpha} = \sum\limits_{\alpha\epsilon\bar\alpha} C_\alpha =
\sum\limits_{(\alpha_1, \cdots, \alpha_n)\epsilon\bar\alpha}
P^n_{\alpha_n} (t_n) \cdots P^1_{\alpha_1} (t_1)
\tag fourthree
$$
These generalized $C_{\bar\alpha}$ need not themselves be chains of
projections and thus we sometimes extend the use of $C_\alpha$ to
denote the operators representing individual histories in these more
general coarse-grained sets.

The second important generalization is to allow the histories to be
branch dependent, that is, for the set of alternatives at time $t_k$ to
depend on the values of earlier labels $\alpha_1, \cdots, \alpha_{k-1}$
[\cite{2, 6}].
Branch dependence is important, because in a quasiclassical domain past
events may determine what is a suitable quasiclassical variable.  For
example, if a quantum fluctuation gets amplified so that it leads to
condensation of a galaxy in one branch and no such condensation in other
branches, then the outcome clearly influences what are suitable
quasiclassical variables in the region where the galaxy would form.

For the general case of branch dependence, a better notation than
\(fourone) for chains of projections  would be
the
following:
$$
C_\alpha = P^n_{\alpha_n} \left(t_n; \alpha_{n-1}, \cdots, \alpha_1
\right)\, P^{n-1}_{\alpha_{n-1}} \left(t_{n-1}; \alpha_{n-2}, \cdots,
\alpha_1\right) \cdots P^1_{\alpha_1}(t_1) \ ,\tag fourfour 
$$
where the $P^k_{\alpha_k} (t_k; \alpha_{k-1}, \cdots, \alpha_1)$ define
an exhaustive set of orthogonal projection operators as $\alpha_k$ runs
over all values for fixed $\alpha_1, \cdots, \alpha_{k-1}$, corresponding
to an exhaustive set of mutually exclusive alternatives for the closed
system.\footnote{$^3$}{In
Ref.~[\cite{2}], Section X, we unnecessarily
eliminated the possibility of branch-dependent chains of the form
\(fourfour) and
restricted attention to sets of $\alpha_k$'s that were independent of
one another.  We did that in order to safeguard a special derivation
 of the weak decoherence condition for assigning probabilities to
alternative coarse-grained histories.  In fact, that special derivation
is unnecessary, and as we shall see below
weak decoherence can easily be seen to be the
necessary and sufficient condition for the probability calculus to apply
to histories that are chains or sums of chains,
whether or not the choice of the set $P^k_{\alpha_k}$ is
branch-dependent.}
In limiting ourselves to projections that depend only on
previous alternatives rather than future ones we have incorporated a
notion of causality consistent with the usual arrow of time in quantum
mechanics.  Further generalizations to formulations without an arrow of
time are possible  [\cite{24}].

In the models that we treat in detail in this article, we do not make
use of branch dependence, since we assume a fixed division of
co\"ordinates into those distinguished, $x^a$, and those ignored,
$Q^A$, and the $\Delta$'s (intervals of $x$ values) are taken to be
branch-independent. We can thus employ the simplified notation \(fourone).
It should be borne in mind, though, that realistic coarse-grained
histories do involve branch dependence.

Having in hand 
this discussion of the possible sets of alternative coarse-grained
histories of a closed system, we can now turn to the various
notions of their decoherence.  The central quantity,
the decoherence functional, is defined generally for pairs of histories
in a coarse-grained set by 
$$
D(\alpha^\prime, \alpha) = Tr
\bigl[C_{\alpha^\prime}\rho\, C^\dagger_\alpha\bigr] \tag fourfive
$$
for a density matrix $\rho$ representing the initial condition and
operators $C_\alpha$ representing the individual histories.  

The necessary and sufficient condition for probability sum rules to be
satisfied  is
$$
Re D(\alpha^\prime, \alpha)  = 0\ ,\ \alpha^\prime \not= \alpha\ .
\tag foursix
$$
In previous work [\cite {33}]
  we have called this the {\it weak decoherence} condition\footnote{$^4$}{As
pointed out to us by Bob
Griffiths, we have
incorrectly attributed the weak decoherence condition
(4.6)
to him (in reference [\cite{2}] and elsewhere).
In fact, Griffiths [\cite {15}] and Omn\`es [\cite{6}] employ
a
weaker condition than (4.6) as the
necessary and sufficient condition for the ``consistency of histories''.
That is because they require, in the notation of \(fourone), that
two chains of projections $C_\alpha$ and $C_{\alpha^\prime}$ must have
the
real part of their interference term vanish {\it only if
$C_\alpha + C_{\alpha^\prime}$ is another chain} of projections,
whereas we require it
in the case of {\it all} the chains $C_\alpha$ and $C_{\alpha^\prime}$
(see below).
Our weak decoherence applies to coarse grainings that are allowed to
unite any two histories in the set being studied, while the ``consistent
histories'' condition of Griffiths and Omn\`es applies only to some of
those coarse grainings.
For example, if $\alpha$ and $\alpha^\prime$ differ in only one index,
our approach and theirs give the same conditions, but if $\alpha$ and
$\alpha^\prime$ differ in more than one index, then Griffiths and
Omn\`es do not always require that $Re[Tr(C_{\alpha^\prime} \rho
C^\dagger_\alpha)] = 0$ but rather a weaker condition.
In essence Griffiths and Omn\`es restrict themselves to
histories defined by independent alternatives at a sequence of times.
Each history corresponds to a sequence of such alternatives and
therefore to a chain of projections.  They do not therefore incorporate
branch dependence, at least in the sense of \(fourfour). In the models
studied in this paper, the stronger conditions are, in fact, satisfied.
The whole of any off-diagonal element of the decoherence functional
approximately vanishes --- not just the real part.  Also, when more of
the indices in $\alpha$ and $\alpha^\prime$ differ, the decoherence
condition (4.8) is satisfied more strongly, not more weakly.  This gives
us some confidence that our stronger conditions are physically
realistic.}
 to
distinguish it from stronger decoherence conditions we shall discuss
below.
When a set of
histories weakly decoheres, the probability of a history $\alpha$ is the
corresponding ``diagonal'' element of the decoherence functional
$$
p(\alpha) = D (\alpha, \alpha)\ .
\tag fourseven
$$
Eq.~\(fourone) is the necessary and sufficient condition 
that these numbers obey the sum rules of
probability theory. All that is needed to show this is to notice that
the probability that either of two histories will happen is the sum of
the probabilities of the two individual histories if and only if the sum
of the interference terms represented by \(foursix) vanishes.
 Weak decoherence is the criterion by which quantum
mechanics discriminates between those sets of histories that can be
assigned
probabilities and those that cannot.

A stronger notion of decoherence is provided by the {\it medium decoherence}
condition [\cite{33}]
$$
D(\alpha^\prime, \alpha) = 0\ ,\ \alpha^\prime \not= \alpha\ .
\tag foureight
$$
Clearly medium decoherence implies weak decoherence, but not the other
way around.
Medium decoherence is a consequence of realistic mechanisms that are
widespread in the universe.  It is, therefore, a natural condition to
impose in characterizing a quasiclassical domain.  It is not, by itself,
sufficient to single out a quasiclassical domain.  The necessary further
criteria are a large part of the subject of this paper.

When the initial state is pure, exact medium decoherence is equivalent
to the existence of generalized records for each history in the
decohering set [\cite {33}].  To see this, notice that
for a
pure initial condition, $\rho=|\Psi \rangle \langle \Psi |$,
$$
D(\alpha^\prime, \alpha) = \bigl(\langle\Psi | C^\dagger_\alpha\bigr)
\left(C_{\alpha^\prime} | \Psi \rangle\right)\ . \tag fournine
$$

To every set of alternative histories there corresponds a resolution of
the pure initial state into branches
$$
|\Psi \rangle = \sum_\alpha \left(C_\alpha |\Psi\rangle\right) =
\sum\nolimits_{\alpha_1,\cdots, \alpha_n} P^n_{\alpha_n} (t_n) \cdots
P^1_{\alpha_1} (t_1) |\Psi\rangle\ . \tag fourten
$$
For a set of histories obeying exact medium decoherence, the branches are
orthogonal, as \(foureight) and \(fournine) show.  Therefore, there exists
at least one  set
of orthogonal projection operators $\{R_\alpha\}$ that project onto
these branches
$$
R_\alpha |\Psi\rangle = C_\alpha |\Psi\rangle \ ,\tag foureleven
$$
where
$$
R_\alpha R_{\alpha^\prime} = \delta_{\alpha\alpha^\prime} R_\alpha
\quad {\rm and} \quad \sum\limits_\alpha R_\alpha = 1 \ ,\tag fourtwelve
$$
so that the projections $\{R_\alpha\}$ are exclusive and exhaustive.

Now we can also assign projection operators $R_\alpha$ to coarse grainings of
the histories $\{\alpha\}$, that is, sums of chains of projections
$\{C_\alpha\}$.  To every such coarser-grained history, representing the
union of a subset of the histories $\{\alpha\}$ or the sum of the
corresponding $\{C_\alpha\}$, we assign the projection operator that is
the sum of the corresponding $\{R_\alpha\}$.  That is perfectly
consistent with the extension of \(foureight) to the coarser-grained
histories.  The resulting $R$'s have the property that progressive fine
graining of the coarser-grained
 histories results in a sequence of {\it nested} $R$'s, projecting
onto smaller and smaller subspaces of Hilbert space, where each such
space is a subspace of the preceding one.

When the branches $C_\alpha | \Psi \rangle \not= 0$ do not form a {\it
complete} set of orthogonal states for the Hilbert space, there can be
many sets of projections $\{R_\alpha\}$ that obey the conditions
\(foureleven) and \(fourtwelve).  When the branches do form a complete set,
then the $\{R_\alpha\}$ are unique; they are just the projections onto
the single states $C_\alpha | \Psi\rangle$.  The set of histories
$\{\alpha\}$ is then said to be {\it full} [\cite{33}].  The $R$'s for
coarse grainings of the histories $\{\alpha\}$ are then also unique: for
the further coarse-grained history that corresponds to the union of 
 a given subset of the
$\{\alpha\}$, the corresponding $R$ is just the sum of the relevant
$R_\alpha$ and projects onto the space spanned by the corresponding
vectors $C_\alpha |\Psi\rangle = R_\alpha | \Psi\rangle$.

When assigned a time after $t_n$ in the sequence $t_1 < t < \cdots <
t_n$, the $R_\alpha$'s may be thought of as representing  
generalized records of the histories.  They may not represent  records in the
usual sense of being constructed from quasiclassical variables
accessible to us, but the condition \(foureleven) means that at any time
there is  complete information somewhere in the universe about the
histories $\{\alpha\}$. 

The above discussion shows that medium decoherence in a pure initial
state implies the existence
of generalized records.  The converse is also true.  The existence of
orthogonal generalized record projections satisfying \(foureleven)
ensures the medium decoherence of the corresponding
set of histories through
\(foureight) and \(fournine).  Exact medium decoherence can thus be
characterized by records, and the physical formation of records is a way
to understand mechanisms by which medium decoherence occurs.  In the
example implicit in the work of Joos and Zeh\footnote{$^5$}{For a more
detailed analysis in terms of histories, see [\cite{3}].} [\cite{7}],
histories describing successive
 alternative positions of a dust grain, initially in
a superposition of positions about a millimeter apart, very accurately
satisfy the condition of medium decoherence
 simply by virtue of the scattering by
cosmic background radiation photons.  The successive scatterings of
these photons effectively create records of the histories of positions of the dust grain in the
electromagnetic degrees
of freedom.  The commuting
records of successive positions
 are stored independently in the vastness of cosmological space
as the photons move off at the speed of light.  They may not be
accessible to us, but their existence is a way of understanding how this
mechanism of medium decoherence works.

The permanence of the past is a feature of the quasiclassical domain 
that is naturally explained by medium decoherence
when there is a pure initial condition.  By permanence of the past we mean the
feature of a quasiclassical domain that what has happened in the past is
independent of any information expressed by a future projection.
Neither the
decoherence of past alternatives nor the selection of a particular past
alternative is threatened by new information. ``The Moving Finger
writes; and having writ, Moves on: nor all thy Piety nor Wit Shall lure
it back to cancel half a Line, Nor all thy Tears wash out a Word of
it'' [\cite{38}].

In other words, we are discussing the property of the
decohering coarse-grained histories $\{\alpha\}$ that, at any of the
times $t_k$, there is, for each history up to that time, an effective
density matrix [\cite{3}]
$$
\frac{P^k_{\alpha_k} (t_k) \cdots P^1_{\alpha_1} (t_1) \rho
P^1_{\alpha_1} (t_1) \cdots P^k_{\alpha_k} (t_k)}{Tr\left[P^k_{\alpha_k}
(t_k) \cdots P^1_{\alpha_1} (t_1) \rho P^1_{\alpha_1} (t_1) \cdots
P^k_{\alpha_k} (t_k)\right]}\ , \tag fourthirteen
$$
which can be utilized for all future predictions without
concern about the other values of $\alpha_1, \cdots, \alpha_k$.
For a pure state, this corresponds to the ``reduction of the state
vector''
$$
|\Psi \rangle\rightarrow N^{-\half} P^k_{\alpha_k} (t_k) \cdots
P^1_{\alpha_1} (t_1) |\Psi\rangle\ , \tag fourfourteen
$$
where $N$ is the trace in \(fourthirteen).

When the past is permanent, we may still
lose the ability to retrodict the probabilities of alternatives in the
past through the impermanence or inaccuracy of present
records 
{\it but not from the failure
of those past alternatives to decohere in the face of the projections
that describe information we acquire as we advance into the future.}
Yet we know that such continued
decoherence of the past is not guaranteed {\it in general} by quantum
mechanics.\footnote{$^6$}{For further and less informal discussion see
[\cite{3}], Section
II.3.2.}
  Adjoining future alternatives to a set of histories
is a fine graining of that set and in general
 a fine graining of a decoherent set
of histories may no longer decohere. 
Verifying the continued decoherence of all the past alternatives
 as we fine-grain our set of
histories to deal with the future would in general require significant
computation.  We would have to 
check that the branches corresponding to {\it every}
alternative past that {\it might} have happened continue to be orthogonal
in the presence of  
their newly adjoined sets of projections.  Yet we adjoin sets of
projections onto ranges of quasiclassical operators without making this
calculation, secure in the faith that previous alternatives will continue
to decohere despite this fine graining.  It is this assumption of
continued decoherence of the past that permits the focus for future
predictions on the one
branch corresponding to our particular history and
the discarding of all others.  In other words, we pointed out above, it is 
the permanence of the past that permits the ``reduction of the state
vector''. 

If we consider, instead of the set of histories $\{\alpha\} =
\{(\alpha_1, \cdots, \alpha_n)\}$, the set of abbreviated histories
$\{(\alpha_1, \cdots, \alpha_k)\}$ with $k<n$, running up to time $t_k <
t_n$, that is an example of a coarser graining of the set of
coarse-grained histories $\{\alpha\}$.  As these abbreviated histories
get further $\alpha$'s ($\alpha_{k+1}$ then $\alpha_{k+2}$, etc.)
adjoined to them, that represents a sequence of fine grainings of the
coarser-grained histories $\{(\alpha_1,\cdots, \alpha_k)\}$.

If $\rho$ is pure, there are nested records $R$ corresponding to these
abbreviated histories.  When further $\alpha$'s are adjoined, as the
histories unfold from ($\alpha_1, \cdots, \alpha_k$) to $(\alpha_1,\cdots,
\alpha_{k+1})$, etc., up to $(\alpha_1,\cdots, \alpha_n)$, the nested
record operators are projections onto subspaces of Hilbert 
space that progresively narrow.  In such a situation, the past always
continues to decohere as the histories advance into the future.  In
fact, the physical formation of nested, generalized record operators $R$
guarantees, in general, the permanence of the past, including not only
the permanence of
its decoherence, but also the permanence of the
 selection of particular past alternatives
as well.

If we relax the condition of exact decoherence and consider approximate
medium
decoherence, defined by the approximate validity of eq.~\(foureight),
then it is possible to understand more about the formation
of records and the origins of medium decoherence for the kind of model 
coarse grainings 
studied in this paper, which distinguish 
particular co\"ordinates and ignore a large number of others. The
Hilbert space  is a tensor product
of a Hilbert space of functions of 
the distinguished co\"ordinates, ${\cal H}^x$,
 and one for the ignored co\"ordinates,
${\cal H}^Q$. The coarse-grained histories consist of chains of 
projections at times $t_1, t_2, \cdots$ 
that, in the Schr\"odinger picture, act only on
${\cal H}^x$ and represent partitions of some 
complete set of
states in ${\cal H}^x$ at each time. 
In the models, these states are, in fact, just the
localized states in $x^a$ at each time, 
but for greater generality we shall consider different complete
orthogonal sets at each time represented by wave functions
$\{\phi^1_{r_1} (x)\}$ at time $t_1$, $\{\phi^2_{r_2} (x)\}$ at time
$t_2$, etc.  In the subsequent discussion we shall augment this notation
to indicate the branch dependence of the possible orthogonal sets.

In order to make a connection with the thinking of some authors, we
shall take a brief excursion into the Schr\"odinger picture, while
assuming a pure initial state represented by a wave function
$\Psi(x,Q,t_0)$.
Let us follow for three steps 
the Schr\"odinger evolution of this initial $\Psi$. It may be
evolved to the time $t_1$ of the first set of alternatives and expanded
in the first complete set  $\{\phi^1_r(x)\}$ as 
$$
\Psi (x, Q, t_0) \rightarrow \sum_{r_1} \phi^1_{r_1} (x) \chi_{r_1} (Q)
\ .
\tag fourfifteen
$$
The right-hand side is $\Psi(x,Q,t_1)$ in the Schr\"odinger picture and 
$\langle x(t_1) = x, Q(t_1) = Q|\Psi\rangle$ in the Heisenberg picture.
The coefficients $\chi_{r_1}$ are in general neither normalized nor
orthogonal for different $r_1$.  A coarse graining divides the
$r_1$ up into exclusive and exhaustive
 sets $\{\alpha_1\}$.  The sum on the right of
\(fourfifteen) may be similarly decomposed:
$$
\Psi (x, Q, t_1) = \sum_{\alpha_1}
\sum_{r_1\epsilon\alpha_1} \phi^1_{r_1} (x) \chi_{r_1} (Q)\ .
\tag foursixteen
$$

The result of the evolution of each branch $\sum\limits_{r_1\epsilon\alpha_1}
\phi^1_{r_1}(x) \chi_{r_1} (Q)$  in \(foursixteen) to the time $t_2$
of the next set of alternatives may again be expanded in a new complete
set of functions in the $x$'s.  In general this set will be branch
dependent, that is dependent on $\alpha_1$.  We use the notation
$\{\phi^{2\alpha_1}_{r_2} (x)\}$ to indicate this,
 and the expansion then has the
form
$$
\sum_{r_1\epsilon\alpha_1} \phi^1_{r_1} (x) \chi_{r_1} (Q) \rightarrow 
\sum_{\alpha_2}
\sum_{r_2\epsilon\alpha_2} \phi^{2\alpha_1}_{r_2} (x)
\chi^{\alpha_1}_{r_2} (Q)\ . \tag fourseventeen
$$
Similarly for the evolution from $t_2$ to $t_3$:
$$
\sum_{r_2\epsilon\alpha_2} \phi^{2\alpha_1}_{r_2} (x) 
\chi^{\alpha_1}_{r_2} (Q) \rightarrow
\sum_{\alpha_3} \sum_{r_3\epsilon\alpha_3}
\phi^{3\alpha_2\alpha_1}_{r_3} (x) \chi^{\alpha_2\alpha_1}_{r_3} (Q) 
\ ,\tag foureighteen
$$
and so forth.  Thus,
$$
\Psi\left(x,Q, t_3\right) = \sum_{\alpha_3\alpha_2\alpha_1}
\left(\sum_{r_3\epsilon\alpha_3} 
\phi^{3\alpha_2\alpha_1}_{r_3} (x)
\chi^{\alpha_2\alpha_1}_{r_3} (Q) \right) \ .\tag fournineteen
$$
The term in the brackets is the branch wave function corresponding to
the history\hfill\break
 $\alpha=(\alpha_1, \alpha_2, \alpha_3)$. The overlap of branches
gives the decoherence functional according to \(fournine) and
\(fourten).  Thus
$$
D\left(\alpha^\prime, \alpha\right) = \sum_{r^\prime_3\epsilon\alpha^\prime_3}
\sum_{r_3\epsilon\alpha_3} 
\left(\phi^{3\alpha_2\alpha_1}_{r_3},
\phi^{3\alpha^\prime_2\alpha^\prime_1}_{r^\prime_3}\right)
\left(\chi^{\alpha_2\alpha_1}_{r_3}\ ,
\ \chi^{\alpha^\prime_2 \alpha^\prime_1}_{r^\prime_3}\right)\ . 
\tag fourtwenty 
$$ 
If the complete set at time $t_3$ is independent of previous
alternatives, so that it is not branch dependent, then the decoherence
of the last alternative is automatic.  The scalar product $(\phi^3_{r_3},
\phi^3_{r^\prime_3}) = \delta_{r_3 r^\prime_3}$ and so the right-hand side of
\(fourtwenty) is diagonal in $r_3$.
Otherwise it is a non-trivial
condition because $\{\phi^{3\alpha_2\alpha_1}_{r_3}\}$ and
$\{\phi^{3\alpha^\prime_2\alpha^\prime_1}_{r\prime_3}\}$ 
are {\it different}
orthogonal sets
when $\alpha$'s do not coincide.
Suppose the Hilbert space ${\cal H}^Q$ is effectively
very large compared to ${\cal
H}^x$, in the sense that at each time $t_k$  a one-to-one correspondence is
established between ${\cal H}^x$ and a tiny portion of ${\cal H}^Q$.
 Then we may expect 
the scalar products between 
$\chi$'s
differing by any index to be typically
very small\footnote{$^7$}{Of course, as more and more times are added
to the histories, and the set of $\alpha$'s grows larger accordingly, we
expect that eventually even the space ${\cal H}^Q$ will be exhausted and
$\chi$'s no longer orthogonal.},
 leading to small values of $D(\alpha, \alpha^\prime)$ for
$\alpha^\prime \not=\alpha$. 
The summations over unorganized phases between different
values of $r_3$ may make the off-diagonal elements of $D(\alpha^\prime,
\alpha)$
even smaller. That is approximate medium decoherence.
In the approximation in which {\it all} the $\chi$'s are really
orthogonal to one another, 
the records $R_\alpha$ that accomplish projections onto branch wave
functions such as those in \(fournineteen) may then be taken to be
projections onto the corresponding sets of $\chi$'s:
$\{\chi^{\alpha_2\alpha_1}_{r_3}|r_3\epsilon\alpha_3\}$.

The reduced density matrix in $x$ may also be constructed.  It is
$$
\tilde \rho\left(x^\prime, x\right) = \sum_{\alpha^\prime_3
\alpha^\prime_2 \alpha^\prime_1} \sum_{\alpha_3\alpha_2\alpha_1}
\sum_{r^\prime_3\epsilon \alpha^\prime_3} \sum_{r_3\epsilon\alpha_3}
\phi^{3\alpha^\prime_2\alpha^\prime_1}_{r^\prime_3} (x^\prime)
\phi^{3\alpha_2\alpha_1*}_{r_3} (x)
\left(\chi^{\alpha_2\alpha_1}_{r_3}
\ ,\ \chi^{\alpha^\prime_2\alpha^\prime_1}_{r^\prime_3}\right) 
\ ,\tag fourtwentyone
$$
involving a sum over all branches.  If indeed all the $\chi$'s are
approximately orthogonal to one another because of the largeness of the
Hilbert space ${\cal H}^Q$, then $\tilde\rho(x^\prime, x)$ is
approximately diagonal in the histories $\{\alpha\}$.

Many
authors have considered, not the full density matrix \(fourtwentyone),
but 
the reduced effective density matrix for each branch [\cf
\(fourthirteen)], relevant when the decoherence of earlier
alternatives may be assumed.  The applicable portion of
$\tilde\rho(x^\prime, x)$ [\(fourtwentyone)] is 
$$
\rho^{\alpha_2\alpha_1}\left(x^\prime, x\right) =
\sum_{\alpha_3} \sum_{\alpha^\prime_3}
\sum_{r^\prime_3\epsilon\alpha^\prime_3} \sum_{r_3\epsilon\alpha_3}
\phi^{3\alpha_2\alpha_1}_{r^\prime_3} \left(x^\prime\right)\,
\phi^{3\alpha_2\alpha_1*}_{r_3} (x) \left(\chi^{\alpha_2\alpha_1}_{r_3}
\ ,\ \chi^{\alpha_2\alpha_1}_{r^\prime_3}\right)\ .
\tag fourtwentytwo
$$
Approximate orthogonality of these $\chi$'s for different $\alpha_3$
 leads to approximate
diagonality of $\rho^{\alpha_2\alpha_1}$ in the $\alpha_3$'s. 
However, a much stronger
condition on the $\chi$'s is needed to ensure the decoherence of whole
histories as in \(fourtwenty) than is needed to ensure the diagonality
of \(fourtwentytwo). We need the approximate orthogonality of the
$\{\chi^{\alpha_2\alpha_1}_{r_3}\}$ not only when the $r_3$ belong to
different sets $\alpha_3$, but also between functions corresponding to
different values of $\alpha_1$ and $\alpha_2$.  The strength of these
conditions may be appreciated by noting two facts: First, since  any
coarse graining of a decoherent set is also decoherent, the $\chi$'s
must be approximately orthogonal in all previous steps.  That is, 
 the $\{\chi^{\alpha_1}_{r_2}\}$ in the example
 must be approximately orthogonal
for different $\alpha_1$'s and when the $r_2$ lie in different
$\alpha_2$'s. There is a similar condition for the $\{\chi_{r_1}\}$.
Since $x$ and $Q$ interact between the times $t_1, \cdots, t_n$ none
of these conditions is a simple consequence of the others.  Second,
nothing in our discussion has fixed the choice of the times $t_1,
\cdots, t_n$.  Of course, they must be sufficiently separated for the
interactions to disperse the phases, but once that is satisfied, we
expect decoherence to hold for a range of times giving rise effectively
to even more conditions.  The decoherence of histories is a much
stronger requirement than the diagonality of density matrices.

Some authors [\cite{17, 42, 18}]
have discussed how, in the case of a {\it pure}
initial density matrix $\Psi(x^\prime, Q^\prime) \Psi^* (x,Q)$, one can
use the Schmidt decomposition to achieve some of the objectives of
decoherence.  In the notation we have been using for the Schr\"odinger
picture in eqs.~\(fourfifteen) -- \(fournineteen), we can choose, at
each time
$t_k$, the functions $\phi^{k\alpha_{k-1}\cdots\alpha_1}_{r_k}(x)$ to be
Schmidt functions, which means that their coefficients
$\chi^{\alpha_{k-1}\cdots \alpha_1}_{r_k}(Q)$ are {\it orthogonal} for
different values of $r_k$ (although still not normalized). In that way
the reduced density matrix for each branch
 at time $t_k$ becomes diagonal in $r_k$, as
we can see for the case $k=3$ in \(fourtwentytwo).  It is then, of
course,
also diagonal in $\alpha_k$ ($\alpha_3$ for the case $k=3$). However,
this Schmidt construction does not guarantee the decoherence of even the
final alternative as defined by the decoherence functional
\(fourtwenty) (unless we have $\alpha_1 = \alpha^\prime_1$ and $\alpha_2
= \alpha^\prime_2$).  That is because the Schmidt functions are {\it
necessarily branch-dependent} and the orthogonality of the $\chi$'s for
one branch does not guarantee the orthogonality between different branches
that would lead to decoherence of the final alternative.  In any event,  
the main thing
for the decoherence of {\it histories} is to have $D(\alpha^\prime,
\alpha)$ diagonal in $\alpha_2$ and $\alpha_1$ as well.  That is not
guaranteed at all by the Schmidt procedure, which does not imply the
orthogonality of $\chi$ functions for different values of the
$\alpha$'s. In particular, the Schmidt procedure does not guarantee the
permanence of the past discussed above.
The relation between diagonalization of the
reduced density matrix at successive times and the decoherence of
histories will be further discussed in [\cite{19}].

Now let us consider the situation with an impure initial state
represented by a density matrix $\rho$. An impure initial state 
could be fundamental, representing the initial condition of the universe.
However, 
 even if the cosmological initial condition is pure, a reduced
density matrix, in which some co\"ordinates already have been 
traced over, can be a useful description of local physics. A familiar
example is the cosmic background radiation. Imagine that the 
background photons 
have all been pair produced in a pure cosmological initial state. In
that case, for every photon near us, there would be a correlated photon with
equal but opposite momentum on the other side of the universe. The local
physics, however, would be accurately described by a nearly 
thermal reduced density matrix
in which the distant photons had been traced over. 

In [\cite{33}]
 we discussed a ``strong decoherence'' condition that is the analog 
of \(foureleven) for density matrices. A set of histories was said to
decohere strongly when, for each history in the set, there exists
record projections satisfying \(fourtwelve) such that 
$$
C_\alpha \rho =
R_\alpha \rho \ .\tag fourtwentytwo
$$  
However, we shall see that for highly impure states this is too strong
a condition to characterize usefully a quasiclassical domain. 

We can express an impure density matrix $\rho$ in terms of its
eigenstates and eigenvalues:
$$
\rho=\sum\nolimits_\mu \pi_\mu |\Psi^\mu\rangle\,\langle\Psi^\mu |\ ,
\tag fourtwentythree a
$$
or, in the special case of $x$ and $Q$ variables,
$$
\rho\left(x^\prime, Q^\prime; x, Q\right) = \sum\nolimits_\mu \pi_\mu
\Psi^\mu (x^\prime, Q^\prime) \Psi^{*\mu} \left(x, Q\right)\ .
\tag fourtwentythree b
$$
Here the $\pi_\mu$ are the probabilities of the initial states
$|\Psi^\mu\rangle$.

If $\rho$
has non-zero probabilities for many states, then \(fourtwentytwo) 
is difficult to
satisfy. If $\rho$ has non-zero probabilities for an orthogonal
 set of states $|\Psi^\mu\rangle$, then \(fourtwentytwo) would imply
$$
C_\alpha |\Psi^\mu\rangle = R_\alpha |\Psi^\mu\rangle
\quad {\rm for\ all}\ |\Psi^\mu\rangle 
\ . \tag fourtwentyfour
$$
If the set $|\Psi^\mu\rangle$ is {\it complete}, then \(fourtwentyfour)
implies that $C_\alpha = R_\alpha$.  That 
can be satisfied only in the trivial case in which all the $P$'s
for all times commute with one another.  We shall henceforth ignore
``strong decoherence'' for impure density matrices.

We could define, when $\rho$ is impure, a kind of ``medium strong
decoherence'', in which we would have, for each $|\Psi^\mu\rangle$ with
non-zero probability $\pi_\mu$, a generalized record projection operator
$R^\mu_\alpha$ such that
$$
C_\alpha |\Psi^\mu\rangle = R^\mu_\alpha |\Psi^\mu\rangle\ ,
\tag fourtwentyfive
$$
where for {\it each} $\mu$ the $R^\mu_\alpha$ are exclusive and exhaustive
projections.
This requirement would mean medium decoherence separately for each
$|\Psi^\mu\rangle$ (with $\pi_\mu \not=0$) with respect to the same set
of histories $C_\alpha$.  While not so difficult to satisfy as strong
decoherence, it is still a very stiff requirement.  For example, 
we shall see, in the
linear oscillator models discussed below, that it is not
 very well satisfied there. 
However, if satisfied, medium strong decoherence would supply, in the case
of an impure $\rho$, the same attractive features that medium
decoherence yielded for the pure case, including the permanence of the
past.

Approximate medium decoherence for density matrices continues to be
 defined by the approximate satisfaction of the
condition \(foureight) and can be discussed for the type of model coarse
grainings considered in this paper, where  a fixed set of
co\"ordinates is distinguished and the other co\"ordinates included in
the model are ignored.  The mechanism of the formation of correlations
between distinguished co\"ordinates and ignored ones can continue to
operate for each state in the density matrix. 
Now, however, 
there is the possibility for improvement in the effectiveness of
approximate
decoherence from summations over the states in the impure density matrix.

To make this explicit, consider the Schr\"odinger evolution of
 an initial density matrix of the form \(fourtwentythree b).
The evolution described by eqs.~\(fourfifteen) -- \(fournineteen) is as before
except that each $\chi$ acquires an index $\mu$.  In particular \(fourtwenty)
 and \(fourtwentyone) become
$$
D\left(\alpha^\prime,\alpha\right) = \sum_{r_3\epsilon\alpha_3} 
\sum_{r^\prime_3\epsilon\alpha_3}
\left(\phi^{3\alpha_2\alpha_1}_{r_3}\,
,\,\phi^{3\alpha^\prime_2\alpha^\prime_1}_{r^\prime_3}\right)
 \sum\nolimits_\mu \pi_\mu 
\left(\chi^{\mu \alpha_2\alpha_1}_{r_3}\ ,
\ \chi^{\mu\alpha^\prime_2\alpha^\prime_1}_{r^\prime_3}\right)\ , 
\tag fourtwentysix
$$
$$
\tilde\rho\left(x^\prime, x\right) =
\sum_{\alpha^\prime_1\alpha^\prime_2\alpha^\prime_3}
\sum_{\alpha_1\alpha_2\alpha_3} \sum_{r^\prime_3\epsilon\alpha^\prime_3}
\sum_{r_3\epsilon\alpha_3}
\phi^{3\alpha^\prime_2\alpha^\prime_1}_{r_3^\prime} \left(x^\prime\right)
\phi^{3\alpha_2\alpha_1*}_{r_3} (x) \sum\nolimits_\mu \pi_\mu
\left(\chi^{\mu\alpha_2\alpha_1}_{r_3}\ ,
\ \chi^{\mu\alpha^\prime_2\alpha^\prime_1}_{r^\prime_3}\right)\ .
\tag fourtwentyseven
$$
The additional sum over $\mu$ can lead to further phase
 cancellations and more
effective decoherence.  We will see an illustration of this in the
discussion of the oscillator models discussed below.
\vskip .13 in
\centerline{\sl (b) Decoherence in the Linear Oscillator Models}

Medium decoherence can be treated explicitly in the kind of oscillator
models worked out by Feynman and Vernon [\cite{12}] and Caldeira and
Leggett [\cite{11}], which we described in Section III. Generally these
models assume an impure initial density matrix factored as in
\(eight), with the ignored co\"ordinates representing a continuum
of oscillators in a thermal state.

An explicit illustration of the medium decoherence of a pure state can
be found in the zero temperature limit of the linear oscillator model.  In that
limit all the ignored oscillators are in a pure state and the initial
state of the distinguished oscillators may be taken to be pure. Denote
the initial ground state of the bath by $\chi_0(Q_0)$ and the initial state
of the distinguished oscillators by $\psi_0 (x_0)$, so that the pure
initial wave function of the whole system is $\psi_0(x_0)\chi_0(Q_0)$.

The Schr\"odinger evolution of this state can be represented in path
integral form.  At time $t$ we have
$$
\Psi\left(x, Q, t\right) = \int \delta x\delta Q\ \exp(iS[x(\tau),
Q(\tau)]/\hbar) \psi_0 (x_0) \chi_0 (Q_0) \tag fourtwentyeight
$$
The integral is performed
first over paths that start at $x_0$ and $Q_0$ at
time $t_0$ and end at $x$ and $Q$ at time $t$, and then 
over $x_0$ and $Q_0$.  The integral over the
distinguished co\"ordinates $x$ may be written as a sum over coarse
grainings $\alpha_1, \cdots, \alpha_k\, ,\ t_k <t$ 
and an integral over the paths restricted by the
coarse graining, in the form
$$
\Psi(x, Q, t) = \sum_{\alpha_1,\cdots,
\alpha_k}\int\nolimits_\alpha \delta x\,\delta Q\,\exp
\bigl(iS\bigl[x(\tau), Q(\tau)\bigr]/\hbar\bigr) \psi_0 (x_0) \chi_0 
\left(Q_0\right)\ .
\tag fourtwentynine
$$
This can be rewritten in the form
$$
\Psi (x,Q,t) = \sum_{\alpha_1, \cdots, \alpha_k}
\int^{+\infty}_{-\infty} dr\,\delta(x-r)\,\chi^{\alpha_k \cdots
\alpha_1}(r,Q)
\tag fourthirty
$$
where
$$
\chi^{\alpha_k \cdots \alpha_1} (r, Q) = \int\nolimits_\alpha \delta
x\delta Q\, \exp \bigl(iS\bigl[x(\tau), Q(\tau) \bigr]/\hbar\bigr) \psi_0
(x_0) \chi_0 (Q_0)
\tag fourthirtyone
$$
and the integral is over paths, consistent with the coarse graining,
that start at $(x_0, Q_0)$ and end at $(r, Q)$, including an integral
over the values of $x_0$ and $Q_0$.
This is evidently the analog of \(fournineteen) with $r$ being a continuous
index and $\phi^k_r (x) = \delta(x-r)$, branch independent and the same
at each time.
The overlap that occurs in \(fourtwenty), of course,
gives the decoherence functional. 

In the zero temperature limit, Caldeira and Leggett find for the
imaginary part of the influence phase [\cf (6.8)]
$$
{\cal I}m W \left[x^\prime(\tau), x(\tau)\right] = \frac{1}{4}
 \int\nolimits^T_0 dt
\int\nolimits^T_0 dt^\prime \xi (t) k_I(t-t^\prime) \xi (t^\prime)
\ .\tag fourthirtytwo
$$
Here, $\xi(t) = x^\prime(t) - x(t)$ and
$$
k_I(\tau) = \frac{4M\gamma}{\pi}  \int^\Omega_0 d\omega\,\omega
\,\cos(\omega\tau)\ ,\tag fourthirtythree
$$
where $\Omega$ is the cutoff of the oscillator spectrum.  This imaginary
part of the influence phase favors contributions to the functional
integral defining the decoherence functional \(five) from values of
$\xi$ near zero, or $x^\prime(t)$ close to $x(t)$.

A crude estimate of the time intervals by which coarse-grained
alternatives of position $\{\Delta^k_{\alpha_k}\}$ must be spaced
in order to ensure decoherence may be obtained as follows: Approximate
$\xi(t)$ by a constant value $d$ that is characteristic of the sizes of
the intervals $\{\Delta^k_{\alpha_k}\}$.  Find the time interval
 $t_{\rm
decoherence}$ such that the integral in  \(fourthirtytwo) evaluated over
that time interval is of order of magnitude unity. One then has  
a rough
estimate of a time interval long enough for ${\cal I}mW/\hbar$ 
to be large enough
for $\exp(-{\cal I}mW/\hbar)$ to be small.  The answer is
$$
t_{\rm decoherence} \sim \frac{1}{\Omega}\ \exp\left(\frac{\hbar}{M\gamma
d^2}\right)\ , \tag fourthirtyfour
$$
assuming $\Omega t_{\rm decoherence}>>1$. Note that $t_{\rm decoherence}$
decreases as the coupling $\gamma$ is made stronger or the graining is
made coarser or the number of oscillators becomes larger.  In the
present case of a pure initial state, the time scale for decoherence is
essentially set by the cut-off, for fixed coupling and coarse graining.

The improvement in decoherence from the sum over states in an impure
density
matrix is well illustrated in the linear oscillator models.  In the high
temperature Fokker-Planck limit, where many states contribute with nearly
equal probability to the density matrix $\rho_B$, the imaginary part of
the influence phase is given by [\cf \(ten)]
$$
{\cal I}m W \left[x^\prime(\tau), x(\tau)\right] = \frac{2M\gamma k
T_B}{\hbar}\ \int\nolimits^T_0 dt \left[x^\prime(t) - x(t)\right]^2
\ .\tag fourthirtyfive
$$
The large value of ${\cal I}mW$ suppresses contributions to the
functional integral \(five) defining the decoherence functional when
$x^\prime(t)$ is significantly different from $x(t)$, provided enough
time has elapsed between successive alternatives so  that significant
values of ${\cal I}m W/\hbar$ are built up, 
yielding approximate  medium decoherence.
  Further, in this limit of very large $T_B$ such that
$kT_B>>\hbar \Omega$, there is more efficient medium decoherence than
is provided in the pure ground state example of \(fourthirtyfour).

Another feature that can be illustrated in the linear oscillator models is the
permanence of the past.  We have seen how the successive narrowing of the
records implied by medium decoherence in the case of a pure state gives
a natural explanation of the permanence of the past.  For an impure
state there is no satisfactory
 corresponding notion of generalized record, but in the
oscillator models it is still possible to show how the past becomes
permanent.

History is approximately permanent for a suitably restricted class of
coarse grainings in the oscillator model in the Fokker-Planck
approximation.  The reason is that
decoherence there is essentially local in time.  More precisely, consider the
following integral:
$$
\int\nolimits_{\alpha^\prime} \delta x^\prime \int\nolimits_\alpha \delta
x\, M\left(x^\prime_{k+1}, x_{k+1}\right)\, \exp \Bigl\{i\bigl(S_{\rm free}
[x^\prime(\tau)]
$$
$$ - S_{\rm free} [x(\tau)] + W [x^\prime(\tau),
x(\tau)]\bigr)/\hbar\Bigr\} N\left(x^\prime_k, x_k\right)\ ,
\tag fourthirtysix
$$
where the path integrals, as well as the integrals defining the actions and the
influence phase, are over the interval of time from the time $t_k$ of a
set
of alternatives $\{\alpha_k\}$ to the time of the next set $t_{k+1}$.  
The values $x^\prime_k, x_k$ are the
endpoints at time $t_k$ and $x^\prime_{k+1}, x_{k+1}$ are the 
endpoints at time $t_{k+1}$.
We consider
any functions $M$ and $N$.  The expression \(fourthirtysix) for the
 imaginary part of the influence phase implies
 that, if the intervals $\{\Delta^k_{\alpha_k}\}$ by which
the paths $x(t)$ are coarse grained have a characteristic size $d$ and
the interval between $t_k$ and $t_{k+1}$ 
 is larger than the characteristic decoherence time scale
[\cite{20a}]
$$
t_{\rm decoherence}\sim\frac{1}{\gamma} 
\ \left[\frac{\hbar}{\sqrt{2MkT_B}}\cdot\frac{1}{d}\right]^2\ ,
\tag fourthirtyseven
$$
then the ``off-diagonal'' terms in \(foureight) will be very small and
approximate
decoherence will be achieved for the alternatives $\{\alpha_k\}$ for a
large class of functions $M$ and $N$.

Consider a set of coarse-grained histories defined by regions
$\{\Delta^k_{\alpha_k}\}$ at times $t_1, \cdots, t_n$ separated by time
intervals longer than the characteristic decoherence time
\(fourthirtyseven).   Fine-graining this set by adjoining {\it further}
sets of intervals at similarly spaced times greater than $t_n$ does
not affect the decoherence of those already present because the
mechanism of decoherence exhibited by \(fourthirtysix) is operative over a
time scale of the order of $t_{\rm decoherence}$ about the time of each
set of alternatives.  Physically, that is reasonable.  In this model,
phases are carried away by interactions, local in time,
 of the distinguished variables
$x^a$ with the rest.  Once dispersed among the
continuum of oscillators described by the $Q^A$, they cannot be recovered
by finer grainings beyond  $t_n$
that involve the $x^a$ {\it alone}. 
To recover the phases, one
would need a much finer graining that involved the whole set of
variables.
\taghead{5.}
\vskip .26 in
\centerline{\bf V. Distributions for Decoherence Functionals}
\vskip .13 in
As the review presented in 
 the preceding two Sections makes clear, quantum theory can
be organized into two parts:  First, there is the calculus of amplitudes
for histories or the bilinear combinations that are the decoherence
functionals.  The rules of this calculus derive ultimately from the
principle of superposition.  Second, there are the rules for deriving
probabilities from these amplitudes, most generally, the several notions
of decoherence of sets of alternative histories of a closed system.  In
this section, we show how the first part --- the calculus of decoherence
functionals for histories --- may be usefully reexpressed in terms of
distribution functionals analogous to the Wigner distribution for
alternatives at a single moment of time.

In familiar quantum mechanics the probability that a determination of
coordinates $x^a$ at one moment of time will yield a result in a
volume $V$ of the reduced configuration space spanned by these
co\"ordinates is
$$
 p (V) = \int\nolimits_V dx \tilde\rho (x,x)\ ,
 \tag tena
$$
where $\tilde\rho (x^\prime, x)$ is the reduced density matrix on the reduced
configuration space [cf. \(four)] and $dx$ is the reduced volume element.
As is well known, the density matrix $\tilde\rho (x^\prime, x)$ that gives such
probabilities may be usefully expressed in terms of the Wigner
distribution on phase space
$$\tilde\rho\left(X + {\xi/2}, X - {\xi/2}\right) = \int dP w(X,P)
e^{i\xi^\dagger P/\hbar}\ . \tag tenb
$$
The probability $p(V)$ is then given by
$$
p(V) = \int\nolimits_V dX \int dP w\left(X,P\right)\ .\tag tenc
$$

It is also true that the probability density for the momentum, $P$,
conjugate to $x$ is given by $\int d X w (X,P)$.  In these properties
$w(X,P)$ is like a classical distribution on phase space. Other ways
in which it is similar have been extensively discussed. (See
e.g., [\cite{20, 21}])~~It
differs from a classical distribution in that it is not   
in general positive 
 and it does not provide analogs for
all probabilities that are defined on classical phase
space.
For example, 
it would be incorrect to think of $w(X,P)$ itself as
a probability for a simultaneous
determination of position {\it and} momentum.  The calculus of
amplitudes must therefore be supplemented by the rules that specify which
classical quantities on phase space can be assigned probabilities.
Ignoring those rules, and considering just the calculus of amplitudes 
for alternatives at a single time, we
see from equations \(tenb) and \(tenc) how  the calculus 
may be reformulated in
terms of a distribution function on phase space, although not, in
general, in terms of a positive one.  In the following we shall give an
analogous formulation for the calculus of amplitudes for time histories.

The decoherence functional defined in equation (3.1) is the bilinear
combination of amplitudes of which the diagonal elements give the probabilities
 of the individual histories in a decoherent set [cf
\(fourseven)].  In this
sense, it plays the role for histories that the reduced density matrix
does for alternatives at a single moment of time.  We now construct a
distribution functional for the decoherence function in much the same
 way 
in which the Wigner distribution was constructed from the reduced density
matrix. 

We consider the partially coarse-grained 
decoherence functional $D[x^\prime(\tau), x(\tau)]$ for
histories that are fine-grained
 in the variables $x$ distinguished by the
initial coarse-graining discussed in the previous section.  Introduce variables
$X(t)$ and $\xi(t)$ that are the average and difference  respectively
of the
arguments of the decoherence functional:
$$
X(t) = {1 \over 2} (x^\prime (t) + x (t))\ , \tag tend a
$$
$$
\xi (t) = x^\prime (t) - x(t)\ . \tag tend b
$$
In terms of these variables the fine-grained decoherence functional defined by
equation \(five) may be written
$$
D \bigl[X (\tau), \xi (\tau); X_0, \xi_0\bigr) = \delta (\xi_f)
\exp \left(\frac{i}{\hbar} A \bigl[X(\tau), \xi(\tau); X_0, \xi_0\bigr)\right)
\tilde\rho\left(X_{0} + \frac{\xi_0}{2}, X_{0} - \frac{\xi_0}{2}\right)
\ , \tag tene
$$
where
$$
A \bigl[X (\tau), \xi (\tau); X_0, \xi_0\bigr) = S_{\rm free} \left[X (\tau) +
\xi (\tau)/2\right]
- S_{\rm free} \left[X(\tau) - \xi (\tau)/2\right] 
$$
$$
+ W\bigl[X(\tau),
\xi(\tau); X_0, \xi_0\bigr)\ . \tag tenf
$$

A distribution $G[R(\tau),X(\tau),X_0, \xi_0)$ may be introduced
for the decoherence functional by taking its functional Fourier
transform with respect to $\xi(t)$.
We define $G$ by the formula
$$
D\bigl[X(\tau), \xi(\tau); X_0, \xi_0\bigr) = \int \delta R \exp 
\left[{i \over\hbar}
\int\nolimits^T_0 dt \xi^{\dagger}
 (t) R (t)\right]  G \bigl[R
(\tau), X (\tau); X_0, \xi_0\bigr)\ . \tag teng
$$
The functional $G$ may be calculated from the decoherence functional by an inverse
functional Fourier transform.  Expressions like \(teng) are to be
interpreted as limits of multiple integrals over paths that are
piecewise  
linear between a discrete set of time slices $t_{0} = 0, t_{1},
\cdots,  t_{N} = T$ as the number of slices, $N$, tends to infinity.  This
is a standard way of defining path integrals; details for the
particular integrals of interest are discussed more fully in 
Appendix (c).  As an aid to the present discussion, however, 
it is useful to note
that we always represent the integral such as that
 in the exponent of \(teng) by 
discrete sums of the form
$$
\int\nolimits^T_0 d\tau \xi^{\dagger} (\tau) R (\tau) = \Sigma^N_{k=1} 
\epsilon
\xi^{\dagger}_{k}R_{k}\ , \tag tenh
$$
where $\epsilon$ is the separation between time slices.  The variable
$\xi_0$ is thus not transformed and appears on both sides of the
equation.

The transform variable $R(t)$ has the dimension of force.  As we
shall see later,\break
\noindent $G[X(\tau), R(\tau); X_0, \xi_{0})$ may be regarded as
a classical distribution of $R(t)$, given the path $X(\tau)$ and
$X_0$ and $\xi_0$, although in general a non-positive one.  It is
therefore instructive to consider its moments.  We define
$$
<R(t_1)\cdots R(t_n)>_{c} = \frac{\int \delta R\ R(t_1) ...
R(t_n) \ G [R (\tau), X (\tau); X_0, \xi_0)}
{\int \delta R G [R (\tau), X (\tau); X_0, \xi_0)}\ .  \tag tenj
$$
Clearly, $<R(t_1) \cdots R (t_n)>_c$ is a functional of the path
$X(\tau)$ and a function of $X _0$ and $\xi_0$.

The first moment we {\it define} to be {\it the} average ``total
force'' 
$$
{\cal E}(t, X_0, \xi_0; X(\tau)] \equiv <R(t)>_c \ . \tag tenk
$$
By ``total force'' we mean the force minus the inertial term, so that
${\cal E} = 0$ is the effective or phenomenological classical equation
of motion on the average.
The deviations of the force $R(t)$ from its expected value define the
Langevin force
 ${\cal L}[t, X_0, \xi_0; X (\tau)]$ for a given path $X (\tau)$:
$$
{\cal L}(t, X_0, \xi_0; X (\tau)] \equiv R(t) - {\cal E}(t, X_0, \xi_0; 
X (\tau)]\ .
\tag tenl
$$
The reason for these designations will become clear when we express the
quantum-mechanical probability for an individual history $\alpha$ in a
{\it decoherent} set of histories in terms of the distribution
$G$.

The fully coarse-grained decoherence functional (3.5) is given by
$$
D(\alpha^\prime, \alpha) = \int\nolimits_{\alpha^\prime} \delta x^\prime
 \int\nolimits_\alpha \delta
x D [X(\tau), \xi (\tau); X_0, \xi_0)\ , \tag tenm
$$
where $X$ and $\xi$ are connected to $x^\prime$ and $x$ by  \(tend) 
 and their  range is
 restricted by the coarse-graining through \(tend).  The
partially coarse-grained 
decoherence functional is given by \(tene) and \(tenf).  If the
further graining defined by the successive sets of regions is
coarse enough so that sufficient positive imaginary part of $W$ is
built up between one set of intervals and the next, then there will be a
significant contribution to the integral defining the decoherence
functional only for values of $\xi (t)$ near zero and for $\alpha =
\alpha^\prime$.  (See Figure 1).  That is medium decoherence.  Further, in the
diagonal elements of the decoherence functional, which are the
probabilities of the individual coarse-grained histories, the integral
over $\xi (t)$ may be carried out, to an excellent approximation, as
though unrestricted by the coarse graining, provided the intervals are
sufficiently coarse.  (See Figure 1).
{\topinsert
\centerline{\includegraphics[width=5.0in]{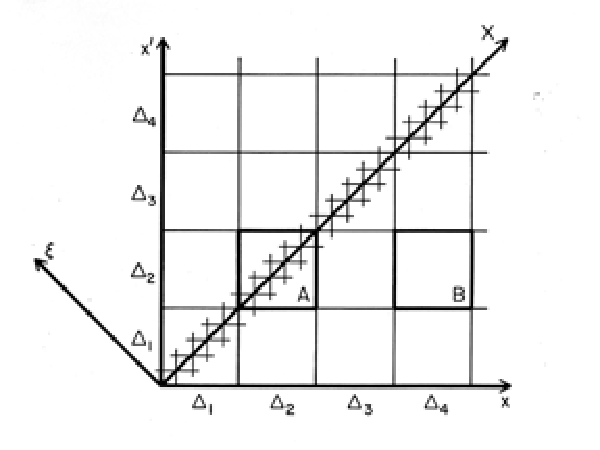}}

\singlespace
\noindent Figure 1.  The decoherence of histories coarse-grained by
intervals of a distinguished set of configuration space co\"ordinates.
The decoherence functional for such sets of histories is defined by the
double path integral of \(five) over paths $x^\prime(t)$ and $x(t)$ that
are restricted by the coarse graining. These path integrals may be
thought of [{\sl cf}. \(hundredtwo)]
as the limits of multiple integrals over the values of
$x^\prime$ and $x$ on a series of discrete time slices of the interval
$[0,T]$.  A typical slice at a time when the range of integration is
constrained by the coarse graining is illustrated. Of course, only
 one of the
distinguished co\"ordinates $x^a$ and its corresponding $x^{\prime a}$
can be shown and we have assumed for illustrative purposes that the
regions defining the coarse-graining correspond to a set of intervals
$\Delta_\alpha, \alpha = 1, 2, 3, \cdots$ of this co\"ordinate.  
On each slice where there is
a restriction from the coarse graining the, integration over $x^\prime$
and $x$ will be restricted to a single box.  For the ``off-diagonal''
elements of the decoherence functional corresponding to distinct
histories, that box will be off the diagonal (e.g. B) for {\it some}
slice.  For the diagonal elements, corresponding to the same histories,
the box will be on the diagonal (e.g. A) for all slices.

If the imaginary part of the influence phase $W[x^\prime(t), x(t)]$
grows as a functional of the difference $\xi(t) = x^\prime (t) - x(t)$,
as it does in the oscillator models [cf \(ten)], then integrand of the
decoherence functional will be negligible except when $x^\prime(t)$ is
close to $x(t)$ a regime
illustrated by the shaded band about the diagonal in the
figure.  When the characteristic sizes of the intervals $\Delta_\alpha$
are large compared to the width of the band in which the integrand is
non-zero   the off-diagonal elements of the
decoherence functional will be negligible because integrals over those
slices where the histories are distinct is negligible (e.g.~over box B).
That is decoherence of the coarse-grained set of histories.  Further
concerns the evaluation of the diagonal elements of the decoherence
functional that give the probabilities of the individual histories in
decoherent set can be simplified.  
If the integrations over $x^\prime$ and $x$ are
transformed to integrations over $\xi = x^\prime - x$ and $X=(x^\prime +
x)/2$ the restrictions on the range of the $\xi$-integration to one
diagonal box may be neglected with negligible error to the probability.
\endinsert}


When re\"expressed in terms of the distribution $G$, the probabilities
for the individual histories $p(\alpha)$ are, in this approximation, 
$$
p(\alpha) = \int\nolimits_\alpha \delta X \int \delta R \int \delta \xi
\  G
[R(\tau),  X (\tau); X_0, \xi_0) \exp \Bigg [{i \over\hbar}
\int\nolimits^T_0 dt \xi^\dagger (t) R (t)\Bigg]\ , \tag tenn
$$
where the integrals over $\xi(t)$, including that over $\xi_0$, are
unconstrained.

The expression \(tenn) for the probabilities of the individual histories
in a coarse-grained set has an especially transparent interpretation in
the special case where the initial density matrix factors as in \(eight)
and a simple kinetic minus potential energy form \(six) is assumed for
the action $S_{\rm free} [x(\tau)]$.  A simplification following from
factorization is that the influence phase $W$ has no direct dependence
on $x^\prime_0$ and $x_0$, and hence on $X_0$ and $\xi_0$, except through
the paths $x^\prime(\tau)$ and $x(\tau)$.  Thus, we write simply
$W[x^\prime(\tau), x(\tau)]$.  More accurately, in a time slicing
implementation of the functional integral in which integrals over the
paths are discretized as in \(tenh), $W$ is independent of $X_0$ and
$\xi_0$.  Further discussion is given in Appendix (b).  When the action
$S_{\rm free} [x(\tau)]$ has a simple kinetic minus potential form the
dependence of the rest of the exponent $A[X(\tau), \xi(\tau); X_0,
\xi_0)$ may be easily isolated.  An integration by parts in \(tenf)
yields
$$
A\bigl[X(\tau), \xi(\tau); X_0, \xi_0\bigr) = - \xi^\dagger_0 M\dot X_0 +
\tilde A \bigl[X(\tau), \xi(\tau)\bigr]\ , \tag tenna
$$
where
$$
\tilde A\bigl[X(\tau), \xi(\tau)\bigr] = \int\nolimits^T_0 dt
\Bigl[-\xi^\dagger(t) M\ddot X (t) - V\left(X(t) +
\xi(t)/2\right) + V\left(X(t) - \xi(t)/2\right)\Bigr]
$$
$$
+ W\bigl[X(\tau), \xi(\tau)\bigr]\ .\tag tennb
$$
Then, under the ground rules for discretizing functionals discussed above,
the functional $\tilde A$ is independent of both $\xi_0$ and $X_0$. In
particular, the only terms in the fine-grained decoherence functional
\(tene) that depend explicitly on $\xi_0$ are the density matrix and the
surface term in \(tenna).  The integral over $\xi_0$ may therefore be
carried out in \(tenn), giving
$$
p(\alpha) = \int\nolimits_\alpha \delta X\int \delta R
\int \delta\xi
\ g\bigl[R(\tau), X(\tau)\bigr]\ w(X_0, M\dot X_0)\exp
\Bigl[\frac{i}{\hbar} \int\nolimits^T_0 dt \xi^\dagger(t) R(t)\Bigr]\ .
 \tag tennc
$$
where $w$ is the Wigner function defined by \(tenb) and the distribution
$g[R(\tau), X(\tau)]$ is defined by
$$
\delta(\xi_f)\exp\Bigl[\frac{i}{\hbar}\tilde A\bigl[X(\tau), \xi(\tau)
\bigr]\Bigr] = \int \delta R\ g\bigl[R(\tau), X(\tau)\bigr]\exp
\Bigl[\frac{i}{\hbar} \int\nolimits^T_0 dt \xi^\dagger(t) R(t)\Bigr]\ .
\tag tennd
$$
These expressions acquire precise meaning in the time slicing
implementation of the path integrals discussed in Appendix (b). The
important point for the present discussion is that the assumptions that
the initial $\rho$ factorizes and that $S_{\rm free}$ has a simple kinetic
minus potential energy form  lead to a factorization of the general
distribution $G$ into a distribution, $w$, of initial values $X_0$ and
$M\dot X_0$ and a
distribution, $g$,  involving forces $R(t)$ and paths $X(t)$.  
Further, the equation of motion ${\cal E}(t, X(\tau)]$ and the Langevin
force ${\cal L}(t, X(\tau)]$ become independent of the initial
conditions and calculable just from the distribution $g$ of force $R(t)$
through expressions of the same form as \(tenj), \(tenk), and \(tenl)
with $G$ replaced by $g$. These features allow a simple interpretation,
which we shortly describe.

The Wigner distribution by which the initial conditions are distributed
in the expression for the probabilities \(tennc) is not generally
positive and neither is the distribution $g$.  However, the
probabilities $p(\alpha)$ must be positive.  It is not difficult to
see how this comes about.  Were the restriction on the range of
integration arising from the coarse graining restored in expressions
\(tennb) or \(tennc), the numbers $p(\alpha)$ they define would be
manifestly positive.  That is because they are expressions for a
diagonal element of a decoherence functional which is always positive
[\cf \(fourfive)].  This is not unlike other smearings of the Wigner
distribution [\cite{44}] which are known to give generally positive
results.  Of course, the approximation in which the restrictions of the
coarse graining on the integration are ignored may result in small negative
probabilities, but, to the extent the approximation is good, these are
equivalent to zero for physical purposes. 

The representation \(tennc) allows the probabilities of decohering
 coarse-grained
histories $p(\alpha)$ to be thought of as the probabilities of the
histories of a system moving classically under the action of a
stochastic force.  The Wigner function gives the distribution of initial
conditions.  The distribution $g[R(\tau), X(\tau)]$ may be thought of as
the distribution of total force $R(t)$ acting on a system that describes
the path $X(t)$.  Alternatively, if re\"expressed in terms of ${\cal
L}(t, X(\tau)]$ and ${\cal E}(t, X(\tau)]$ defined by \(tenm) and
\(tenl), $g\Bigl[{\cal E}(t, X(\tau)] + {\cal L}(t, X(\tau)]; X(\tau)
\Bigr]$ 
may be
thought of as the distribution of ${\cal L}(t)$ given the path $X(t)$.
In the approximation we have discussed, the 
unconstrained integration over $\xi(t)$ in \(tennc) leads to a
functional $\delta$-function that enforces the condition, $R(t)=0$, that
the ``total force'' on the system vanish.  That is, it enforces the
effective
classical equation of motion, corrected by the Langevin force:
$$
{\cal E}\bigl(t, X(\tau)\bigr] + {\cal L}\bigl(t, X(\tau)\bigr] = 0\ .
\tag tenne
$$
In this equation the average total force 
 ${\cal E}(t, X(\tau)]$ is the  known functional of
$X(\tau)$ defined by \(tenk). The force ${\cal L}(t, X(\tau)]$ is
distributed according to the distribution $g$.  It is for this reason that we
have called ${\cal L}$ the Langevin force; it can be thought of as
 noise.  

The probabilities $p(\alpha)$ for decohering coarse-grained histories
are thus obtained from a mathematical description of 
classical dynamical system characterized by an
equation of motion ${\cal E}(t, X(\tau)]$ together with distributed
initial conditions and distributed noise.  We should stress that this
does not mean that quantum mechanics is equivalent to some kind of
classical physics.
  For one thing, the distributions of noise and initial
conditions are not generally positive and that is certainly a
non-classical feature.  For another thing, there is the coarse graining
needed for 
decoherence --- an entirely non-classical requirement that must be
satisfied before the $p(\alpha)$ may be considered as the probabilities
of histories.  

The non-positivity that distinguishes quantum-mechanical distributions
from classical ones may be regarded as the reason that Bell's classical
inequalities [\cite{37}] are violated in quantum mechanics, leading to
important experimental tests of the theory.  Of course, the Wigner
distribution considered here is for continuous variables while
Bell's discussion of the EPRB problem was for a discrete spin system.
However, Feynman [\cite{32}] showed that there is an analog of the
Wigner distribution for spin-$\half$ systems and that the departure of
quantum mechanics from Bell's inequality is traceable to the fact that
the analogous distribution is not generally positive.
At the level of the calculus of amplitudes, the difference between
classical and quantum mechanics is just the possibility of negative
distributions.  That possibility alone  does not completely
characterize the
difference when we go beyond the calculus of  amplitudes,
because in quantum mechanics
 we also have the requirement of decoherence of
histories.

When the initial $\rho$ does not factor, or when $S_{\rm free}$ is not of
simple kinetic minus potential energy form, the interpretation of the
general expression \(tenn) for the probabilities is less direct.
However, building on the analogy of the special case, we may still think
of $G[R(\tau), X(\tau); X_0, \xi_0)$ as a combined distribution of total
force and initial conditions given the path $X(\tau)$.  Now, however,
the distribution of the force $R(\tau)$ is not independent of the
initial conditions but depends on them.  Further, the distribution of
initial momenta is not given directly but only implicitly through the
integral over $\xi_0$.  Finally, the equation of motion and noise depend
on the initial values of $X_0$ and $\xi_0$.  The integral over $\xi(t)$
in \(tenn) continues to enforce the
classical condition that the total force $R(t) = 0$ vanish.  These
features complicate the interpretation of \(tenn) but they do not
vitiate its validity.

When the noise ${\cal L}(t, X(t)]$ is small compared the equation of
motion term ${\cal E}(t, X(t)]$ in \(tenne), we expect approximate
classical determinism.  More precisely we expect significant
probabilities for histories correlated in time by the classical equation
${\cal E}(t, X(\tau)] = 0$ with small deviations produced by the noise.
In the following Sections we shall analyse the circumstances where this
is so.
\taghead{6.}
\vskip .26 in
\centerline{\bf VI. Linear Systems}
\vskip .13 in
\centerline{\sl (a) Equations of Motion}

We begin our discussion of the derivation of classical equations of
motion for quantum systems by considering the simplest possible example
--- linear systems. 
  This is the class of models studied by Feynman and Vernon [\cite {12}], 
Caldeira and Leggett [\cite {11}], and many others and for which there is a
wealth of information available on the specific forms of the influence
functional, its dependence on the initial condition, etc.  
Either
implicitly or explicitly, equations of motion have been considered 
for these models by several authors.
  We are thus on familiar territory.

Linear systems may be characterized precisely, following Feynman and
Vernon, by the following two requirements:

\item{1.} A free action for the variables distinguished by the coarse
graining with a kinetic energy that is quadratic in the co\"ordinates
and their velocities
$$
S_{{\rm free}}[x(t)] =\int\nolimits^T_0
dt\Big[{1\over 2}~\dot x^\dagger (t)M \dot x(t) -
{1\over 2}~x^\dagger(t)Kx(t)\Big]. \tag eleven
$$
We use here and throughout an obvious vector notation so that $M$ and
$K$ are positive constant matrices and $(1/2)(\dot x^{\dagger} Mx) = (1/2) \Sigma_{ab} \dot
x^a M_{ab} \dot x^b$, etc.,~in this case where the variables are real.

\item{2.} An influence phase that is at most quadratic in the variables
$x(t)$.  Its most general form has been deduced by Feynman and
Vernon [\cite {12}] from general symmetries and quantum-mechanical 
causality. (See also the exposition in [\cite {22}].) It consists of
terms linear in the distinguished variables $x(t)$ and $x^\prime(t)$ and
terms quadratic in them.  The linear terms may be eliminated by a
time-dependent shift in $x(t)$ and for simplicity we imagine this has
been done.  The general form is then 
$$
W \left[x^\prime(t), x(t)\right] = 
{1\over 2}\int^T_0 dt \int^t_0
 dt^\prime\left[x^\prime(t)-x(t)\right]^{\dagger} 
\left[k(t,t^\prime)x^\prime(t^\prime)
+ k^*(t,t^\prime)x(t^\prime)\right]
$$
$$
={1\over 2} \int^T_0 dt \int^t_0 dt^\prime
[x^\prime(t)-x(t)]^\dagger
\Bigl\{k_R(t,t^\prime)[x^\prime(t^\prime) + x(t^\prime)] + i
k_I(t,t^\prime) [x^\prime(t^\prime) - x(t^\prime)]\Bigr\}\ ,
\tag twelve
$$
where $k(t,t^\prime)$ is a complex matrix kernel with real and imaginary parts $k_R(t,t^\prime)$ and
$k_I(t,t^\prime)$ respectively. We shall explicitly assume that $W$
depends on $x_0$ and $x^\prime_0$ only implicitly through its dependence
on the paths $x^\prime(t)$ and $x(t)$ and not explicitly as in the
general case. 

Only under very restrictive conditions will the influence phase be
exactly quadratic as in \(twelve)  with no explicit dependence on
$x^\prime_0$ and $x_0$. Such an influence phase will
certainly follow if~~(i) the action of the ignored variables,
$S_0[Q(t)]$, is quadratic in the $Q$'s; ~(ii) the interaction between
the distinguished and ignored variables is exactly linear in each,
giving
$$
S_{{\rm int}} \left[x(t), Q(t)\right] = -\int\nolimits^T_0 dt
 x^{\dagger}(t)f(Q(t))\ , \tag thirteen
$$
where the force $f$ is homogeneous and
 linear in the $Q$'s;  and~~(iii) the initial density matrix factors as
in \(eight) with a density matrix $\rho_B (Q^\prime_0, Q_0)$ that is of the
form
$$
\rho_B (Q^\prime_0, Q_0) = \exp\left[-B(Q^\prime_0, Q_0)\right]
\ ,\tag thirteena
$$
where $B(Q^\prime_0, Q_0)$ is quadratic in its arguments.  Under these
conditions the integral in \(three) is a Gaussian and a quadratic
influence phase results.  The models defined by \(eleven), \(twelve),
\(six), and \(thirteena) are, in fact, just those considered by Feynman
and Vernon [\cite {12}] and Caldeira and Leggett [\cite {11}].

A useful explicit example, studied in [\cite {11, 12}], is the case of a single
co\"ordinate, $x$, interacting with an assembly of ``bath''
 oscillators, with an
initial condition that factors as in \(eight) and  with a state of thermal
equilibrium for the ``bath''
 oscillators at temperature $T=(k\beta)^{-1}$.
The oscillators are described by a free action
$$
S_0[Q(t)] = \sum_A \int^T_0\ dt\ \left[\half m\left(\dot Q^A(t)\right)^2 -
\half m\omega^2_A\left(Q^A(t)\right)^2\right]\ .\tag thirteenb
$$
The force $f$ entering \(thirteen) may be written explicitly as
$$
f\bigl(Q(t)\bigr) = \sum_A C_A Q^A\tag thirteenc
$$
with coupling  constants $C_A$, and the function $B$ is independent of
$x^\prime_0$ and $x_0$ and given, up to an additive constant
(determinable through normalizing $\rho_B$), by the expression
$$
B\left(Q^\prime_0, Q_0\right) = \sum_A
\ \frac{m\omega_A}{2\hbar\sinh(\hbar\beta\omega_A)}
\ \Bigl\{\left[(Q^{\prime A}_0)^2 +
(Q^A_0)^2\right]\cosh(\hbar\beta\omega_A) - 2Q^{\prime A}_0 Q^A_0\Bigr\}
\ .
\tag thirteend
$$
Then, as in [\cite {11}] or [\cite {12}],
$$
k_R(t,t^\prime) = -\sum_A\ \frac{C^2_A}{m\omega_A}\ \sin \omega_A
(t-t^\prime)\tag thirteene a
$$
and
$$
k_I (t,t^\prime) = \sum_A\ \frac{C^2_A}{m\omega_A} \coth \left(\half
\hbar\beta\omega_A\right)\cos\omega_A(t-t^\prime)\ .\tag thirteene b
$$

  The  influence phase quoted in eq.~\(ten) for a
distinguished oscillator interacting with a high temperature thermal
bath is a special case of these expressions in which a continuum limit of
oscillators with the special couplings
$$
\rho_D(\omega)C^2(\omega) = 4mM\gamma\omega^2/\pi
\tag thirteenf
$$
below the cutoff was assumed, where $\rho_D$ is the density of states.
With this coupling,
we have the following, with the high temperature limit needed only
in (6.10b) and the high cut-off, which facilitates phase dispersal, used
in both equations:
$$
\eqalignno{
k_R(t,t^\prime)& =- 4M\gamma\delta^\prime(t-t^\prime)\ ,&(fourteen
a)\cr
k_I(t,t^\prime)& = {8M\gamma
kT_B \over \hbar}\delta(t-t^\prime)\ , &(fourteen b)\cr
}
$$
where the $\delta^\prime$ function occurring in retarded integrals
 is a distribution such
that
$$
\int\nolimits^t _{-\infty}
 g(t^\prime)[\delta^\prime(t-t^\prime)]\  dt^\prime = {1\over
2}~g^\prime (t)\ .\tag fourteen c
$$
[Eq.~\(fourteen a) gives the renormalized value of $k_R(t,t^\prime)$.  Infinite
terms in \(twelve) proportional to $\delta(t-t^\prime)$ that arise from the
continuum limit taken by Caldeira and Leggett have been absorbed in a
renormalization of the frequency of the distinguished oscillator in its
free action.]
Although the influence phase is exactly quadratic only under restrictive
circumstances, the linear cases supply  useful models for more general ones,
as we shall see.  

With these preliminaries, we can now give a derivation of the classical
equation of motion for these linear systems.  The equation we shall
derive is, of course, the same as considered, for example, by Caldeira and
Leggett.
Even in this linear case, however, 
we believe that there are several important new features of this 
derivation. It is consistent with the general discussion of the
average equation of motion and noise in Section V. 
The probabilities that coarse-grained histories of the distinguished
particle are correlated in time by equations of motion are explicitly
considered.  The form of the phenomenological equation of motion is
derived from a consideration of these probabilities.  
The amount of coarse graining necessary for the  decoherence
of histories and their 
classical behavior is discussed quantitatively, and the connection 
between decoherence
and quantum noise is made explicit.  Most importantly, the derivation
suggests how the generalization to the nonlinear case is to be carried
out.  

The imaginary term in the influence phase \(twelve) gives rise to
decoherence between the trajectories of the distinguished variables,  provided
that the coarse graining is such that the integral of this term is
sufficiently large for different coarse-grained histories so that the
corresponding ``off-diagonal'' elements of the decoherence functional
are exponentially small.
To exhibit this decoherence explicitly it is useful to change variables,
in the integral \(five) defining the decoherence functional, from
$x^\prime(t)$ and $x(t)$ to the average and difference $X(t)$ and
$\xi(t)$ defined by \(tend).  The exponent in the decoherence functional
\(five) can be explicitly expressed in terms of $X(t)$ and $\xi(t)$ using
\(twelve). Denoting this exponent by $A[X(\tau), \xi (\tau)]$ as in
\(tenf), one finds after a few integrations by parts that
$$
D(\alpha^\prime, \alpha) = \int\nolimits_{(\alpha^\prime, \alpha)}
\delta X \delta\xi\ \delta(\xi_f)\ \exp\bigl(iA[X(\tau),
\xi(\tau)]/\hbar\bigr)\tilde\rho\left(X_0 + \xi_0/2, X_0 -
\xi_0/2\right)\ ,\tag fifteen
$$
where
$$
A[X(\tau), \xi(\tau)]
= -\xi^\dagger_0 M\dot X_0
+\int^T_0 dt \xi^\dagger(t) e(t) + {i\over 4} 
\int^T_0
dt \int^T_0\ dt^\prime\xi^{\dagger}(t) k_I(t,t^\prime)\xi(t^\prime)
\ .\tag sixteen a
$$
Here, $e(t)$ is the average ``total force''
$$
e(t) = -M\ddot X(t) - KX(t)+ \int^t_0\
dt^\prime k_R(t,t^\prime)X(t^\prime)\ ,\tag sixteen b
$$
and $k_I$ is defined to be symmetric in its argument, so that the limit of integration on $t^\prime$
can be extended to $T$. 
The quantities $\dot X_0$, $\ddot X_0$, etc. are understood in the 
usual path integral
sense as finite difference expressions in a time-sliced implementation
of the path integral. (See Appendix (b).) The integrals over $X(t)$ and
$\xi(t)$ are constrained by the coarse graining defining the histories
$\alpha^\prime$ and $\alpha$.

The imaginary term in (6.12a) leads to decoherence. We shall give below,
in the discussion of noise, an explicit construction of the
kernel $k_I$ that shows it to be a positive kernel.  The
imaginary part of $W$ is proportional to $k_I$ and occurs in
the expression $\exp(iW/\hbar)$, giving a decreasing exponential.  If the
 graining
defined by the successive sets of regions $\{\Delta^k_{\alpha_k}\}$
is coarse enough so that sufficient positive imaginary part of $W$ is
built up between one set of intervals and the next, then there will be a
significant contribution to the integral defining the decoherence
functional only for values of $\xi(t)$ near zero and for 
$\alpha=\alpha^\prime$.  (See Figure 1).  That is medium decoherence.  

If only
values of $\xi(t)$ near zero contribute significantly to the integral
\(fifteen), then  
in the diagonal elements of the decoherence functional, which are the
probabilities of the individual coarse-grained histories, the integral
over $\xi(t)$ may be carried out, to an excellent approximation, as though
 unrestricted by the coarse
graining when the intervals are
sufficiently coarse.  (See Figure 1). The integral over $\xi_0$ leads to
the Wigner distribution \(tenb) as in \(tennc).  The result of the
unrestricted Gaussian integrals over the rest of $\xi(t)$ is again a
Gaussian functional.  We thus obtain for the probabilities of the
individual histories in
the coarse-grained set, the following expression\footnote{$^8$}{Dowker
and Halliwell [\cite{40}] have obtained analogous expressions in linear
models for the probabilities of histories defined by a finite number of
``Gaussian slits''.}:
$$
p(\alpha) \cong \int_\alpha \delta X
\Bigl[det (k_I/4\pi)\Bigr]^{-{1\over 2}}\exp\biggl[-{1\over \hbar}
\int^T_0 dt \int^T_0 dt^\prime e^\dagger
(t) k^{inv}_I(t,t^\prime) e(t^\prime)\biggr]
 w(X_0, M\dot X_0)\ , \tag eighteen
$$
where $k^{inv}_I$ is the inverse kernel to $k_I$.
The integral is over all paths that proceed from $t=0$ to $t=T$ and lie
in the class corresponding to the coarse-grained history $\alpha$.  The
integral includes an integration over the initial and final endpoints
$X_0$ and $X_f$ respectively.  
Again, for further details and an
explicit representation of \(eighteen), see Appendix (b).

The Gaussian exponential in \(eighteen) means that, for given $X_0$ and $M\dot
X_0$, the histories with the largest probabilities will be those with
$e(t)=0$, that is, those for which the time evolution is correlated 
according to
the effective average classical equation of motion
$$
e(t) = -M\ddot X(t) - KX(t)+ \int^t_0\
dt^\prime k_R(t,t^\prime)X(t^\prime)=0\ . \tag nineteen
$$
This is, of course,  not the equation of motion following from the free action
 of the
distinguished oscillator.  It differs 
by the additional force, non-local in time, that arises from the
 interactions of the distinguished variables with the rest.
  The  presence of such a force will in general mean that energy
is not conserved, leading sometimes to dissipation.
 Although non-local in time, the additional force in \(nineteen) is
retarded, expressing classical causality.  The origin of this
retardation can be traced to the retarded form of the general influence
phase \(twelve).  That, in turn, follows from quantum-mechanical causality
the fact that
 the decoherence functional has a trace in the future and a density
matrix differing from the unit matrix in
the past.  Causality in quantum mechanics thus implies the causality of
classical physics.

A special case of a linear system is the Fokker-Planck limit of the
oscillator model, for which the influence phase is exhibited in \(ten). 
 With the
corresponding $k_R$ of 
eq.~(6.10a), the equation of motion away from $t=0$ becomes\footnote{$^9$}
{We are thus for simplicity ignoring the terms proportional to
$x(0)$ that arise when the integral in \(sixteen b) is carried out using
\(fourteen).  For further discussion see [\cite{36}]. We thank J.P.~Paz
for a discussion of this point.}
$$
e(t) = - M\ddot X(t) - KX(t) - 2M\gamma\dot X(t) = 0\ .
\tag twenty
$$
This is local in time, but that is a special property of the 
way in which the limit of a continuous spectrum of oscillators was taken
in the Caldeira-Leggett model, not a general one. In that limit,
  eq.~\(twenty) explicitly exhibits the familiar form of  frictional
dissipation, not necessarily a general characteristic of the additional
force in \(nineteen).

The individual classical histories in \(eighteen) are distributed according to
the probabilities of their initial conditions $X_0$ and $P_0=M\dot X_0$
given by the Wigner function $w(X_0, P_0)$.  Although the Wigner
function is not generally positive, we know, as discussed in Section V,
that
 apart perhaps from small errors introduced by the approximation in
which the constraints of the coarse graining on the $\xi$-integrations
were neglected, the result of the integral \(eighteen) must be positive
even though the Wigner function is not.\footnote{$^{10}$}{For explicit
examples of this see [\cite{45}].}
\vskip .13 in
\centerline{\sl (b) Noise and Predictability}

The distribution of probabilities for histories \(eighteen) predicts the
largest probability for histories obeying the classical equations of
motion but also predicts probabilities for deviations from classical
predictability.  Those give the noise, including quantum noise.  The
same
interactions (of the
distinguished variables with the others) that carry away phase
information to produce decoherence also produce the quantum and
classical-statistical 
buffeting of the trajectory of the distinguished variables
that constitutes the noise.

In Section V we showed how the probabilities of coarse-grained histories
$p(\alpha)$ 
could be thought of as the probabilities of a classical dynamical
problem with (generally non-positive) distributions of force and initial
conditions.  We now specialize that discussion to the linear systems of
the present Section.

Since we have assumed factorization of the initial $\rho$, we may compute
separate distributions for the initial conditions and for the total
force as in \(tennc).  The initial conditions are distributed according
to the Wigner distribution as shown by that equation or directly from
\(thirteen).  The total force is then distributed according to the
distribution function $g[R(\tau), X(\tau)]$ defined by \(tennd).  Using
(6.12a) for $A$ and  \(tenna) to define $\tilde A$, we may calculate
$g[R(\tau), X(\tau)]$ directly.  It is
$$
g[R(\tau), X(\tau)] = [{\rm det}(k_I/2\pi)]^{-\half}
$$
$$
\times\exp\left[-\frac{1}{\hbar} \int\nolimits^T_0 dt \int\nolimits^T_0
dt^\prime \bigl(R(t) - e(t)\bigr)^\dagger k^{\rm inv}_I (t, t^\prime)
\bigl(R(t^\prime) - e(t^\prime)\bigr)\right]\ ,
\tag twentya
$$
where $e(t)$ is given by (6.12b) and the precise meaning of the
inverse kernel $k^{inv}_I (t, t^\prime)$ is discussed in the Appendix.

Evidently, we have, for this case, 
$$
{\cal E}(t, X(\tau)] \equiv \langle R(t)\rangle_c = e(t)\ .
\tag twentyb
$$
The equation of motion {\it defined} as the expected value of $R$ in
Section Vb, therefore, coincides with the equation of motion $e(t)$
whose correlations are favored by the probabilities \(eighteen). The
Langevin force ${\cal L} (t, X(\tau)]$ that governs the deviations from
classical predicability is distributed according to $g[e(t) + {\cal L}
(t, X(\tau)], X(\tau)]$. 
As is easily seen from \(twentya) for the
linear models under consideration this noise is distributed with a {\it
positive} Gaussian probability distribution that is independent of the
path $X(\tau)$.  To emphasize this we write
$$
{\cal L}(t, X(\tau)] = \ell(t)\tag twentyc
$$
in the linear case, and $\ell(t)$ is then distributed according to
$$
\bigl[{\rm det}(k_I/2\pi)\bigr]^{-\half}\ \exp\left[-\frac{1}{\hbar}
\int\nolimits^T_0 dt \int\nolimits^T_0 dt^\prime \ell^\dagger(t) k^{\rm inv}_I
(t, t^\prime) \ell(t^\prime)\right]\ .\tag twentyd
$$
The spectrum of the Gaussian noise is summarized by the formula
$$
\left\langle\ell(t)\ell(t^\prime)\right\rangle_c 
= (\hbar /2) k_I(t,t^\prime) 
\ .\tag twentyfive
$$
In the Fokker-Planck limit of the oscillator model, we have
$$
\left\langle\ell(t)\ell(t^\prime)\right\rangle_c = 4M\gamma
kT_B \delta(t-t^\prime)\ ,\tag twentysix
$$
giving rise to the model for Brownian motion
discussed by Caldeira and Leggett[\cite{11}].

For linear systems instructive, explicit  expressions for the functions
 $k_R(t,t^\prime)$, and $k_I(t,t^\prime)$, which describe the
influence of the ignored variables on the distinguished ones, may be
obtained in terms of quantum-mechanical expectation values of the force
$f(Q)$ defined by \(thirteen).  As these are straightforwardly derived as
special cases of the similar formulae, applicable to non-linear
situations, to be discussed in Section VI, we shall just quote the
results here.

Consider the Hilbert space of square-integrable functions in the ignored
$Q$'s.  Define an expected value, $\langle\ \rangle_0$, of an operator $A(t)$
evolving by the Hamiltonian of $S_0$ by
$$
\left\langle A(t)\right\rangle_0 = Sp \left[A(t)\rho_B\right]\ ,
\tag twentyseven
$$
where $Sp$ denotes the trace over the ignored variables.
The subscript zero  
means that the time dependence of the operators inside
the expected value is calculated using the Hamiltonian $H_0$ of the
$Q$'s alone neglecting interactions with the $x$'s.

As a consequence of our convention that the influence phase has no terms
linear in $x(t)$, the expected value of $f(Q(t))$ vanishes: 
$$
\Bigl\langle f(Q(t))\Bigr\rangle_0 = 0\ .\tag twentynine
$$
The real and imaginary parts of the kernel $k(t,t^\prime)$ may be
expressed in terms of expected values of fluctuations in the force as
follows
$$
\hbar k_R(t,t^\prime)= i\Bigl\langle\Bigl[f(Q(t)),
 f(Q(t^\prime))\Bigr]\Bigr\rangle_0\ ,\tag thirty
$$
$$
\hbar k_I(t,t^\prime)=\Bigl\langle\Bigl\{f(Q(t)),
 f(Q(t^\prime))\Bigr\}\Bigr\rangle_0\ ,\tag thirtyone
$$
where $[\ ,\ ]$ and $\{\ ,\ \}$ denote the commutator and anticommutator
respectively and the matrix elements of $k$ are understood to be the
tensor product of the $f$'s.
Expression \(thirtyone) shows explicitly that the kernel
$k_I(t,t^\prime)$ is positive in the sense that
$$
\int^T_0 dt \int^T_0 dt^\prime \xi^{\dagger}(t)
k_I(t,t^\prime)\xi(t^\prime)\geq 0\tag thirtytwo
$$
for any real vector $\xi(t)$.
The same equation and \(twentyfive)
 demonstrate that the spectrum of the random Gaussian force
in the Langevin equation $e(t) + \ell(t) =0$ is directly given by the quantum
correlation function of the fluctuation in the force $f(Q)$, {\it viz.}
$$
\left\langle\ell(t)\ell(t^\prime)\right\rangle_c = \half
\Bigl\langle\Bigl\{f(Q(t)), f(Q(t^\prime))\Bigr\}\Bigr\rangle_0~.
\tag thirtythree
$$

Expressions \(thirty) and \(thirtyone) lead to the
essential content of the fluctuation-dissipation
theorem.  If $\rho_B$ is diagonal in the energy
representation defined by the Hamiltonian of $S_0$ as it is  for
equilibrium distributions, then $k_R(t,t^\prime)$ and $k_I(t,t^\prime)$ are
functions of the time difference $t-t^\prime$, and their spectral
weights are simply related. More specifically, it follows from the
symmetries of commutator and anticommutator that we could write
$$
\eqalignno{
k_R(t,t^\prime) & = \int^\infty_0 d\omega\ \tilde k_R(\omega) \sin
\omega(t-t^\prime)\ , & (thirtyfour a)\cr
k_I(t,t^\prime) & = \int^\infty_0 d\omega\ \tilde k_I (\omega) \cos
\omega(t-t^\prime)\ . & (thirtyfour b)\cr
}
$$
If $p_i$ is the probability of an energy eigenstate $|i\rangle$ with
eigenvalue $E_i$ in the density matrix $\rho_B$, we have 
$$
\eqalignno{
\hbar \tilde k_R(\omega)& = \sum_{ij}\ i(p_i-p_j) \Bigl|\Bigl\langle
i\Bigl| f(Q(0))\Bigr|j\Bigr\rangle\Bigr|^2 \delta(E_j-E_i-\hbar\omega)
\ ,& (thirtyfive a) \cr
\hbar \tilde k_I(\omega) & = \sum_{ij}\ (p_i + p_j)\Bigl|\Bigl\langle i
\Bigl| f(Q(0))\Bigr| j \Bigr\rangle\Bigr|^2 \delta(E_j-E_i-\hbar\omega)
\ .& (thirtyfive b) \cr}
$$
Then, in a thermal bath where $p_i = \exp (-\beta E_i)/Z$, we recover the
famous relation (see, \eg [\cite {14}])
$$
\hbar\tilde k_I(\omega) = \hbar~\coth\left({\beta \omega 
\hbar\over 2}\right)\tilde k_R
 (\omega)\ .
\tag thirtysix
$$
This connects the kernel $k_R(t, t^\prime)$ governing the effective
force in \(nineteen) with the kernel $k_I(t, t^\prime)$ governing the
fluctuations in \(twentyfive).  This connection is the
fluctuation-dissipation theorem.

An important fact that emerges clearly from these linear models is that
the same coarse graining and
 interactions that accomplish decoherence also lead to
dissipation and noise.  The fluctuation-dissipation theorem derived
above
 is a well
known example of the connection between noise and dissipation.  The
connection between decoherence and these two phenomena appears to have
been less widely stressed.  We now consider it.

There is, in effect, a competition between decoherence and classical
predictability.  Consider, for example, a model of an oscillator
interacting with a thermal bath.
Increase the temperature of the thermal bath, $T_B$, and
decoherence is more effective.  The characteristic time $t_{\rm decoh}$,
by which successive intervals of typical length $d$ must be spaced to
give decoherence, {\it decreases} with $T_B$ according to \(fourseven).
However, deviations from
classical predictability expressed by \(eighteen)  
also increase.  To
ensure {\it both} classical predictability and decoherence we must consider a
further limit, the limit of high inertia.  In the present model, 
that is the limit of 
 large $M$, and the exponent in \(eighteen) can be written 
$$
-{M\over 8\gamma kT_B}\ \int^T_0\ dt\ \left[\ddot X +
M^{-1} V^\prime(x) + 2\gamma\dot X\right]^2\ .\tag thirtyseven
$$
Assuming that $V$ does not increase faster
than $M$, we see that the probabilities for histories
will become sharply peaked about the certainties implied by a classical
equation of motion in the limit of large $M/T_B$ even as $T_B$ itself is
becoming large to ensure efficient decoherence.

Of course, in realistic situations 
the parameters of a given system, such as the mass, are fixed.  The same kind
of limit of high inertia can be achieved, however, by considering
coarser and coarser graining of a kind that keeps increasing the inertia of the
variables distinguished by the coarse graining.  Take the case of coarse
grainings defined by hydrodynamic variables that are integrals, over
suitable volumes, of
densities of exactly or 
 approximately conserved quantities such as mass, energy,
or momentum.  By making the size of these
volumes larger, the resistance to noise can be increased.  In the present
model, decoherence and classical predictibility can be achieved only by
varying the parameters of the model.  In realistic situations they are 
achieved by a suitably coarse graining.

Classical behavior requires sufficient coarse graining and
 interactions for decoherence but
sufficient ``inertia'' to resist the deviation from predictability that
the coarse graining and
interactions produce.  Traditionally other descriptions have been
given of the requirements for classical behavior of measured subsystems.
Large action or high quantum numbers are often mentioned. While such
criteria are not as precise or as complete as those deduced here, it can
be seen from simple dimensional arguments that in typical situations an
action {\it will} be large compared to $\hbar$
when the two requirements of decoherence and
sufficient inertia are satisfied.  Let us consider a one-dimensional
oscillator model of the kind just discussed in the high temperature
limit.  Let $t_{\rm dyn}$ be the shortest dynamical time scale of
interest  and assume that the coarse graining is characterized by sets
of intervals of characteristic size $\Delta$ separated in time by
$t_{\rm dyn}$. From \(ten), it follows that decoherence requires
$$
\frac{M\gamma kT_B}{\hbar^2}\left(t_{\rm dyn} \Delta^2\right) >> 1\ .
\tag sixthirtytwo
$$
>From \(thirtyseven), the requirement of sufficient inertia is
$$
\frac{M}{\gamma k T_B}\ t_{\rm dyn}\ \left(\frac{\Delta}{t^2_{\rm
dyn}}\right)^2 >> 1\ ,\tag sixthirtythree
$$
if we assume $t_{\rm dyn} << t_{\rm relaxation} \equiv 1/\gamma$.

These two requirements may be re\"expressed in terms of the
characteristic scale of classical actions
$$
S \sim t_{\rm dyn}\ M\ \left(\frac{\Delta}{t_{\rm dyn}}\right)^2
\tag sixthirtyfour
$$
and the thermal correlation time $t_{\rm thermal} \equiv \hbar/kT_B$.
One finds
$$
(S/\hbar) >> (1/\eta)\qquad , \qquad (S/\hbar) >> \eta \tag
sixthirtyfive
$$
for the requirements \(sixthirtytwo) and \(sixthirtythree) respectively,
where $\eta$ is the ratio
$$
\eta = \left(t_{\rm dyn}/ t_{\rm thermal}\right)\left(t_{\rm dyn}/t_{\rm
relaxation}\right)\ .\tag sixthirtysix
$$
Whatever the size of $\eta$, the relations \(sixthirtyfive) imply that
$S/\hbar >>1$ in the classical limit.
\taghead{7.}
\vskip .26 in
\centerline{\bf VII. Equations of Motion for Non-Linear Systems}
\vskip .13 in
In this Section we generalize the results of the preceding two Sections
to the more realistic situation where the action $S_{\rm free}$
$[x(\tau)]$,  and the influence phase $W [x^\prime(\tau), x(\tau)]$ 
are
not necessarily quadratic functionals of their arguments.  We begin in
(a) by deriving some useful general properties of the influence phase.
These are used in (b) to derive the form of the phenomenological equations
of motion and analyze the restrictions on the coarse graining 
that permit the histories to stay close to solutions of those equations
with high probability. In (c) we derive the classical causality of these
equations of motion from quantum-mechanical causality.  The linear
theory, discussed in the preceding two Sections, is recovered in (d).
\vskip .13 in
\centerline{\sl (a) General Relations for the Influence Phase}

The influence phase $W[x^\prime(\tau), x(\tau)]$ is defined by the
functional integral (3.5) over the ignored variables $Q^A$.  
We continue to assume a factored initial condition as in \(eight), so that
$W$ has no explicit dependence on $x^\prime_0$ and $x_0$. 
A useful
operator expression for $W[x^\prime(\tau), x(\tau)]$ may be derived by
noting the following:  The integrals over the $Q(t)$ in (3.3) are over
paths between $t=0$ and $t=T$ that are {\it unrestricted} except at their
initial and final endpoints.  They may, therefore, be thought of as
defining the unitary evolution of a family of operators
$\rho(x^\prime_0, x_0)$ in the Hilbert space ${\cal H}_Q$ of
square-integrable functions of the $Q$'s. 
The dynamics of this
evolution of the $Q$'s is specified by the action
$$
S_Q[x(\tau), Q(\tau)] = S_0[Q(\tau)] + S_{\rm int} [x(\tau), Q(\tau)]
\ ,\tag thirtyeight
$$
which depends on the path $x(\tau)$. There is a corresponding Hamiltonian
operator on
${\cal H}_Q$.  If we assume that the interaction is local in time, 
specifically such that
$$
S_{\rm int} [x(\tau), Q(\tau)] = \int^T_0\ dt\ L_{\rm int} \bigl(x(t),
Q(t)\bigr)\ ,\tag thirtynine
$$
then that Hamiltonian at time $t$ depends only on the instantaneous
value of $x(t)$, {\it viz.}:  
$$
H_Q\bigl(x(t)\bigr) = H_0 + H_{\rm int} \bigl(x(t)\bigr)\ .\tag forty
$$
Here $H_0$ is the Hamiltonian of the $Q$ variables, omitting their
interaction with the $x$ variables, corresponding to the action $S_0[Q]$.
The operator effecting the unitary evolution generated by this
Hamiltonian between times $t^\prime$ and $t^{\prime\prime}$ is
$$
U_{t^{\prime\prime}, t^\prime} [x(\tau)] = {\bf T}\  \exp
\Bigl[-\frac{i}{\hbar}\ \int^{t^{\prime\prime}}_{t^\prime}\ dt
\ H_Q\bigl(x(t)\bigr)\Bigr]\ ,\tag fortyone
$$
where ${\bf T}$ denotes the time ordered product.

Write $\rho_B$ for the density operation on ${\cal H}_Q$ whose matrix
elements are  
$$
\left\langle Q^\prime_0 \left|\rho_B
\right|Q_0\right\rangle = \rho_B
(Q^\prime_0, Q_0)\ ,
\tag fortytwo
$$
where $\rho_B(Q^\prime_0, Q_0)$ is the factor of the initial density
matrix \(eight) referring to the $Q$'s. Utilizing the $U$'s defined by
\(fortyone) and the $\rho_B$ defined by \(fortytwo) the path integral 
relation \(three)
defining the influence phase $W[x^\prime(\tau)]$
may be
re\"expressed [see Appendix (d)] as
$$
\exp \Bigr(i\ W\left[x^\prime(\tau), x(\tau)\right]/\hbar\Bigr)=
Sp\Bigl\{U_{T,0} [x^\prime(\tau)]
 \rho_B U^\dagger_{T,0} \bigl[x(\tau)]
\Bigr\}\ ,
\tag fortythree
$$
where $Sp$ denotes the trace operation on the Hilbert space ${\cal
H}_Q$.  We shall now use this relation to derive some useful general
properties of $W[x^\prime(\tau), x(\tau); x^\prime_0, x_0)$.

First, it is an immediate consequence of $Sp (A^\dagger) = (S
p (A))^*$ and the form of right-hand side of \(fortythree) that 
interchanging
$x^\prime(t)$ and $x(t)$ on the left-hand side is equivalent to complex
conjugation.  (This elementary result also follows directly from \(three),
as noted in [\cite {12, 22}].)
Thus,
$$
{\cal R}e\ W\left[x^\prime(\tau), x(\tau)\right] = -{\cal R}e\ W 
\left[x(\tau), x^\prime(\tau)\right]\ ,\tag fortyfour a
$$
$$
{\cal I}m\ W\left[x^\prime(\tau), x(\tau)\right] = +{\cal I}m\ W
\left[x(\tau), x^\prime(\tau)\right]\ .\tag fortyfour b
$$
In particular, if the influence phase is written as a functional $W
\ [X(\tau), \xi(\tau)]$ of the average of $x^\prime(t)$ and $x(t)$ and
the difference between them [eq.~\(tend)], then ${\cal R}e(W)$ is an
{\it odd} functional of $\xi(t)$ while ${\cal I}m(W)$ is an {\it even}
functional.

As shown by Brun [\cite{27}] in the following paper, an elementary
application of Schwarz's inequality shows that
$$
\exp\Bigl(-{\cal I}mW\big[X(\tau), \xi(\tau)\bigr]\Bigr)
     \leq \sum\nolimits_i p_i\bigl| \bigl\langle\psi_i\bigl| U^\dagger_{T,0}
[x(\tau)]\, U_{T,0} [x^\prime(\tau)]\bigr| \psi_i\bigr\rangle\bigr|
     \leq \sum\nolimits_i p_i = 1\ , \tag fortyfoura
$$
where $p_i$ are the eigenvalues and $|\psi_i\rangle$ the eigenvectors of
$\rho_B$. Thus ${\cal I}mW[X(\tau), \xi(\tau)]$ is positive, which is essential
for the convergence of the functional integral defining the decoherence
functional as well as decoherence itself.

The
expression \(fortythree) may be used to find convenient operator 
expressions for
the coefficients of the expansion of $W[X(\tau), \xi(\tau)]$ 
in powers of $\xi(t)$. Generally, 
$$
W\left[x(\tau), \xi(\tau)\right] = W \left[x(\tau),0\right] +
\int^T_0\ dt\ \xi^\dagger(t)\left(\frac{\delta W}{\delta\xi(t)}
\right)_{\xi(t)=0}
$$
$$
+\half\ \int^T_0\ dt\ \int^T_0\ dt^\prime\
\xi^\dagger(t)\left(\frac{\delta^2
W}{\delta\xi(t)\delta\xi(t^\prime)}\right)_{\xi(t)=0}\xi(t^\prime)
 +\cdots \ .
\tag fortyfive
$$
For the leading term in \(fortyfive), we have, evaluating the right-hand
side of \(fortythree) at $\xi(t) = x^\prime(t) - x(t) =0$,
$$
\exp\Bigl(iW\left[X(\tau), 0\right]/\hbar\Bigr)
=Sp\Bigl\{U_{T,0} \left[X(\tau)\right] \rho_B\ U^\dagger_{T,0}
\left[X(\tau)\right]\Bigr\}\ .
\tag fortysix a
$$
Using the cyclic property of $Sp$ and the unitarity of
$U_{T,0}[X(\tau)]$ it is easy to see that the right-hand side of
\(fortysix) is unity.  Thus, the leading term in \(fortyfive) vanishes:
$$
W\left[X(\tau), 0\right] = 0\ . \tag fortysix b
$$

To evaluate the next term in the expansion \(fortyfive), we must consider
the derivatives $\delta U_{T,0}[X(t) \pm \xi(t)/2]/\delta\xi(t)$. To
do this we introduce the definition
$$
F\bigl(x(t)\bigr) = -\frac{\partial H_Q(t)}{\partial x(t)} =
\frac{\partial L_{\rm int}\bigl(x(t), Q(t)\bigr)}{\partial x(t)}\ .
\tag fortynine
$$
The operator $F(x(t))$ is an operator in the Schr\"odinger picture 
corresponding to the unitary evolution operators
we have been using.  It is a function of $x$ because $L_{\rm
int}$ is a function of $x$ and it becomes a function of $t$ because $x$ is
a function of $t$.  It is an operator representing the force on the
distinguished co\"ordinates $x(t)$ due to their interaction with the
rest of the system.

Carrying out the indicated differentiations of the $U$'s yields
$$
\eqalignno{
\Bigl(\delta U_{T,0}\bigl[X(\tau) \pm
\xi(\tau)/2\bigr]/\delta\xi(t)\Bigr)_{\xi(t)=0}
& = \pm (i/2\hbar) U_{T,t}\ \left[X(\tau)\right]\ F\left(X(t)\right)\ U_{t,0}
\left[X(\tau)\right]\cr
&\equiv \pm (i/2\hbar)\ F\bigl(t, X(\tau)\bigr]\ .&(fifty)\cr}
$$
The operator $F(t,X(\tau)]$ is the Heisenberg picture representative of
the Schr\"odinger picture operator \(fortynine).  It is  a
function of $t$ but also a functional of the path $X(\tau)$. We
indicate this dependence by writing $f(t,X(\tau)]$, using a parenthesis
on the left to indicate that it is a function and the bracket on the
right to indicate that it is also a functional.

With the result \(fifty), it is only a short calculation to find the
coefficient of the linear term in \(fortyfive).  It is 
$$
\bigl(\delta W/\delta\xi(t)\bigr)_{\xi(t)=0} = Sp
\bigl\{F\bigl(t,X(\tau)\bigr]\ \rho_B \bigr\}
\ .\tag fiftyone
$$
If we define the Heisenberg picture expected value by
$$
\langle A\rangle = Sp \left(A\ \rho_B\right) 
\ ,\tag fiftytwo
$$
eq.~\(fiftyone) may be written in the compact form
$$
\bigl(\delta W/\delta\xi(t)\bigr)_{\xi(t)=0} = \bigl\langle F\bigl(t,
X(\tau)\bigr]\bigr\rangle\ .\tag fiftythree
$$
As required by \(fortyfour), this contribution to the part of $W$ that
is odd in $\xi(t)$ is purely real.  

The coefficient $(\delta^2
W/\delta\xi(t)\delta\xi(t^\prime))_{\xi(t)=0}$ of the quadratic term in
the expansion of $W$ in powers of $\xi(t)$ is similarly evaluated.
The expansion of the $U$'s on the right-hand side of \(fortythree) 
will result in products of first derivatives, such as those
 in \(fifty), but also in
second derivatives.  Those second derivatives yield a term in the
expansion proportional to
$$
\delta (t-t^\prime)\ Sp \left(\ \Bigl[H^{\prime\prime}_{\rm
int}\bigl(t,X(\tau)\bigr],\ \rho_B\Bigr]\ \right),\tag fiftyfour
$$
where $H^{\prime\prime}_{\rm int}(t,X(\tau)]$ is the Heisenberg picture
representative of the operator $\partial^2 H_{\rm int}/\partial x(t)^2$.
A term like \(fiftyfour) vanishes,  of course, for a linear interaction
like \(thirteen).  It formally vanishes in the non-linear case  too because
it is the trace of a commutator.  Of course, that is a delicate
issue in the case of unbounded operators, as the non-vanishing value of
$Sp(\ [\Pi,Q]\ )$ shows. 
However, in the present case, where $\rho_B$ is bounded and $H_{\rm
int}$ is a function of the $Q$'s and not of their conjugate momenta, we
may reasonably assume that  \(fiftyfour) vanishes and we shall do so in what
follows.

The remaining contribution to the coefficient of the quadratic term in 
the expansion of $W[X(\tau), \xi(\tau)]$ comes from products of first
derivatives of $U$'s such as those
 in \(fifty) and is straightforwardly evaluated.
One finds
$$
\bigl(\delta^2W/\delta\xi(t)\delta\xi(t^\prime)\bigr)_{\xi(t)=0}
=(i/2\hbar) \Bigl\langle\bigl\{\Delta F\bigl(t,X(\tau)\bigr],\ \Delta
F\bigl(t^\prime, X(\tau)\bigr]\bigr\}\Bigr\rangle\ ,\tag fiftyfive
$$
where $\{,\}$ denotes the anticommutator and $\Delta F$ is the operator
$$
\Delta F\bigl(t,X(t)\bigr] = F\bigl(t,X(\tau)\bigr] - \bigl\langle
F\bigl(t,X(\tau)\bigr]\bigr\rangle\tag fiftysix
$$
representing fluctuations in the force $F$ about its mean.  
We note that, as required by
\(fortyfour), this contribution to the even part of $W[X(\tau),
\xi(\tau)]$ in $\xi(t)$ is purely imaginary.  When divided by $i$,
it is also manifestly
positive in the sense of \(thirtytwo).

With these preliminaries we may now derive the non-linear equations of
motion and discuss their form.
\vskip .26 in
\centerline{\sl (b) Non-Linear Equations of Motion}

We consider a set of alternative coarse-grained histories specified 
at a sequence  at times $t_1, \cdots,t_n$ by sets of exhaustive and 
exclusive regions of the $x$'s
which we denote by $\{\Delta^1_{\alpha_1}\}, \{\Delta^2_{\alpha_2}\},
 \cdots, \{\Delta^n_{\alpha_n}\}$.
 The decoherence functional for such sets of histories is given
by \(five).  We assume that the regions and times are chosen so that
there is a negligible contribution to the path integrals 
in the decoherence functional except when $\xi(t)
= x^\prime(t) - x(t)$ is small.  We expect to have such coarse
  grainings, for example, if
the imaginary part of the influence phase $W [X(\tau), \xi(\tau)]$ 
has its minimum at  $\xi(t)=0$.  Our assumption about the
integrals in \(five) implies the decoherence of a set of
alternative coarse-grained histories (cf.~Figure 1) and the following
formula for their probabilities:
$$
\eqalignno{
p(\alpha)&\cong \int_\alpha\ \delta X\, \delta\xi\ \delta (\xi_f)
\ \exp \Bigl\{i\bigl(S_{\rm free}\left[X(\tau) + \xi(\tau)/2\right]\cr
         &-S_{\rm free}\left[X(\tau)-\xi(\tau)/2\right] + W\left[
X(\tau), \xi(\tau)\right]\bigr)/\hbar\Bigr\}
          \tilde\rho \left(X_0 + \xi_0/2, X_0-\xi_0/2\right)\ .
&(fiftyseven)\cr}
$$
The functional integral is over paths in both $X$ and $\xi$ as
restricted by the coarse-grained history $\alpha$.

If only small values of $\xi(t)$ contribute to the integrals in
\(fiftyseven), we may make a further approximation by expanding the
exponent in powers of $\xi(t)$, utilizing the expansion in \(fortyfive),
up to the quadratic terms.  We then have an integral for the probability
of a history that is of precisely the same form as the one occurring in the
discussion of the linear theory, \(eighteen), with $e(t)$ replaced by
$$
{\cal E}(t) = \frac{\delta S_{\rm free} \left[X(\tau)\right]}{\delta
X(t)} + \bigl\langle F\bigl(t, X(\tau)\bigr]\bigr\rangle\tag fiftyeight
$$
and with $k_I(t,t^\prime)$ replaced by
$$
K_I(t, t^\prime) = \hbar^{-1} \Bigl\langle\bigl\{\Delta F\bigl(t,
X(\tau)\bigr], \Delta F(t^\prime, X(\tau)\bigr]\bigr\}\Bigr\rangle\ .
\tag fiftynine
$$
In writing out these identifications we have made use of \(fortysix b),
\(fiftythree), and \(fiftyfive) for the expansion coefficients of
$W[X(\tau), \xi(\tau)]$.  Eq.~\(fiftynine) shows the kernel,
$K_I(t, t^\prime;
X(\tau)]$ is
manifestly positive --- a necessary condition for the mechanism of
decoherence being discussed, not to mention the convergence of the
integral \(fiftyseven) with the expanded exponent.

Under the assumption that only a narrow range of $\xi(t)$ near zero
contributes to the integral \(fiftyseven), it is a good further
approximation to neglect the constraints on the ingegration range of
$\xi(t)$ arising from the coarse graining.  The resulting Gaussian
integrals can then be carried out, yielding an expression for $p(\alpha)$
that is the generalization of \(eighteen):
$$
\eqalignno{
p(\alpha)&\cong\int_\alpha\ \delta X\ \bigl[ {\rm det}
\left(K_I/2\pi\right)\bigr]^{-\half}\cr
         &\times \exp \Bigl[-\frac{1}{\hbar}\ \int^T_0\ dt
\ \int^T_0\ dt^\prime\ {\cal E}^\dagger(t)\ K^{\rm inv}_I
(t,t^\prime; X(\tau)]\ {\cal E}(t^\prime)\Bigr]\ w(X_0, P_0)\ ,
&(sixty)\cr}
$$
where $K^{\rm inv}_I(t, t^\prime; X(\tau)]$ is the inverse kernel to $K_I(t,
t^\prime; X(\tau)]$ and $P_0$ is the momentum of the free action expressed in
terms of $X_0$ and $\dot X_0$. The measure $\delta X$ is discussed in the
Appendix (b).

The derivation and analysis of the equations of motion now proceed as in
the linear example, with important differences that we shall mention.
The Gaussian form of the exponent in \(sixty) means that for given $X_0$
and $P_0$ the most significant contribution comes 
from the histories with ${\cal E}(t)\approx 0$, that is, those whose
evolution in time nearly follows the effective classical equation of motion
$$
{\cal E}(t) = \frac{\delta S_{\rm free} \left[X(\tau)\right]}{\delta
X(t)} + \bigl\langle F\bigl(t, X(\tau)\bigr]\bigr\rangle = 0\ .
\tag sixtyone
$$
The probabilities predicted by \(sixty) are, therefore, those of an
ensemble of classical histories individually correlated in time by the
equation of motion \(sixtyone) and with initial conditions distributed
according to the Wigner function $w(X_0, P_0)$.

The first term in \(sixtyone) is the equation of motion of the distinguished
co\"ordinates $x^a$ in the absence of any interaction with the
remaining 
co\"ordinates $Q^A$.  The second term is expected
value of the force arising from that interaction.  This is a {\it
functional} of the trajectory and will, in general,  be non-local in
time.  As we shall show below, it is retarded as a 
consequence of quantum-mechanical 
causality.  It typically leads to dissipation, although under some conditions 
the energy might actually increase.  The phenomenological force is also
generally dependent on the initial condition $\rho_B$ through
\(fiftyone).  The familiar, phenomenological equations describing, say,
{\it dissipative} friction, are  characterized by a few parameters {\it
independent} of initial conditions and are the result of further
approximations to \(sixtyone). These are typically good in situations
where there is a significant contribution only from retarded times that
are short compared to the relaxation times of that part of the bath that
interacts significantly. 
If the distinguished system has  energy large
compared to $kT_B$, it will lose energy to the bath
on the average.  The result is a dissipative phenomenological equation,
local in time, with parameters independent of initial conditions, like
\(twenty).

Viewed as a generalization of the 
linear case, the
important point about the equation of motion
\(sixtyone) is that both the free part of the
equation of motion and the contribution from the interaction of
the $x$'s with the $Q$'s are, in general, non-linear in $X(t)$.  For the
special coarse grainings in which the variables are divided into  a set
distinguished by the coarse graining and a set ignored, we therefore
have a general derivation of the form of the phenomenological equations
of motion.  We now discuss in more detail the implications of 
 quantum-mechanical causality and quantum noise.
\vskip .26 in
\centerline{\sl (c) Quantum-Mechanical Causality
Implies Classical Causality}

Feynman and Vernon [\cite{12}] used path integral arguments to show that if
$\xi(t)$ is set to zero for $t>t_*$, then $W[X(\tau), \xi(\tau)]$ is
independent of $X(t)$ for $t> t_*$.  This result could be used to show
that the force $\langle F(t, X(\tau)]\rangle$ is retarded, that is
independent of the path $X(\tau)$ for values of $\tau$ greater
than $t$.  However the result also follows easily from the definitions
\(fifty), \(fiftytwo), and \(fiftythree).  In the Schr\"odinger picture
$$
\bigl\langle F\bigl(t, X(\tau)\bigr]\bigr\rangle = 
Sp\Bigl\{U_{T,t} [X(\tau)]\  F\bigl(X(t)\bigr)\ U_{t,0}\ [X(\tau)]\ \rho_B
\ U^\dagger_{T,0}\ [X(\tau)]\Bigr\}\ .\tag sixtytwo
$$
Using the cyclic property of the spur, the composition law, and unitarity
of the evolution operators defined by \(fortyone), we may write
 eq.~\(sixtytwo) in the form  
$$
\bigl\langle F\bigl(t, X(\tau)\bigr]\bigr\rangle = Sp\bigl\{F\bigl(X(t)\bigr)
\ U_{t,0} [X(\tau)]\ \rho_B
\ U^\dagger_{t,0}[X(\tau)]\bigr\}\ .
\tag sixtythree
$$
Since $U_{t,0}[X(\tau)]$ depends on $X(\tau)$ only for $0<\tau<t$,  
[cf.~\(fortyone)], this shows that $\langle F(t, X(\tau)]\rangle$ is
retarded.

The expression \(one) for the decoherence functional incorporates a
quantum-mechanical notion of causality.  At one end of the histories,
information about the specific closed
system in the form of a density matrix must be supplied.  
At the other end is a unit matrix in the form
of a $\delta$-function representing a condition of indifference with
respect to states at the end of the histories.  This asymmetry between
the two ends of the histories is the arrow of time in quantum
mechanics. (See, \eg [\cite {43, 2}].)
It is by {\it convention} that we call the extremity of the
histories next to the density matrix $\rho$ {\it the past} and consider time as
increasing from it.  

To predict the future in quantum mechanics, we need the initial $\rho$
and information about histories up to the present.  We need no
information about the future.  That is quantum-mechanical causality.
The retardation of $\langle F(t, X(\tau)]\rangle$ expressed by 
\(sixtythree) shows that we need know the trajectory of the
distinguished system only from the time of the initial $\rho$ to the
present to predict the system's future.  That is classical causality.
Eq.~\(sixtythree) thus shows that quantum mechanical causality implies
classical causality.

The origin and implications of the arrow of time in quantum mechanics
may be usefully discussed using a hypothetical
 generalization of the quantum
mechanics of closed systems that employs both initial and final
conditions [\cite {5, 3, 24}].  In this generalization the decoherence
functional would be given by
$$
D(\alpha^\prime \alpha) = N \int_{\alpha^\prime}\ \delta q^\prime
\ \int_\alpha\ \delta q\ \rho_f (q^\prime_f, q_f)
\ \exp\Bigl\{i\bigl(S\left[q^\prime(\tau)\right]-S\left[q(\tau)\right]\bigr)
/\hbar\Bigr\}\rho_i\left(q^\prime_0, q_0\right)\ ,
\tag sixtyfour a
$$
where
$$
N^{-1}=\int dq \int dq^\prime\ \rho_f (q^\prime, q)\ \rho_i\
(q,q^\prime)\ .\tag sixtyfour b
$$
This generalization of quantum mechanics would permit 
the future and the past to be treated
similarly.  Arrows of time would arise in particular
universes where $\rho_f$ is different from the time-reversed version
 of $\rho_i$.
In particular, usual quantum mechanics, represented by \(one)
 and its associated causality  and arrow of time arise 
for those universes, like ours, where $\rho_f \propto I$ 
is a good representation of the final condition.  In more
general situations, with $\rho_f$ not proportional to $I$,
there would be neither  a notion of quantum mechanical
causality nor a notion of ``state of the system at a moment of time''.
Nor would the argument described above succeed in deriving classical
causality.  There would be advanced as well as retarded effects.
\vskip .26 in
\centerline{\sl (d) Quantum Noise}
 
In Section V we showed how the probabilities of decoherent sets of
coarse-grained histories of the type under discussion could be thought
of as the probabilities of a classical system in which the distinguished
co\"ordinates $x^a$ evolve from probabilistically distributed initial
conditions according to an equation of motion in the presence of noise.
In the case of a factored initial density matrix, 
initial conditions and noise are separately
distributed and the system may be thought as obeying the Langevin equation
\(tenne),
$$
R(t) = {\cal E}\bigl(t, X(\tau) \bigr] + {\cal L} \bigl(t, X(\tau)\bigr] = 0
\ , \tag sixtyfive
$$
where the total force $R(t)$  is
distributed according to the generally non-positive distribution
$g[R(\tau), X(\tau)]$ constructed from the decoherence functional
according to \(tennd).  The analysis of this section provides an
explicit form for ${\cal E}(t, X(\tau)]$ in \(sixtyone) and a systematic
approximation scheme for the spectrum of the noise.

When decoherence is good enough that the restriction on the range of the
$\xi$-integration in \(fiftyseven) arising from the coarse graining can
be neglected, a systematic perturbation scheme for the approximate
probabilities $p(\alpha)$ can be obtained by expanding the exponent in
powers of $\xi(t)$.  In the leading approximation \(sixty), an explicit
expression for the distribution of the Langevin force is obtained, which
is 
$$
\left[{\rm det}(K_I/2\pi)\right]^{-\half}\ \exp\left[-\frac{1}{\hbar}
\int\limits^T_0 dt\int\limits^T_0 dt^\prime {\cal L}^\dagger 
(t, X(\tau)] K_I^{\rm inv}
(t, t^\prime; X(\tau)]{\cal L} (t, X(\tau)]\right]\ .\tag sixtysix
$$
In this leading approximation, the noise is distributed with a {\it
positive} Gaussian distribution function whose spectrum is fixed by the
correlation function
$$\eqalignno{
\left\langle{\cal L}(t, X(\tau)] {\cal L} (t^\prime,
X(\tau)]\right\rangle_c& = (\hbar/2) K_I (t, t^\prime ; X(\tau)]\cr
                       & = \half \left\langle\left\{\Delta F(t,
X(\tau)], \Delta F (t^\prime, X(\tau)]\right\}\right\rangle \ .& 
(sixtyseven) \cr}
$$
For the linear problem, that approximation is exact [cf.~\(twentyd)], with
the further simplifying feature that the spectrum of Langevin force is
independent of the path $X(t)$. 

Higher order terms in the expansion of the exponent of \(fiftyseven) may
be regarded as providing corrections to the Gaussian noise.  The general
expression for the correlation functions is \(tenk). Other corrections
to the Gaussian noise arise 
from the coarse-graining restrictions on the range of
$\xi$-integration.
\taghead{8.}
\vskip .26 in
\centerline{\bf VIII. Examples and Comparison with Classical Cases}
\vskip .13 in
In this Section we specialize the general non-linear theory of the
preceding Section to some particular cases considered by ourselves and
other authors.  We begin with the linear models described in Sections VI.
\vskip .13 in
\centerline{\sl (a) Recovery of the Linear Theory}

In Section  VI we derived the equations of motion for linear
systems defined by an influence phase \(twelve) that was quadratic in
the co\"ordinates, $x$, distinguished by the coarse graining but
containing no
linear terms.  An influence phase of this form will arise when the free
action is quadratic in the co\"ordinates distinguished by the coarse
graining as in \(eleven) when the interaction with the ignored
co\"ordinates $Q$ is linear as in \(thirteen), and when the density
matrix has the special form \(thirteena). 
To recover the linear equation of motion \(nineteen) from the non-linear
\(fiftyeight), we need to evaluate $\langle F(t,X(\tau)]\rangle$ 
under these conditions.  Eq.~\(sixtythree) gives a
general expression for $\langle F(t, X(\tau)]\rangle$.  We know from the
general arguments described in Section VI that, when the interaction
between distinguished and ignored variables is linear as in \(thirteen),
the influence phase is quadratic in the $x$'s and $\langle F(t,
X(\tau)]\rangle$ is linear in $X(\tau)$.  It therefore suffices to
evaluate \(sixtythree) to linear order in a perturbation expansion in
$X(\tau)$; the higher orders must cancel.  The result is
$$
\bigl\langle F\bigl(t, X(\tau)\bigr]\bigr\rangle = \bigl\langle
f(t)\bigr\rangle_0 + (i/\hbar)
\ \int^t_0 dt^\prime \bigl\langle \left[f(t), f(t^\prime)\right]\bigr\rangle_0
\ X(t^\prime)\ .\tag sixtyeight
$$
Here, the forces $f(t)$ are the Heisenberg picture representatives of 
the homogeneous, linear functions of the ignored
co\"ordinates $f(Q(t))$ [cf. \(fortynine)], 
and the expected values $\langle ~~~\rangle_0$ are
computed using the time dependence of operators provided by Hamiltonian
$H_0$ of the ignored co\"ordinates $Q$ alone, neglecting their
interaction with the $x$'s [cf. \(thirteenb)].  
If we assume, as we did in Section VI, that $\langle f(t)\rangle_0 = 0$
(this may always be achieved by a time-dependent shift in $x(t)$) then 
we recover both the form of the linear
equation of motion \(nineteen) and the expression for the additional force
anticipated in \(thirty).

The derivation of the expression \(thirteenb) for the spectrum of
quantum noise in the linear problem is even more straightforward.  From
the form \(fiftynine) we see that $K_I(t, t^\prime)$ is equal to  $k_I(t,
t^\prime)$ of the linear case.  Since we assumed $\langle f (t)\rangle$
vanished for the linear problem [cf.~\(twentynine)], eq.~\(fiftynine)
involves only the expected value of $\{f(t), f(t^\prime)\}$.  However,
general arguments for the linear problem show that when the
interaction is of the linear form \(thirteen) this expected value,
proportional to $k_I(t, t^\prime)$, is {\it independent} of the $x$'s.
It may thus be evaluated in zeroth order in perturbation theory in $x(t)$;
that is, the time dependence of the operators in \(fiftynine) is
provided by the Hamiltonian $H_0$ with no interactions between $Q$'s and
$x$'s.  This is the time dependence we denoted by a subscript zero in Section
VI, and thus we recover \(thirtyone).
\vskip .13 in
\centerline{\sl (b) Semi-Linear Systems}

Simple expressions for the spectrum of quantum noise and the kernel of
the associated fluctuating force analogous to eqs \(thirty) and
\(thirtyone) have been obtained in more general cases than strict
linearity.  A particularly simple case occurs when  the  
dependence of the total action 
 on the ignored variables (the $Q$'s) is restricted to be
quadratic, but arbitrary dependence on the distinguished variables (the $x$'s)
is allowed in the potential energy and interaction terms.  This case was
discussed\footnote{$^{11}$}{We thank T.~Brun for these references.}
 in purely classical situations by Zwanzig [\cite {25}],
and in particular field theory examples by Ryang and Saito [\cite {26}], and
more recently in a
general survey of such problems by Brun[\cite {27}], who applies our
methods.  

Let us consider a specific class of problems. For simplicity,
we assume a single
distinguished variable, $x$, to avoid matrix notation.
Suppose that \ (i) the action of the ignored variables
$S_0[Q(t)]$ is a quadratic functional of the $Q$'s;  (ii) the interaction
of the $Q$'s and $x$'s is of the form
$$
S_{\rm int} \left[x(t), Q(t)\right] = -\int^T_0 dt\  a
\bigl(x(t)\bigr) f \bigl(Q(t)\bigr)\ ,\tag sixtynine
$$
where $f$ is homogeneous and linear in the $Q$'s, but $a$ is 
not necessarily a linear
function of $x$; and \ (iii)  the density matrix factors as in
\(eight) and has the special form \(thirteena).
Under these
conditions the influence phase will be a functional that is at most
quadratic in $ a (x(t))$, and its form is given by \(twelve) with $x(t)$ and
$x^\prime(t)$ replaced by $a(x(t))$ and $a(x^\prime(t))$ respectively,
assuming, as before, that linear terms vanish.  
The equation of motion will then be
of the form \(fortysix), with the force arising from the interaction
given by 
$$
\bigl\langle F\bigl(t, X(\tau)\bigr]\bigr\rangle = 
a^\prime\left(X(t)\right)\ \int^t_0 dt^\prime\ k_R
(t,t^\prime)\ a\left(X(t^\prime)\right)\ ,\tag seventy
$$
where $a^\prime = da/dX$.
Here, $k_R$, as before in \(thirty), is proportional to the expected value of
the commutator of $f(Q(t))$ evolved in time without the
interaction with the $x$'s. Linear terms in the influence phase would
give rise to a non-zero value of $\langle f(t)\rangle_0$ and an
additional term in \(seventy) of the form $a^\prime (X(t)) \langle
f(t)\rangle_0$.  With the addition of this term, the result \(seventy)
 is
equivalent to that of Brun [\cite{27}], who does not eliminate linear
terms in the influence phase.
The Langevin force ${\cal L} (t)$ is given by the formula
$$
{\cal L} (t ; X(\tau)] = a^\prime (X(t))\ell(t)\ ,\tag seventyone
$$
where $\ell(t)$ is a Gaussian random force with a spectrum given by
\(seventyone) and 
\(twentyfive).
The Langevin equation describing both dissipation and fluctuations in
this approximation is then
$$
{\cal E}(t) = \frac{\delta S_{\rm free} [X(\tau)]}{\delta X(t)}
            + a^\prime(X(t))\left[\int\nolimits^t_0 dt^\prime k_R
(t,t^\prime)  a\left(X(t^\prime)\right)
            + \ell(t)\right] =0\ .  \tag seventya 
$$
A fluctuation-dissipation theorem of the form given by \(thirtysix)
continues to hold.
\vskip .13 in
\centerline{\sl (c) Equations of Motion for Semi-linear
Classical Coarse-Grained
Systems}

The evolution of a classical Hamiltonian system is, of course,
deterministic and essentially
reversible when followed in all detail.  In a
coarse-grained description, however, such systems will, in general,
approximately obey {\it effective} equations of motion including the effects
of  dissipative
forces and with deviations from these equations produced by classical noise.
The classical problem analogous to that considered in this paper would
be to derive the Langevin equation for the motion of some followed
variables $x(t)$ that are interacting with some ignored variables $Q(t)$
whose initial conditions are probabilistically distributed according to
some given rule.  To our knowledge, this kind of problem has not been
considered classically for the non-linear situations discussed in
Section VI.  However, it has been worked out by R. Zwanzig [\cite {25}]
for the
semi-linear systems treated in the preceding subsection.  T.~Brun [\cite
{27}]
has shown that the Langevin equation deduced from quantum mechanics
coincides with Zwanzig's result in the limit $\hbar\to0$ as it must.
We briefly review Zwanzig's derivation and Brun's demonstration here.

Again, we assume for simplicity a single distinguished variable $x(t)$
interacting with an assembly of oscillators according to \(sixtynine).  
The free action $S_0[Q(t)]$ and the function $f(Q(t))$ are given
explicitly by \(thirteenb) and \(thirteenc) respectively.
The classical equation of motion for the $Q$'s following from
\(thirteenb) and \(sixtynine) may be solved explicitly with the result
$$
Q^A(t) = \ell^A(t) - \frac{1}{m\omega_A} \int\nolimits^t_0 dt^\prime \sin
\left[\omega_A (t-t^\prime)\right] C_A a(X(t^\prime))\ .
\tag seventyoneb
$$
Here, $\ell^A(t)$ is the solution of the free oscillator equations
following from the action \(thirteenc) with the same initial position and
momentum as $Q^A(t)$.  This result for $Q^A(t)$ may be substituted into
the classical equation of motion for $X(t)$. 
The result is an equation of motion for $X$ of the form
$$
\frac{\delta S_{\rm free} [X(\tau)]}{\delta X(t)} + a^\prime (X(t))
\left[\int\nolimits^t_0 dt^\prime k_R (t, t^\prime) a(X(t^\prime)) +
\ell(t)\right]=0\ , \tag seventyonec
$$
where
$$
\ell(t) = \sum_A C_A \ell^A (t)\tag seventyoned
$$
and we have made use of the fact that the sum of retarded Green's
functions from \(seventyoneb) enters \(seventyonec) as the combination
we called $k_R(t,t^\prime)$ in \(thirteene a).

If the initial conditions for the $Q$'s are probabilistically
distributed, then the motion $X(t)$ will be probabilistically distributed
as well.  The initial conditions of the $Q$'s are the initial conditions
of the free oscillator motions $\ell^A(t)$. In \(seventyonec),
 the only way  
 these initial conditions enter is through the function $\ell (t)$
defined in \(seventyoned).  Equation \(seventyonec)
may, therefore, be interpreted as a Langevin equation with a stochastic force
$a^\prime (X(t) \ell (t)$.  The time dependence of this force is known because
the $\ell^A(t)$ in \(seventyoned) satisfy the harmonic oscillator equations
of motion.  When the values of the $Q$'s and their conjugate momenta are
distributed thermally, the correlation functions of this noise are easily
calculated.  The classical phase space distribution analogous to the
thermal bath used in [\cite {11}] and [\cite {12}] is
$$
\rho^{cl} (\Pi_0, Q_0) \propto \exp\left[ - H_0(\Pi_0,
Q_0)/kT_B\right]\ ,\tag seventyonee
$$
where $H_0$ is the classical Hamiltonian corresponding to $S_0$. (A
slightly different distribution was assumed by Zwanzig in [\cite {25}] with
slightly different results.) The result is Gaussian noise with $\langle
\ell(t)\rangle_c=0$ and
$$
\langle \ell(t)\ell(t^\prime)\rangle_c = kT \sum_A \frac{C^2_A}{m\omega^2_A}
\cos \left[\omega_A (t-t^\prime)\right]
\equiv \half k^{cl}_I (t, t^\prime)\ . 
\tag seventyonef
$$
Eq \(seventyonec) and \(seventyonef) are essentially the results of
Zwanzig [\cite {25}].  We note that the spectral weights of the spectrum
of the fluctuations \(seventyonef) and of the kernel of the dissipative
force term \(thirteene) as defined by \(thirtyfive) are related by
$$
\tilde k^{cl}_I(\omega) = (kT_B/\omega) \tilde k_R(\omega)\ . \tag
seventyoneg
$$
This is the classical fluctuation-dissipation theorem.

As discussed by Brun, the Langevin equation of the quantum-mechanical
problem \(seventya) coincides in form with \(seventyonec) derived
classically by Zwanzig [\cite {25}].  The only difference is that the
 noise spectrum  $\hbar k_I(t, t^\prime)$ is given by \(thirteene b)
in the quantum-mechanical case and by \(seventyonef) in the classical one.
Indeed, the two expressions coincide in the classical limit as they must.
The classical noise is entirely thermal.  When quantum mechanics is
taken into account there is quantum noise as well.
\taghead{9.}
\vskip .26 in
\centerline{\bf IX. More General Coarse Grainings}
\vskip .13 in
The coarse grainings discussed in the previous sections are limited to
those that distinguish a fixed subset of the co\"ordinates $q^i$.
Coarse grainings that realistically describe a quasiclassical domain are
not of this simple type.  As we have discussed elsewhere [\cite {2}], it is
likely that a quasiclassical domain will be described by, among other
things, coarse graining with respect to ranges of values of the averages
of densities of conserved or approximately conserved quantities over
suitably small volumes.  Examples are the densities of energy, momentum,
charge, current, nuclear species,  etc. Together with field averages,
these {\it are} the ``hydrodynamic'' variables that enter into the
differential equations of classical physics.  Sufficiently large volumes
would give these variables enough ``inertia'' to enable them to resist
the deviations from predictability caused by the interactions that
effect decoherence, as we have described in the earlier sections.

The coarse grainings discussed in this paper must be generalized in two
ways to discuss such variables.  They must be generalized to allow
the original fine-grained description to involve 
momenta as well as co\"ordinates.  They must also  be 
generalized to permit
coarse grainings by ranges of values of averaged densities, These 
correspond to no particular fixed subset of co\"ordinates.  In this section we
introduce the machinery necessary to consider such coarse grainings,
although we do not carry out an analysis of the circumstances in which
they decohere or behave quasiclassically.
\vskip .13 in
\centerline{\sl (a) Phase-Space Coarse Grainings}  
 
A fairly general class of coarse grained histories may be obtained by
considering partitions of the co\"ordinates $q^i$ at some times and
their conjugate momenta $\pi_i$ at others. More specifically, we 
consider partitions of configuration space into an exhaustive set of
exclusive regions $\{\Delta^k_{\alpha_k}\}$ at some of the times $t_k$ and
partitions by partitions of momentum space into an exhaustive set of
regions of momentum space $\{\widetilde\Delta^k_{\alpha_k}\}$ at other
of the times $t_k$.  In this Section, concerned only with this type of coarse
graining, we will reserve the notation $\{P^k_{\alpha_k} (t_k)\}$ for
projections of configuration space alternatives and introduce the
notation $\{\widetilde P^k_{\alpha_k}(t_k)\}$ for a set of orthogonal
projections onto a set of exclusive momentum space regions.  The sets of
histories we are considering consists of sequences of sets of either
$P$'s or $\widetilde P$'s at times $t_1, \cdots, t_n$. 
The decoherence functional for such a set of histories is given
generally by \(fourfive).  We now show that it has a sum-over-histories
representation by path integrals in {\it phase-space}.

Utilizing complete sets of co\"ordinate eigenstates, 
we may write the decoherence functional \(fourfive) in the form
$$
D(\alpha^\prime, \alpha) = \int dq^\prime_f \int dq_f \int dq^\prime_0
\int dq_0\ \delta(q^\prime_f - q_f)
\langle q^\prime_f T  | C_{\alpha^\prime} |
q^\prime_0 0\rangle \rho \left(q^\prime_0, q_0\right)\langle
q_0 0 |C^{\dagger}_\alpha | q_f T\rangle\ .\tag seventyfive
$$
The matrix elements of the $C_\alpha$ may be written as the compositions
of sequences of
 propagators between definite co\"ordinate or momentum eigenstates.
For example, if $\widetilde P^1_{\alpha_1}(t_1)$ is a projection onto a
 momentum
region $\widetilde\Delta^1_{\alpha_1} (t_1)$ and $P^2_{\alpha_2}
(t_2)$ onto a co\"ordinate region $\Delta^2_{\alpha_2} (t_2)$ we can
write
$$
\langle q_f T  | C_\alpha  | q_0 0 \rangle =
\langle q_f T | P^2_{\alpha_2} (t_2) \widetilde P^1_{\alpha_1} (t_1) 
| q_0 0 \rangle\tag seventysix
$$
in the Schr\"odinger picture as
$$
\int\nolimits_{\Delta^2_{\alpha_2}} dq_2
\int\nolimits_{\widetilde\Delta^1_{\alpha_1}} d\pi_1 \left\langle q_f T
 | q_2t_2\right\rangle\left\langle q_2t_2 |\pi_1t_1\right\rangle
\left\langle\pi_1t_1| q_0 0\right\rangle\ .\tag seventyseven
$$
The propagators in \(seventyseven) may be represented as phase-space
path integrals if the Hamiltonian $H(\pi,q)$ associated with the action
$S[q(t)]$ is of a suitably simple form [\cite {28, 29}].  In particular, 
they can be so
represented if $H(\pi,q)$ is a sum of a function of the
$\pi$'s and a function of the $q$'s.  For example,
$$
\left\langle q^{\prime\prime} t^{\prime\prime}|\pi^\prime
t^\prime\right\rangle = \int \delta\mu\ \exp\bigl\{
iS\bigl[\pi(t),
q(t)\bigr]/\hbar\bigr\}\ ,\tag seventyeight
$$
where $S[\pi,q]$ is the canonical form of the action:
$$
S[\pi,q] = \int\nolimits^{t^{\prime\prime}}_{t^\prime} dt
[\pi_i(t) \dot q^i(t) - H\bigl(\pi(t), q(t)\bigr)]\ .\tag seventynine
$$
The integral is over phase-space paths between $t^\prime$ 
and $t^{\prime\prime}$
weighted by the invariant Liouville measure and restricted 
by the conditions that
they intersect $\pi^\prime$ at $t^\prime$ and $q^{\prime\prime}$ at
$t^{\prime\prime}$.  The details of these integrations are spelled
out in the Appendix.

By inserting \(seventynine) into \(seventyeight) and 
\(seventyeight) into \(seventysix), one arrives at a
sum-over-histories form for the decoherence functional for phase-space
coarse grainings
$$
D(\alpha^\prime, \alpha) = \int\nolimits_{\alpha^\prime}
\delta\mu^\prime\ \int\nolimits_\alpha\ \delta\mu\ \delta(q^\prime_f -
q_f) \exp \Bigl\{i\bigl(S[\pi^\prime(\tau), q^\prime(\tau)] - S[\pi(\tau),
q(\tau)]\bigr)/\hbar\Bigr\}\rho(q^\prime_0, q_0)\ .\tag eighty 
$$
The integral is over phase-space paths restricted by the coarse
graining.  For example, the integral over $\pi(t)$ and $q(t)$ is over the
phase-space paths that thread the intervals in either co\"ordinate or
momentum space corresponding to the history $\alpha$.  There is no
integration over the initial momenta $\pi^\prime_0$ and $\pi_0$
but unrestricted
integrations over the final momenta $\pi^\prime_f$ and $\pi_f$.  Again,
the details of this and the measure are in the Appendix.

In usual cases where the Hamiltonian in \(seventynine) is quadratic 
in all momenta,
the momentum dependence in the integrand in \(eighty) is 
that of a Gaussian.
For those coarse grainings that restrict only the co\"ordinates $q^i$ and
ignore the momenta $\pi_i$, the integrals over momenta may be carried out
explicitly.  The result is the Lagrangian path integral for the
decoherence functional \(one).  Indeed, it is by this route that the
 measure in that path integral is usually 
derived from the canonical, Liouville, ``$dpdq/(2\pi\hbar)$'' measure on
paths in phase-space.

There is no obstacle to letting the time of a coarse graining by
momenta coincide with that of a coarse graining by co\"ordinates.  Even in the
quantum mechanics of measured subsystems it is possible to consider a
measurement of position followed after an arbitrarily short time interval
by a measurement of momentum.  Care must be taken, however, to specify
the order of the coarse grainings when two such times coincide.  Since the
corresponding operators do not commute, a projection on a range of
momentum at one time followed immediately by a projection on a range of
position defines a different history from one with the operators in the
opposite time order.  

With this machinery in hand, we may now consider phase-space
coarse grainings, analogous to those of Section III, in which the 
 phase-space co\"ordinates $(\pi_i, q^i$) are divided into cannonically
conjugate pairs $(p_a, x^a)$ that are distinguished by the coarse
graining while the remaining pairs $(\Pi_A, Q^A)$ are ignored.  A
simple, interesting, and important class of models is obtained by
assuming that the action decomposes according to (3.2) with $S_{\rm
free}$ and $S_0$ quadratic in time derivatives and the interaction
independent of all time derivitives.  (A slight generalization would
then be needed to deal with a system of particles interacting
electromagnetically.)  The decomposition of the Hamiltonian
corresponding to these assumptions is
$$
H(\pi, q) = H_{\rm free} (p, x) + H_0 (\Pi, Q) + H_{\rm int} (x, Q)
\ ,\tag (eightyone)
$$
where $H_{\rm free}$ and $H_0$ have a quadratic momentum dependence.
The simplifying consequence of these assumptions is that the Gaussian
integrals over the momenta $\Pi_A$ in \(eighty) may all be carried out
explicitly.  The remaining integrals over the $Q^A$ have the same form
as they do in the Langrangian path integral and may be summarized by 
a single
influence phase $W[x^\prime(t), x(t)]$ defined by \(three).  The decoherence
functional for phase-space coarse grainings may then be written
$$
D(\alpha^\prime, \alpha) = \int\nolimits_{\alpha^\prime} d \mu^\prime
\int\nolimits_\alpha d\mu\  \delta
(x_f-x^\prime_f)
$$
$$
\times \exp \biggl\{ i\Bigl(S_{\rm free} \left[p^\prime(\tau),
x^\prime(\tau)\right] - S_{\rm free} \bigl[p(\tau), x(\tau)\bigr]
+W \left[x^\prime(\tau), x(\tau)\right]\Bigr)/\hbar\biggr\} \tilde\rho
(x^\prime_0, x_0)\ ,\tag eightytwo
$$
with $\tilde\rho$ defined by \(four).

Eq.~\(eightytwo) shows that coarse grainings in which a fixed set of 
 co\"ordinates or their conjugate 
momenta are followed at a sequence of times while all  
others are ignored may be
studied by path integral techniques.  The form of \(seventynine), 
however, allows
an immediate and important distinction to be drawn between co\"ordinate 
coarse grainings and momentum coarse grainings.  Co\"ordinate alternatives
will decohere if $W$ has a positive imaginary part that becomes large
as $x^\prime(t)$ and $x(t)$ are increasingly distinct.  However,
there is no corresponding mechanism leading to the decoherence of
momentum alternatives for this class of models.  

In the case of the
linear models discussed in Section VI, this conclusion may be made more
precise by following a few steps that led to the derivation of the
equations of motion.  Introduce variables for the momenta analogous to
those for the co\"ordinates in \(tend)
$$
\varpi(t) = p^\prime(t) - p(t)\ ,\tag eightythree a
$$
$$
P(t) = \half \left[p^\prime(t) + p(t)\right]\ .\tag eightythree b
$$
The exponent in \(eighty) may now be re\"expressed in terms of 
the variables
of \(tend) and \(eighty) and, after a few integrations 
by parts, put in the form
$$
S_{\rm free} \left[p^\prime(\tau), x^\prime(\tau)\right] + S_{\rm free}
\bigl[p(\tau), x(\tau)\bigr] + W \left[x^\prime(\tau), x(\tau)\right]
$$
$$
=-\xi^\dagger_0 P_0 + \int\nolimits^T_0
dt\biggl\{\varpi^\dagger(t)\Bigl[\dot X(t)-\frac{\partial H}{\partial
P(t)}\Bigr]+\xi^\dagger(t) \Bigl[-\dot P(t) + \frac{\partial H}{\partial
X(t)} + \int\nolimits^T_0 dt^\prime k_R (t, t^\prime) X (t^\prime) \Bigr]
\biggr\}
$$
$$
+\frac{i}{4} \int\nolimits^T_0 dt \int\nolimits^T_0 dt^\prime
\xi^\dagger(t) k_I (t, t^\prime) \xi(t^\prime)\ .\tag eightyfour
$$
The terms in square brackets on the right hand side of \(eightyfour) are
Hamilton's equations of motion augmented by terms describing the
additional forces arising from the interaction of the $(p,x)$ subsystem
with the rest.  However, we cannot conclude that these equations of
motion are valid for phase-space coarse grainings.  The last term in
\(eightyfour) makes the integrand of \(eightytwo) small when $\xi\not=0$ and
thereby enforces the decoherence of co\"ordinate alternatives.  However,
the absolute value of that integrand is {\it uniformly} distributed in
$\varpi$. Unless the integration over the other variables makes
\(eightyfour) small for $\varpi\not= 0$ there will be 
no decoherence of momentum
alternatives.\footnote{$^{12}$}{Such cancellation leading to the
decoherence of momentum alternatives occurs, for example, in the case of
a free, non-relativistic particle. 
The conservation of momentum means that projections onto ranges of
momentum at different times commute and a history composed exclusively
of such projections will automatically decohere. We
thank J.~Halliwell for pointing this out to us.}
  In cases when momentum alternatives cannot even be assigned
probabilities there is {\it a fortiori} no issue of whether the probability 
is high for
their correlation in time by an effective equation of motion.
The origin of this distinction between co\"ordinates and momenta in this
model is, as has been remarked by many authors, that the interaction
Hamiltonian is local in co\"ordinates but not in momentum.

This analysis of both co\"ordinate and momentum coarse grainings
stresses an important if familiar point.  Two coarse grainings that
would be essentially indistinguishable classically may have very
different properties quantum-mechanically.  The present example 
illustrates this with coarse grainings by momentum, $p(t)$, 
and by the difference
in co\"ordinates at two nearby times $M[x(t+\epsilon)-
x(t)]/\epsilon$.  As we have seen, coarse grainings by the
momentum even at one time do not decohere, while coarse grainings by the
co\"ordinates at times separated by intervals longer than the decoherence
time-scale set by $k_I$ [\cf \(fourthirtyseven)] 
do decohere.  Thus, these two types
of coarse graining have essentially different properties with respect to
decoherence even though they would be indistinguishable on the basis of
classical physics when the decoherence time-scale is much shorter than
characteristic dynamical time-scales of the classical motion. Evidently,
considerable care is required in identifying the variables through which
quasiclassical behavior is to be defined in quantum mechanics.  At the
present moment it is, therefore, an open question for the
purpose of defining classicality whether a quantity like momentum density is
best defined in terms of time derivatives of fields or in terms of
 differences of
fields over a macroscopically negligible interval.
\vskip .13 in
\centerline{\sl (b) Densities}

The most important generalization of the coarse grainings studied in
this paper is to cases where the variables that are distinguished 
are not limited to
a fixed set of
fundamental co\"ordinates.  
To coarse-grain by the value of the baryon number in
a small spatial volume, for example, is not the same as following
 some particular
subset of the fields of a fundamental field theory.  We shall discuss how, 
in principle,
more general and realistic coarse grainings can be treated 
by techniques analogous to those used to discuss the special cases
of this paper.
Our considerations are essentially formal and we have
not pushed the analysis far enough to derive equations of motion in
these realistic cases.  Our discussion, however, indicates a route by
which that might be accomplished.  

A very general and useful class of
coarse grainings is obtained by partitioning the fine-grained histories
according to the values of functionals of them.  
To illustrate an interesting case in a
manageable notation, we consider a field theory with a single charged
scalar field $\phi(\vec x, t)$.  A set of functionals leading to coarse
grainings 
relevant for the present discussion consists of the values of the 
charge density at time $t$
averaged over a small spatial volume $v$ at different spatial points
$\vec x$:
$$
J^0(\vec x, t) = \frac{1}{v} \int\nolimits_v d^3\xi\ j^0(\vec x + \vec
\xi, t)\tag eightyfive
$$
where $j^0(\vec x, t)$ is the charge density expressed in terms of
$\phi(\vec x, t)$.  The fine-grained histories, $\phi(\vec x, t)$, may
be partitioned into exhaustive and exclusive classes by the values of
$J^0(\vec x, t)$. (Partition by {\it ranges} of values of these averaged
densities would be a further coarse graining.)   
Consider a particular value $\upsilon(\vec x, t)$.  (The
conventional use of $\rho$ for charge density is precluded by its use
here for the density matrix!).  The coarse-grained
history corresponding to $\upsilon(\vec x, t)$ consists of all 
those $\phi(\vec x, t)$
for which the integral \(eightyfive) has this value.

The decoherence functional for a pair of histories coarse-grained 
by particular values of the charge density  is
$$
D\left[\upsilon^\prime (\vec x, t), \upsilon(\vec x, t)\right] = \int
\delta\phi^\prime \int \delta\phi\  \delta \left[\phi^\prime (\vec x, T) -
\phi^\prime (\vec x, T)\right]
$$
$$
\times \delta\left[J^{0\prime} (\vec x, t) - \upsilon^\prime (\vec x,
t)\right] \exp \Bigl\{\frac{i}{\hbar} \left(S[\phi^\prime (\vec x, t)]
- S[\phi (\vec x, t)]\right)\Bigr\} \delta \left[J^0 (\vec x, t) - \upsilon
(\vec x, t)\right]
 \rho \left[\phi^\prime(\vec x, 0), \phi (\vec x, 0)\right]\ .
\tag eightysix
$$
The first $\delta$-functional in \(eightysix) enforces the coincidence of the
histories at the final time $T$ as in \(seventyfive).  
The other
two $\delta$-functionals restrict the fields in
 the integral to the coarse-grained
histories labeled by $\upsilon^\prime(\vec x, t)$ and $\upsilon(\vec x, t)$.
  Of
course, making precise sense of a formal expression like \(eightysix) raises
many mathematical issues that we shall not pursue here.

The decoherence functional $D[\upsilon^\prime (\vec x, t), \upsilon(\vec x,
t)]$ is the generalization of $D[x^\prime(t), x(t)]$ considered in the
simple models of the earlier sections of this paper.  In those cases
a fixed set of variables describing the fine-grained histories of a
single oscillator were distinguished by the coarse graining while the
variables describing the other oscillators were ignored.  In the case of
a coarse graining by the average value of a density over a volume, one
could loosely say that the ``variables'' describing the different field
configurations internal to the volume and consistent with a given average
value are ignored.  Hence there is no subset of the fundamental 
fields or linear
transformation of them that describes these ``internal variables''.
That is why $D[\upsilon^\prime(\vec x, t), \upsilon(\vec x, t)]$ is a
generalization of $D[x^\prime(t), x(t)]$.

The form of $D[\upsilon^\prime(x,t), \upsilon(x,t)]$ may be brought in closer
analogy with \(five) by
representing some of the $\delta$-functionals in \(eightysix) as 
exponential integrals
$$
\delta [J^0 (\vec x,t) - \upsilon(\vec x,t)] = \int \delta y \exp
\ \Bigl\{i\int d^4x\ y(\vec x,t)\  \bigl[J^0(\vec x,t) -  
\upsilon(\vec x,t)\bigr]\Bigr\}\ .\tag eightyseven
$$
Then, eq.~\(eightysix) becomes:
$$
D\left[\upsilon^\prime (\vec x, t), \upsilon (\vec x, t)\right] = \int \delta
y^\prime \int \delta y \int \delta\phi^\prime \int \delta\phi
 \delta [\phi^\prime (\vec x, T) - \phi(\vec x,T)]
\exp\biggl\{\frac{i}{\hbar}\Bigl(S\left[\phi^\prime (\vec x, t)\right] -
S\left[\phi(\vec x, t)\right]
$$
$$
+ \int d^4 x\ y^\prime(\vec x, t) \left[J^{\prime 0} (\vec x, t) 
-\upsilon^\prime
(\vec x, t) \right] - \int d^4 x\ y(\vec x, t) \bigl[J^0 (\vec x, t) -
\upsilon(\vec x,t)\bigr]\Bigr)\biggr\}
\rho \left[\phi^\prime (\vec x, 0), \phi (\vec x,
0)\right]\ .\tag eightyeight
$$
The integral in \(eightyeight) is of a familiar form in a field theory with
sources and should be accessible to standard approximation techniques.
Of course, to study the putative ``hydrodynamic'' variables of a
quasiclassical domain, coarse graining by the densities of other
approximately conserved quantities such as energy, momentum, baryon
number, as well as the averages of long-range fields, must be
considered.  Once the decoherence functional is calculated from
expressions like \(eightyeight), the circumstances in which such coarse-grained
sets of histories decohere can be investigated and their equations of
motion derived by the methods of Section VII.  An important question will
be the closure of any such set of equations of motion.  However, we have
not yet progressed beyond this formal sketch of the route to a
derivation of the classical equations of motion for realistic quantum
systems.
\eject
\centerline{\bf X. Conclusions and Program}

Instead of merely summarizing the results obtained in this article and
the future directions of research sketched in the last Section, let us
relate them to the program (of understanding quantum mechanics and its
relation to quasiclassical experience) in which we have been engaged,
and which involves many elements elucidated by other authors over the
last thirty-five years.  We shall briefly review that program and where
the present work fits into it.

We start with the quantum mechanics of a closed system representing the
universe (with a deliberately simplified treatment of the complications
caused by quantum gravity).  There is a dynamical theory of all the
elementary particles and their interactions, which we take to be
described by an action or a Lagrangian or a Hamiltonian $H$, and also a
Heisenberg density matrix (or initial density matrix) $\rho$.  The
question is how the quasiclassical domain of familiar experience comes
to be an emergent feature of the system characterized by $H$ and $\rho$.

A ``measurement situation'' can often be characterized as one in which
some variable comes into strong correlation with the quasiclassical
domain.  The quasiclassical domain also permits, for certain probability
tracks, certain spatial regions, and certain epochs of time, enough
classical predictability for the evolution of complex adaptive systems
that learn, observe, and record, and also utilize, in some
approximation, the probabilities assigned by quantum mechanics, on the
basis of $\rho$ and $H$, to different alternative coarse-grained
histories. (Presumably, it is the quasiclassical correlations,
representing near certainties, that are easiest to utilize.)
Observations by such complex adaptive systems, which we call
IGUSes (information gathering and utilizing
systems) when they  are functioning as observers,
are then considered to be actual measurements. 

A quasiclassical domain is defined by decoherence, a measure of
classical predictability, and some sort of maximality (such as what we
have called ``fullness'' [\cite{3}]).  The first requirement is decoherence,
that is, enough coarse graining of the alternative histories of the
universe so that there is, exactly or to a very good approximation, no
interference between the alternative coarse-grained histories, as
measured by the decoherence functional.

Decoherence requires coarse graining that goes far beyond the modest
requirement imposed at each instant of time by the uncertainty
principle.  The mechanism of decoherence involves the loss of phase
information as a result of the coarse graining and is associated with
noise that inextricably combines quantum fluctuations and classical
statistical fluctuations, both necessary for the decoherence.

All of that is most obvious, of course, in the instructive, although
oversimplified models in which the coarse grainings include an
average over some fixed set of
variables (the ignored ones ) while following others
(the distinguished ones) more or less coarsely. 
The ignored variables, through their
interactions with the distinguished ones, carry away the phases and are
responsible for the fluctuations.

The fluctuations, of course, cause departures from any effective (or
phenomenological) classical equations of motion for the distinguished
variables.  But a high degree of decoherence requires very large
fluctuations, which threaten to produce very great departures from
classical predictability.  Thus, for the coarse graining to yield a
quasiclassical domain, it is essential that the distinguished variables
carry very high inertia so as to resist most of the large fluctuations
and follow the effective equations of motion with only small deviations
over long stretches of time and with
only occasional large ones.  That high
inertia is achieved by even much coarser graining than was required for
the decoherence itself.

The requirement of fullness, a kind of maximality, was discussed in
[\cite{33}]. For the coarse graining defining a quasiclassical domain to
be an emergent feature of the universe characterized by $H$ and $\rho$
rather than an artifact chosen by some IGUS, it should be as refined a
description of the universe as possible consistent with the requirements
of decoherence and quasiclassicality.  In [\cite{33}] we proposed, for
the case of perfect decoherence, the notion of a ``full'' set of
alternative coarse-grained histories to capture this idea of maximal
refinement.  Any ``full'' set belongs to an equivalence class defined by
a basis in Hilbert space, provided the density matrix $\rho$ corresponds
to a pure state. (If it does not, and especially if it has a great many
eigenstates with non-zero eigenvalues, then the condition of ``strong
decoherence'' discussed in [\cite{33}] is too strong and the associated
discussion of maximality requires modification.)

We have posed the question as to whether there could be various kinds of
essentially inequivalent quasiclassical domains or whether any
quasiclassical domain is more or less equivalent to any other.  The
former case poses some challenging intellectual puzzles, especially if
we imagine IGUSes evolving in relation to each of the essentially
inequivalent quasiclassical domains.

There is, of course, an indication from our everyday experience of some
of the features of a particular quasiclassical domain.  It involves
distinguished variables that are more or less hydrodynamic in character;
they are integrals or averages over small regions of space of conserved
or nearly conserved densities and of fields coupled to those densities.
The regions must be large enough to produce sufficient inertia to resist
most of the fluctuations associated with the coarse graining and small
enough to implement the requirements of ``fullness''.  It is clear,
however, that the distinguished variables cannot be defined once and for
all, but depend on history.  For example, the suitable hydrodynamic
variables under the conditions that prevailed before the condensation of
the solar system involved much bigger volumes than those that were
suitable inside the planets after they were formed.  That is why the
models in which the ignored variables are fixed for all time are only
instructive examples and not general enough for the realistic case.

Our program thus aims at describing a quasiclassical domain with
history-dependent distinguished variables resembling hydrodynamic ones
and obeying effective classical equations of motion apart from small
fluctuations and occasional large ones, some of which result in the need
for redefinition (for later times) of the distinguished variables.

In this article we have, for the most part, confined ourselves to a
model in which the distinguished variables are separated once and for
all from the ignored ones and are also, unlike the hydrodynamic ones,
co\"ordinate variables of the fundamental theory, like modes of a scalar
field, with kinetic energy bilinear in the time derivatives and the rest
of the energy not involving the time derivatives.  We have also
restricted attention to initial density matrices that factor into the
product of a density matrix for the distinguished variables and one for
the ignored ones. With those
simplifying assumptions, we identify the effective classical equations
of motion and the Gaussian part of the noise that disturbs them.

We divided into two parts the quantum-mechanical process of 
prediction for further coarse grainings of the  histories
of the distinguished system.  First, there is the
calculus of amplitudes for histories fine-grained in the distinguished
variables and of the bilinear combinations of amplitudes that
define the decoherence functional.
Second, there are the requirements of decoherence for deriving
consistent probabilities
from these amplitudes
for histories that are further coarse-grained.

We assumed that the requirements of decoherence were satisfied (although
we discussed mechanisms by which this happens)  and examined when the
probabilities for histories favored classical predictability.

We started with a general formulation that applies to fully non-linear
systems.  By performing a  functional Fourier transform on the decoherence
functional with respect to the difference in distinguished co\"ordinates
on the left and right, we introduced a distribution function for the
``total force'' (including the inertial term) that acts on the
distinguished system along its history.  Like the analogous Wigner
distribution at a moment of time, it is not generally positive.

Utilizing the distribution of the ``total force'',
 we were able, even in non-linear cases, to
represent the probabilities of  a decoherent set of coarse-grained
histories as the probabilities of a classical dynamical system governed
by a Langevin equation incorporating history dependent noise.  The
initial conditions of this classical system are distributed according
to the Wigner distribution and the noise according to the distribution
of ``total force'' mentioned above.  Both distributions are generally
non-positive, which distinguishes them from classical ones, 
although the resulting probabilities for decohering sets of
 histories are, of course,
positive.  When the noise is small compared with the inertial term in
the equation of motion, the coarse-grained histories are classically
deterministic.

As suggested by the simple physical picture of phases being carried away
by the interaction of the distinguished subsystem with the ignored
variables, there is a connection between decoherence and noise.  This was
exhibited by writing
the decoherence functional for the distinguished variables 
as an integral of an average density matrix times an
exponential, and expanding the argument of the exponential in
powers of the generalized vector $\xi(t)$ that measures the difference
between the distinguished variables as functions of time on the left-
and right-hand sides of the decoherence functional.  After a partial
integration, the linear term has the form
$i/\hbar\int\xi^\dagger(t){\cal E}(t)dt$ and the equation ${\cal E}(t) =
0$ is precisely the effective classical equation of motion for $X(t)$,
which is the average of the distinguished co\"ordinate variables on the
left and on the right.  The second term in the expansion of the argument
of the exponential is
$-1/4\hbar\int\int\xi^\dagger(t)K_I(t,t^\prime)\xi(t^\prime)dtdt^\prime$,
where $K_I$ is the positive kernel describing the Gaussian noise, and
can be thought of as the self-correlation (depending in general on
$X(t)$) of a Langevin force $\ell(t)$ added to the equation of motion.

There are, in general higher terms in the expansion, corresponding to
the fact that coarse-grained quantum mechanics is not exactly equivalent
to effective classical equations of motion accompanied by Gaussian
noise.  However, we are interested in the case of approximate
decoherence of the coarse-grained histories, meaning that $\xi(t)$ is
mostly confined to very small values.  When $K_I$ is very large, this is
achieved, provided that the higher terms are not somehow still more
important.  However, large $K_I$ means large noise and that is the
connection between decoherence and noise.
We, therefore, also impose the requirement
 of a very large inertia matrix
$M$, so that the huge noise is mostly resisted and the effective
classical equations of motion are followed with high probability by the
distinguished variables. 

For pedagogical reasons, we started by reviewing the completely linear
case, studied extensively by Feynman and Vernon, Caldeira and Leggett,
Unruh and Zurek, etc.  There the noise correlation $k_I(t, t^\prime)$ is
just a numerical function and the effective equation of motion for $X(t)$
is linear, with a frictional force $\int k_R(t,t^\prime)
X(t^\prime)dt^\prime$ that is in general nonlocal in time. When the
effective density matrix is diagonal in the energy of the ignored
variables, then the numerical functions $k_I$ and $k_R$ depend on
$t-t^\prime$ only and are related to each other by the
fluctuation-dissipation theorem.  In the Fokker-Planck limit, $k_I$ is
proportional to $\delta(t-t^\prime)$ and $k_R$ to
$\delta^\prime(t-t^\prime)$ and the co\"efficients are related by a very
simple fluctuation-dissipation relation.

We then generalized to the nonlinear case discussed above. {\it That is the
principal content of this paper, and the work can be applied to many
problems, as in [\cite{1}] and [\cite{27}].}

We also treated more general coarse grainings, in which both generalized
co\"ordinates and generalized momenta are utilized and what had been a
Lagrangian formulation turns into what is essentially a Hamiltonian
formulation.

Finally, we pointed out the desirability of removing the remaining two
unrealistic conditions, that the distinguished variables have their time
derivatives occurring in the action only in a bilinear kinetic energy
term, and that the distinguished variables be defined independent of
history.  We are studying the mathematical problems posed by these
generalizations.  If the first difficulty can be overcome, then one
should be able to treat the effective classical equations of motion and
the Langevin force for distinguished variables that include the
hydrodynamic quantities of familiar quasiclassical experience.

If the second and greater difficulty can be overcome, and history
dependence introduced into the coarse graining, then we may begin to
tackle the deep problem of introducing individuality into quantum
mechanics.  Actual alternative histories deal, of course, in large part
with individual objects such as our galaxy, the sun, the earth,
biological organisms on the earth, and so forth.  Yet discussions of
quantum mechanics up to now have typically treated such individual
objects only as external systems, labeled as ``observers'' and ``pieces
of apparatus''.  If history dependence can be properly introduced into
the explicit treatment of quantum mechanics, then we may be able to
handle individuality with the care that it deserves.
\taghead{A.}
\vskip .26 in
\centerline{\bf Appendix  --- Path Integrals}
\vskip .13 in
In the body of this paper we have formally manipulated 
both configuration space
and phase space path integrals. 
  We define those integrals more carefully in this
Appendix and use the definition to show that the formal manipulations
we have used are legitimate.
\vskip .13 in
\centerline{\sl (a) Phase Space Integrals}
\vskip .13 in
We begin by deriving an explicit expression for the most general path
integral occurring in the decoherence functional for sets of histories
coarse-grained both by regions of configuration space and regions of
momentum space, as discussed in Section IX.  To keep the notation
manageable, we shall assume for the moment that we are considering a
one-dimensional system with co\"ordinate $q$, momentum $\pi$, and a
Hamiltonian of the form
$$
H(\pi, q) = \frac{\pi^2}{2M} + V(q, t)\ .\tag eightynine
$$
The generalization to larger-dimensional configuration spaces  is
obvious.  A set of histories consisting of chains of just two
projections, one on  momentum space region $\{\widetilde
\Delta_{\alpha_1}\}$ at time $t_1$ and the other on  configuration
space region $\{\Delta_{\alpha_2}\}$ at time $t_2$, will suffice to
illustrate the construction in more general situations.  The important
matrix elements for the construction of the decoherence functional are
of the form \(seventysix).  They may be expressed in the Schr\"odinger
picture as
$$
\bigl\langle q_f T|C_\alpha | q_0 0\bigr\rangle \equiv \bigl\langle q_f
T | P^q_{\alpha_2} (t_2) P^\pi_{\alpha_1} (t_1) | q_0 0 \bigr\rangle
$$
$$
=\bigl\langle q_f | e^{-iH(T-t_2)/\hbar} P^q_{\alpha_2}
e^{-iH(t_2-t_1)/\hbar} P^\pi_{\alpha_1} e^{-iHt_1/\hbar} |
q_0\bigr\rangle\ . \tag ninety
$$
Here we have written $P^q_{\alpha_2}$ and $P^\pi_{\alpha_1}$ to recall
explicitly that we are dealing with  projections onto a configuration
space region $\Delta_{\alpha_2}$ and on a momentum space region
$\widetilde\Delta_{\alpha_1}$ respectively. 

We now cast the right-hand side of \(ninety) into a phase space
path integral form.  We shall be brief because the construction is a
standard one [\cite{28, 29}]. We divide the interval [$0,T$] up into $N$ equally
spaced time slices $t_0 = 0, t, t, \cdots, t_{N-1},
t_N = T$ with an interval $\epsilon = T/N$ between them.  We assume
that the times $t_1$ and $t_2$ coincide with two of these slices for a
sequence of $N$'s tending to infinity.  
Let $K_1$ and $K_2$ be the labels of the slices corresponding to $t_1$
and $t_2$ respectively, 
understanding that
these are functions of $N$.  Write the propagators in \(ninety) as
the product of an appropriate number of factors of
$\exp(-iH\epsilon/\hbar)$.  Between these factors, on each time slice
except the first, insert a resolution of the identity of the form
$$
\int d\pi_k \int dq_k | q_k><q_k|\pi_k><\pi_k | = I\ .
\tag ninetyone
$$
The result is the following expression for the matrix element in
\(ninety):
$$
\eqalignno{
&\bigl\langle q_f T|C_\alpha | q_0 0\bigr\rangle= \int \prod\limits^N_{k=1}
d\pi_k dq_k
\bigl\langle q_f | q_N\bigr\rangle\bigl\langle q_N | \pi_N\bigr\rangle
\bigl\langle\pi_N| e^{-iH\epsilon/\hbar}|q_{N-1}\bigr\rangle\cr
&\times \bigl\langle q_{N-1} | \pi_{N-1}\bigr\rangle\bigl\langle \pi_{N-1} |
e^{-iH\epsilon/\hbar} | q_{N-2}\bigr\rangle\cdots
\bigl\langle q_{K_2} | P^q_{\alpha_2} |\pi_{K_2} \bigr\rangle
\bigl\langle\pi_{K_2} | e^{-iH\epsilon/\hbar}|q_{K_2-1}\bigr\rangle\cdots\cr
&\times \bigl\langle q_{K_1} | P^\pi_{\alpha_1} | \pi_{K_1} \bigr\rangle
\bigl\langle \pi_{K_1} | e^{-iH\epsilon/\hbar} | q_{K_1-1}
\bigr\rangle\cdots
\bigl\langle q_1 | \pi_1\bigr\rangle \bigl\langle \pi_1 |
e^{-iH\epsilon/\hbar} | q_0\bigr\rangle\ .&(ninetytwo)\cr}
$$

Now note the following relations:
$$
\langle q|\pi\rangle = e^{i\pi q/\hbar} /(2\pi\hbar)^\half\ ,\tag ninetythree
$$
$$
\langle q | P^q_{\alpha_2} | \pi \rangle = e_{\alpha_2} (q) e^{i\pi
q/\hbar}/(2\pi\hbar)^\half\ ,\tag ninetyfour
$$
$$
\langle q | P^\pi_{\alpha_1} | \pi\rangle = e_{\alpha_1} (\pi) e^{i\pi
q/\hbar} / (2\pi\hbar)^\half\ ,\tag ninetyfive
$$
where $e_\alpha(x)$ is unity for $x$ in the interval $\alpha$ and zero
outside it.  (The symbol $\pi$ stands for momentum unless it occurs in the 
combination
$2\pi\hbar$.) Further, to first order in small $\epsilon$,
$$
\bigl\langle\pi | e^{-iH\epsilon/\hbar} | q\bigr\rangle \approx e^{-iH(\pi,
q)\epsilon/\hbar} e^{-i\pi q/\hbar}/(2\pi\hbar)^\half\tag ninetysix
$$
where $H(\pi, q)$ is the {\it function} given by \(eightynine).
Inserting these relations in \(ninetytwo), noting that there is a
$\delta$-function that identifies $q_N$ and $q_f$, we find the following
expression, which is exact as $N\to \infty$,
$$
\eqalignno{
&\bigl\langle q_f T | C_\alpha | q_0 0 \bigr\rangle =
\lim\limits_{N\to\infty} \int \frac{d\pi_N}{2\pi\hbar}
\ \left(\prod\limits^{N-1}_{k=1}\ \frac{d\pi_k dq_k}{2\pi\hbar}\right)
 e_{\alpha_2} (q_{K_2}) e_{\alpha_1} (\pi_{K_1})\cr
&\times \exp \biggl\{ \frac{i}{\hbar} \sum\limits^N_{j=1} \epsilon
\ \Bigl[ \pi_j \left(\frac{q_j-q_{j-1}}{\epsilon}\right) - H\left(\pi_j,
q_{j-1}, t_{j-1}\right) \Bigr]\biggr\}&(ninetyseven)\cr}
$$
This is the definition of the phase space path integral that we have
written in Section IX as
$$
\bigl\langle q_f T | C_\alpha | q_0 0 \bigr\rangle = \int_\alpha \delta
\pi \delta q
 \exp \biggl\{\frac{i}{\hbar} \int^T_0 dt \Bigl[\pi(t)\dot q(t) -
H\bigl(\pi(t), q(t), t\bigr)\Bigr]\biggr\}\ .\tag ninetyeight
$$
It is an integral over phase space paths in the class specified by the
coarse graining, that is, over paths which pass through the momentum 
space region
$\widetilde\Delta_{\alpha_1}$ at time $t_1$ and configuration space
region $\Delta_{\alpha_2}$ at time $t_2$.
\vskip .13 in
\centerline{\sl (b) Path Integrals for the Decoherence Functional}

When the coarse graining is only by regions of configuration space, and
there is no coarse graining by momentum space, then the Gaussian
integrals over the $\pi_k, k=1, \cdots, N$ may be carried out in expressions
like \(ninetyseven) leading to Lagrangian path integrals for the matrix
elements $\langle q_f T| C_\alpha | q_0 0\rangle$ corresponding to
individual histories.  For example, a history defined by a sequence of
$q$-intervals $\Delta_{\alpha_1}, \Delta_{\alpha_2}, \cdots,
\Delta_{\alpha_n}$ at times $t_1, \cdots, t_n$ would have
$$
\bigl\langle q_f T | C_\alpha | q_0 0\bigr\rangle = \lim\limits_{N\to\infty}
\int \left[\prod\limits^{N-1}_{k=1}\ \left(\frac{M}{2\pi
i\epsilon\hbar}\right)^\half dq_k\right]\ e_{\alpha_n} (q_{K_n}) \cdots
e_{\alpha_1} (q_{K_1})
$$
$$
\times   \exp \left\{\frac{i}{\hbar}\ \sum\limits^{N-1}_{j=0}\epsilon
\left[\half M\left(\frac{q_{j+1}-q_j}{\epsilon}\right)^2 -
V(q_j, t_j)\right]\right\}\ . \tag ninetynine
$$
Here, $K_i$ is the label of the slice corresponding to $t_i$ and, as
before, $q_N=q_f$.

Eq.~\(ninetynine) is the sum-over-paths usually written
$$
\bigl\langle q_f T | C_\alpha | q_00\bigr\rangle = \int_{[q_0\alpha
q_f]} \delta q 
\  \exp \bigl(i S[q(t)]/\hbar\bigr)  \tag hundred
$$
where $S[q(t)]$ is the Lagrangian form of the action corresponding to
the Hamiltonian \(eightynine).
$$
S[q(t)] = \int^T_0 dt \left[\half M\dot q^2 - V(q, t)\right]\ .
\tag hundredone
$$
In an even more compact notation, where, as in \(one),  there is an 
integration over
$q_0$ and $q_f$,  we have denoted the restrictions on the
range of integration arising from the coarse graining by an unadorned
subscript $\alpha$ on the integral sign.

The path integral for the decoherence functional consists of two
multiple integrals like \(ninetynine) over two polygonal paths $\{q_k\}$
and $\{q^\prime_k\}$ with additional integrals over their initial and
final endpoints weighted by the initial density matrix and final
$\delta$-function respectively.  When, as discussed in Section III, the
action is of a suitable form and the coarse graining constrains only a fixed
subset $x^a$ of the variables of configuration space, the integrals over
the remaining $\{Q^A_k\}$ may be carried out yielding a path integral
involving an influence phase \(five).  To better understand how formal
manipulations are carried out on that path integral we shall now write out
the explicit time-slicing implementation of it following from
\(ninetynine). To keep the notation manageable we shall consider the case
where the coarse graining refers to a single co\"ordinate $x$ and limit
attention to the explicitly linear problem discussed in Section VI having
a quadratic influence phase given by \(twelve).  The generalizations of
this case should be obvious.  The integral \(five) for the decoherence
functional is then explicitly:
$$
D(\alpha^\prime, \alpha) = \lim\limits_{N\to\infty} \int
\biggl[\left(\frac{M}{2\pi i\epsilon\hbar}\right)^{N-1}
\prod\limits^N_{k=0}\ dx^\prime_k dx_k\biggr]
$$
$$
\times \delta (x^\prime_N-x_N)\ E_{\alpha^\prime\alpha} (x^\prime_j,
x_j) \exp \bigl[i A (x^\prime_j, x_j)/\hbar)\bigr]
\ \widetilde\rho (x^\prime_0, x_0)\tag hundredtwo
$$
where the functions $E_{\alpha^\prime\alpha}$ and $A$ are defined as
follows:  The function $E_{\alpha^\prime\alpha}$ is
$$
E_{\alpha^\prime\alpha} (x^\prime_j, x_j) = \prod\limits^n_{i=1}
\ e_{\alpha^\prime_i} (x^\prime_{K_i}) e_{\alpha_i} (x_{K_i})\tag
hundredthree
$$
and enforces the constraints of the coarse graining.  The exponent, $A$,
is the discrete form of \(eleven) and \(twelve), viz.
$$
A(x^\prime_j, x_j) = \sum\limits^N_{j=1} \epsilon\biggl[\half
M\left(\frac{x^\prime_j-x^\prime_{j-1}}{\epsilon}\right)^2 - \half
Kx^{\prime 2}_j -\half M\left(\frac{x_j-x_{j-1}}{\epsilon}\right)^2 +
\half Kx^2_j\biggr]
$$
$$
+ \half \sum\limits^N_{j=1} \sum\limits^j_{\ell=1}\
\epsilon^2(x^\prime_j-x_j) \left[k_R (j,\ell) (x^\prime_\ell + x_\ell) +
ik_I (j, \ell) (x^\prime_\ell - x_\ell)\right]\ .\tag hundredfour
$$
Here, $k_R(j,\ell)$ and $k_I(j,\ell)$ are the real and imaginary parts
of the function $k(t^\prime, t)$ evaluated on the discrete time slices.
In passing from \(ninetynine) to \(hundredthree) we have made use of the fact
that in the limit $N\to \infty$ it makes no difference in the Lagrangian path
integral whether the integral $\int V(x(t))dt$ between time slices is
approximated using the value of $V$ at the start or end of the interval.
The form \(hundredfour) is slightly more convenient for what follows.

We now change variables in the multiple integral \(ninetynine) from
$\{x^\prime_k, x_k\}$ to the discrete versions of \(fifteen) 
$$
\xi_k = x^\prime_k - x_k,\ X_k= \half (x_k + x_k)\ .\tag hundredfive
$$
The Jacobian is unity on each slice so that \(hundredtwo) becomes
$$
D(\alpha^\prime,\alpha) = \lim\limits_{N\to\infty}
\int\biggl[\left(\frac{M}{2\pi i\epsilon\hbar}\right)^{N-1}
\prod\limits^N_{k=0}\ dX_k d\xi_k\biggr]
$$
$$
\times \delta(\xi_N) E_{\alpha^\prime\alpha} \left(X_j +
\xi_0/2, X_j-\xi_0/2\right) \exp \left[i A
(X_j, \xi_j)/\hbar)\right]\ \widetilde\rho \left(X_0 + \xi_0/2, X_0 -
\xi_o/2\right)\ .\tag hundredsix
$$
After a little algebra the exponent may be written
$$
A(X_j, \xi_j) = - \xi_0 M\left(\frac{X_1 - X_0}{\epsilon}\right) +
\sum\limits^N_{j=1}\ \epsilon \xi_j\left[- \left(\frac{X_{j+1} -
2X_j-X_{j-1}}{\epsilon^2}\right) +
 KX_j\right]
$$
$$
+\frac{1}{4}\ \sum\limits^N_{j=1}\sum\limits^N_{\ell=1}
\ \epsilon^2\xi_j \left[k_R(j,\ell) X_\ell + ik_I (j,\ell)
\xi_\ell\right]\ .\tag hundredseven
$$
Eq.~\(hundredseven) is the discrete analog of \(sixteen) and shows
precisely how the second and other derivatives of the path in that
expression are to be interpreted in a time-slicing representation.

We next assume decoherence and carry out the integration over the
$\xi_k$'s neglecting the constraints of the coarse graining as discussed
in Section III.  Note that $A$ in \(hundredseven) depends on $\xi_0$ only
through the first term and that $\xi_N=0$ because of the
$\delta$-function in \(hundredtwo).  The result for the diagonal elements
of the decoherence functional is explicitly
$$
p(\alpha) = \lim\limits_{N\to\infty} \int \biggl[\left(\frac{M}{2\pi
i\epsilon\hbar}\right)\left(\frac{M}{2\pi i \epsilon^2}\right)^{N-2}
\prod\limits^N_{k=0}\ dX_k\biggr]\ \left[{\rm det}
\left(\frac{k_I}{4\pi}\right)\right]^{-\half}\ E_{\alpha^\prime\alpha}
(X_j, X_j)
$$
$$
\times\exp\biggl[-\frac{1}{\hbar}
\sum\limits^{N-1}_{j=1}\sum\limits^{N-1}_{\ell=1}\epsilon^2\ e_j(X_m)
k^{\rm inv}_I (j,\ell)\ e_\ell(X_m)\biggr] w\left(X_0,
M\left(\frac{X_1-X_0}{\epsilon}\right)\right)\ .\tag hundredeight
$$
In this expression ${\rm det}\ k_I$ and $k^{\rm inv}_I$ are the
determinant and inverse respectively of the $(N-1)\times(N-1)$ matrix
$k_I(j,\ell),\ j,\ell=1, \cdots, N-1$. The quantity $e_k$ is the discrete
version of the equation of motion \(nineteen), namely
$$
e_j(X_\ell) = -M \left(\frac{X_{j+1} - 2X_j +
X_{j-1}}{\epsilon^2}\right) - KX_j - \sum\limits^N_{\ell=1}
\epsilon\ k_R (j, \ell) X_\ell \tag hundrednine
$$
and $w(X, P)$ is the Wigner function defined in \(tenb).
Eq.~\(hundredeight) contains the precise measure for the path integral
\(eighteen) and the precise meaning of the integral over the Wigner
function in it.
\vskip .13 in
\centerline{\sl (c) Functional Fourier Transforms}

In Section V we utilized a functional Fourier transform of the
decoherence functional to define a distribution functional for the total
force.  Here, we offer a more explicit definition of what such
transforms mean.  We consider the case of one-dimensional paths for
simpilicity. Consider a functional $D[\xi(\tau)]$.  On paths that are
piecewise linear between time slices $t_0, t_1, \cdots t_N = T$ this
defines a {\it function}, $D(\xi_0, \xi_1, \cdots \xi_N)$ of the values
that $\xi(\tau)$ assumes on these slices.  This function may be Fourier
transformed in the following way
$$
G\left(R_1, \cdots R_n; \xi_0\right) = {\cal N}\ \int
\left(\prod\limits^N_{k=1}\ d\xi_k\right)\ \exp\left(-\frac{i}{\hbar}
\ \sum\nolimits^N_{k=1}\ \epsilon\xi_kR_k\right)\ D\left(\xi_0, \xi_1,
\cdots \xi_N\right) \tag hundredten
$$
where $\epsilon = t/N$ is the spacing between the time slices and ${\cal
N}$ is a
normalizing factor.  The
inverse of this is
$$
D\left(\xi_0, \xi_1, \cdots, \xi_N\right) = {\cal N}^{-1}\
\int\left[\prod\limits^N_{k=1}\ \left(\frac{dR_k}{2\pi}\right)\right] 
\ \exp\left(\frac{i}{\hbar}\ \sum\limits^N_{k=1}\ \epsilon
\xi_kR_k\right)\ G\left(R_1, \cdots R_N; \xi_0\right)\ .
\tag hundredeleven
$$
It is this latter expression, in the limit where the number of slices,
$N$, goes to infinity that gives an explicit meaning to the functional
Fourier transform that we wrote in \(teng) as
$$
D[\xi(\tau)] = \int\ \delta R\exp\left[\frac{i}{\hbar}\ \int\nolimits^T_0
\ dt\xi(t)R(t)\right]\ G\left[R(\tau), \xi_0\right]\ .
\tag hundredtwelve
$$
In particular, \(hundredeleven) defines the measure, $\delta R$.  Note
that because of the way that we have differenced the exponent in
\(hundredtwelve), $\xi_0$ remains untransformed and occurs on both sides
of \(hundredeleven).

The normalizing factor ${\cal N}$ is physically arbitrary and will
cancel from physical expressions such as the definition of the equation
of motion in \(tenj) and \(tenk).  However, mathematically, it must be
chosen carefully in order that expressions like \(hundredten) exist in
the limit $N\to\infty$.  For example, when the decoherence functional is
given as in \(fifteen) and \(sixteen) for linear problems the 
normalizing factor would
depend on $K_I$ and certainly on $\epsilon$.  We assume that such a
normalizing factor always exists for interesting cases.
\vskip .13 in
\centerline{\sl (d) An Operator Expression for the Influence Functional}

The influence functional $\exp(iW[x^\prime(\tau), x(\tau)]/\hbar)$ was
defined by the path integral \(three) but can be represented as the
operator expression \(fortythree) when the initial density matrix
factors as in \(six).  We now spell out the details of this
connection.  By inserting complete sets of states of the Hilbert space
${\cal H}_Q$ that are eigenfunctions of the co\"ordinates $Q^A$, the
right hand side of \(fortythree) can be written
$$
Sp\Bigl[U_{T,0} \left[x^\prime(\tau)\right]\ \rho_B\ U^\dagger_{T,0}
\bigl[x(\tau)\bigr]\Bigr\} = \int dQ^\prime_f\int dQ_f\int
dQ^\prime_0\int dQ_0\delta\left(Q^\prime_f-Q_f\right)
$$
$$
\left\langle Q^\prime_f\bigr|U_{T,0}\left[x^\prime(\tau)\right]\bigl|Q^\prime_0
\right\rangle\ \left\langle Q_0 \bigr| U^\dagger_{T,0}\bigl[x(\tau)\bigr]
\bigl| Q_f\right\rangle\ .\tag hundredthirteen
$$
But $\langle Q_f | U_{T,0}\bigl[x(\tau)\bigr] | Q_0\rangle$ is just the
propagator in the Hilbert space ${\cal H}_Q$ corresponding to unitary
evolution by the time-dependent Hamiltonian $H_Q(x(t))$ [cf. \(fortyone)]
over the time interval $[0,T]$.  This propagator has an elementary path
integral representation whose derivation we have reviewed in subsection
(b) of this Appendix [cf. \(hundred)].  It is
$$
\Bigl\langle Q_f|U_{T,0} [x(\tau)] |Q_0\Bigr\rangle =
\int\nolimits_{[Q_0, Q_f]}\delta Q\exp\Bigl(iS_Q\bigl[x(\tau),
Q(\tau)\bigr]/\hbar\Bigr)\tag hundredfourteen
$$
where $S_Q[x(\tau), Q(\tau)]$ is the action \(thirtyeight) that
corresponds to the Hamiltonian $H_Q(x(t))$.  The path integral is over
paths that start at time $t=0$ at $Q_0$ and proceed to $Q_f$ at time $T$
and are other wise unrestricted. Inserting \(hundredfourteen) into the
right hand side of \(hundredthirteen), noting [eq \(fortytwo)] that
$\langle Q^\prime_0|\rho_B|Q_0\rangle \equiv \rho_B (Q^\prime_0, Q_0)$,
and using \(thirtyeight) for $S_Q[x(\tau), Q(\tau)]$, we recover the
expression \(three) that defines the influence functional
$\exp(iW[x^\prime(\tau), x(\tau)]/\hbar)$.  Thus the identity \(fortythree)
is verified.
\vskip .26 in
\centerline{\bf Acknowledgements}
\vskip .26 in
Part of this work was carried out while the authors were in residence at
the Aspen Center for Physics.  The work of MG-M was supported in part by
DOE contract DE-AC-03-81 ER40050 and by the Alfred P.~Sloan Foundation.
That of JBH was supported in part by NSF grant PHY90-08502.

\references

\refis{1} T. Brun, in preparation.

\refis{2} M. Gell-Mann and J.B. Hartle in {\sl Complexity, Entropy,
and the Physics of Information, SFI Studies in the Sciences of
Complexity}, Vol.  
VIII, ed. by W. \.Zurek,  Addison Wesley, Reading (1990) or in 
{\sl Proceedings of
the 3rd 
International Symposium on the Foundations of Quantum Mechanics in the Light of
New Technology} ed. by S. Kobayashi, H. Ezawa, Y. Murayama,  and S. Nomura,  
Physical Society of Japan, Tokyo (1990). 

\refis{3} J.B. Hartle, in {\sl Quantum Cosmology and Baby
Universes}, ed. by S. Coleman, J. Hartle, T. Piran, and S. Weinberg,
World Scientific, Singapore (1991).

\refis{5} R.  Griffiths, \journal J. Stat. Phys., 36, 219, 1984.

\refis{6} R. Omn\'es, \journal J.~Stat.~Phys., 53, 893, 1988; \journal
ibid., 53, 933, 1988;  \journal
ibid., 53, 957, 1988; \journal ibid., 57, 359, 1989; 
\journal Rev. Mod. Phys., 64, 339, 1992.

\refis{7} E. Joos and H.D. Zeh,  \journal Zeit. Phys., B59, 223, 1985.

\refis{8} L.P.~Kadanoff and P.C.~Martin, \journal Ann. Phys. (N.Y.),
24, 419, 1963.

\refis{9} D.~Foster, {\sl Hydrodynamic Fluctuations, Broken Symmetry,
and Correlation Functions}, Addison-Wesley, Redwood City, (1975).

\refis{10} J.~Lebovitz and H.~Spohn, \journal J.~Stat.~Phys., 28, 539,
1982; \journal ibid., 29, 39, 1982.

\refis{11}  A. Caldeira and A. Leggett, \journal Physica, 121A, 587,
1983.

\refis{12} R.P. Feynman  and J.R. Vernon, \journal Ann. Phys. (N.Y.),
24, 118,
1963.

\refis{13} C.W.~Gardiner, {\sl Quantum Noise}, Springer-Verlag, Berlin
(1992).

\refis{14} R. Kubo, M.~Toda, N.~Hashitsume, {\sl Statistical Physics
II}, Springer Verlag, Berlin (1978) p. 167ff.

\refis{15} H.~Zeh, \journal Found. Phys., 1, 69, 1971.

\refis{16} W.~\.Zurek, 
\journal Phys. Rev., D24, 1516, 1981; \journal ibid.
D26, 1862, 1981; and in {\sl Proceedings of the NATO Workshop on the
Physical Origins of Time Assymmetry}, Mazagon Sapin, Sept 30--Oct 4,
1991, ed.~by J.~Halliwell, J.~P\'erez-Mercader, and W.~\.Zurek, Cambridge
University Press, Cambridge (1992).

\refis{17} O.~K\"ubler and H.D.~Zeh, \journal Ann. Phys. (NY), 76, 405,
1973.

\refis{18} A.~Albrecht, {\sl Identifying Decohering Paths in Closed
Quantum Mechanical Systems} (Fermilab preprint).

\refis{19} M.~Gell-Mann and J.B.~Hartle (forthcoming).

\refis{20} S.R.~deGroot and L.G.~Suttorp, {\sl Foundations of
Electrodynamics}, North Holland, Amsterdam (1972) Chap.~VI.

\refis{20a}  W.~\.Zurek,  in {\sl Non-Equilibrium Quantum
Statistical 
Physics}, ed. by G. Moore  and M. Scully,  Plenum Press, New York
(1984). 

\refis{21} M.~Hillery, R.F.~O'Connell, M.O.~Scully, and E.P.~Wigner,
\journal Phys. Reports, 106, 121, 1984.

\refis{22} R.P. Feynman,  and A. Hibbs, {\sl Quantum Mechanics 
and Path
Integrals}, McGraw-Hill, New York (1965).

\refis{24} M.~Gell-Mann and J.B.~Hartle in {\sl Proceedings of the 1st
International A.D.~Sakharov Conference on Physics}, Moscow USSR, May
27--31, 1991, Nova Science Publishers, New York (1992) or in {\sl
Proceedings of the NATO Workshop on the Physical Origins of Time
Asymmetry} Mazagon, Spain, September 30--Oct 4, 1991, ed.~by
J.~Halliwell, J.~Perez-Mercader, and W.~\.Zurek, Cambridge University
Press, Cambridge (1992).

\refis{25} R.~Zwanzig, \journal J. Stat. Phys., 9, 215, 1973.

\refis{26} S.~Ryang and T.~Saito, \journal Prog. Theor. Phys., 71,
1108, 1984.

\refis{27} T.~Brun, Quasiclassical Equations of Motion for Non-Linear
Brownian Systems. (Caltech preprint)

\refis{28} R.P. Feynman, \journal Phys. Rev., 84, 108, 1951.

\refis{29} C.~Garrod, \journal Rev. Mod. Phys., 38, 483, 1966.

\refis{32} R.P.~Feynman in {\sl Quantum Implications: Essays in Honor of
David Bohm} ed.~by B.J.~Hiley and F.D.~Peat, Routledge \& Kegan Paul,
London (1987).

\refis{33} M.~Gell-Mann and J.B.~Hartle, {\it Alternative Decohering
Histories in Quantum Mechanics}, in the {\sl Proceedings of
the
25th International Conference on High Energy Physics, Singapore, August
2--8, 1990}, ed.~by K.K.~Phua and Y.~Yamaguchi (South East Asia
Theoretical
Physics Association
and Physical Society of Japan) distributed by World Scientific,
Singapore (1990).

\refis{34} B.-L. Hu, J.P. Paz, and Zhang, \journal Phys. Rev., D45,
2843, 1992; and {\sl Quantum Brownian Motion in a General Environment
II} (University of Maryland preprint).

\refis{35} W.~\.Zurek, \journal Physics Today, 44, 36, 1991.

\refis{38} E. Fitzgerald, Rub\'aiy\'at of Omar Khayy\'am (1859).

\refis{36} C.~Morais Smith and A.~Caldeira, \journal Phys. Rev. a, 41,
3103, 1990.

\refis{37} J.S.~Bell, \journal Physics, 1, 195, 1964.

\refis{40} H.~Dowker and J.~Halliwell, \journal Phys. Rev., D46, 1580,
1992. 

\refis{41} R.~Balian, Y.~Alhassid, and H.~Reinhardt, \journal Phys.
Rep., 131, 1, 1986.

\refis{42} S.~Kochen, in {\sl Symposium on the Foundations of Modern
Physics}, ed.~by P.~Lahti and P.~Mittelstaedt, World Scientific,
Singapore (1985).

\refis{43} Y.~Aharonov, P.~Bergmann, and J.~Lebovitz, \journal Phys.
Rev., B134, 1410, 1964.

\refis{44} K.~Husimi, \journal Proc. Phys. Math. Soc. Japan, 22, 264,
1940.

\refis{45} J.~Halliwell, \journal Phys. Rev., D46, 1610, 1992.

\endreferences

\endit
\end